\documentclass[11pt,a4paper]{article}
\usepackage{float}
\usepackage{jcappub}
\usepackage{bm}
\usepackage{bbm}
\usepackage{amsmath}
\usepackage{etoolbox}
\usepackage{mhchem}
\usepackage{gensymb}
\usepackage{color}
\usepackage{appendix}
\usepackage{comment}
\usepackage{hyperref}

\newcommand{\GeV}{\,{\rm GeV}}
\newcommand{\TeV}{\,{\rm TeV}}

\def\circa#1{\,\raise.3ex\hbox{$#1$\kern-.75em\lower1ex\hbox{$\sim$}}\,}

\newcommand{\DM}{{\rm DM}}

\newcommand{\beq}{\begin{equation}}
\newcommand{\eeq}{\end{equation}}
\newcommand{\MeV}{\,{\rm MeV}}

\definecolor{CommentGreen}{rgb}{0.08, 0.35, 0.1}

\usepackage{etoolbox}

\makeatletter
\gdef\@fpheader{}
\makeatother

\begin{document}

\makeatletter

\title{Cosmic Ray-Dark Matter Interactions shining as PeV photons}

\author[a,b]{Pedro de la Torre Luque,}
\author[c,d]{Jorge Terol Calvo}

\affiliation[a]{Departamento de F\'{i}sica Te\'{o}rica, M-15, Universidad Aut\'{o}noma de Madrid, E-28049 Madrid, Spain}
\affiliation[b]{Instituto de F\'{i}sica Te\'{o}rica UAM-CSIC, Universidad Aut\'{o}noma de Madrid, C/ Nicol\'{a}s Cabrera, 13-15, 28049 Madrid, Spain}

\affiliation[c]{Departament de Física, Universitat d’Alacant, 03690 Alacant, Spain}
\affiliation[d]{Istituto Nazionale di Fisica Nucleare, Sezione di Torino, via P. Giuria 1, I--10125 Torino, Italy}

\emailAdd{ pedro.delatorre@uam.es}
\emailAdd{jortecal@protonmail.com}

\abstract{Inelastic scattering of Galactic cosmic rays (CRs) off sub-GeV dark matter (DM) produces neutral mesons whose decays yield TeV--PeV gamma rays. We constrain this process with the diffuse Galactic emission measured by LHAASO, H.E.S.S.\ and HAWC, using spatially varying CR distributions and Monte Carlo simulations of deep-inelastic scattering, hadronization and decays. We study Dirac DM with masses from $1$~keV to $100\MeV$, coupled to quarks through scalar, pseudoscalar, vector or axial-vector mediators. The LHAASO inner-Galaxy data give the strongest baseline constraints. Including the conventional CR background, the vector benchmark with $m_V=5\GeV$ reaches an elastic DM--proton cross section of $3\times10^{-41}$~cm$^2$ at $m_\chi=1$~keV, where nuclear-recoil direct detection has no sensitivity. The gamma-ray limits improve on the corresponding ANTARES neutrino limits by 1 to more than 2 orders of magnitude. Laboratory bounds generally remain stronger for flavour-universal couplings. The gamma-ray comparison with other scattering probes is more favourable for a $5\GeV$ vector coupled only to $u,d$ quarks, or an axial vector with its top coupling removed to satisfy the relevant unitarity condition. These observations probe a Galactic DM population directly through its interactions with CRs, complementing laboratory searches for invisible particles. Dedicated spectral and spatial analyses could further improve their sensitivity.}

\maketitle

\section{Introduction}\label{Sec:Introduction}

\begin{figure}[b]
    \centering
    \includegraphics[width=0.85\textwidth]{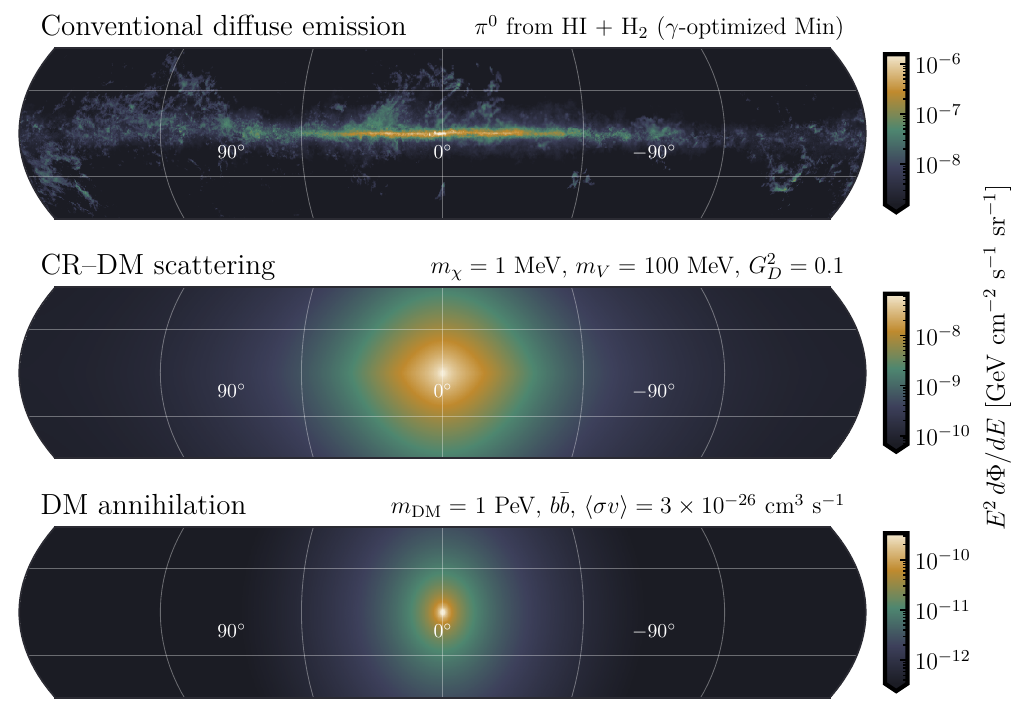}
    \caption{Sky maps of $E_\gamma^2\,d\Phi/dE_\gamma$ at $E_\gamma=100$~TeV for $|b|<30^\circ$ (Mollweide projection in Galactic coordinates, centred on the GC, longitude increasing to the left). \textit{Top:} conventional diffuse emission from $\pi^0$ decays produced by CR interactions with atomic and molecular gas in the $\gamma$-optimized (Min) model~\cite{Luque:2022buq, DelaTorreLuque:2022ats}. \textit{Middle:} CR--DM signal for Dirac DM with $m_\chi=1\MeV$ and a vector mediator with $m_V=100\MeV$, $G_D^2=0.1$. \textit{Bottom:} prompt emission from the annihilation of DM with $m_{\rm DM}=1$~PeV into $b\bar b$, with $\langle\sigma v\rangle=3\times10^{-26}$~cm$^3$\,s$^{-1}$~\cite{Bauer:2020jay}. Both DM maps assume the same NFW profile, and all maps include $\gamma\gamma$ absorption. Each colour scale spans three decades below the brightest $0.01\%$ of the pixels shown.}
    \label{fig:SkyMaps}
\end{figure}

The nature of dark matter (DM) remains one of the central open questions in particle physics and cosmology. Despite extensive searches for its non-gravitational interactions, direct-detection experiments have so far provided strong constraints primarily for DM masses above the GeV scale~\cite{CirelliPPPC, Cirelli:2024yps, Bertone:2004pz, Misiaszek:2024, Navas:2024}. At lower masses, the small kinetic energies transferred to detector targets severely limit the sensitivity of conventional direct-detection techniques, leaving a broad region of sub-GeV parameter space comparatively unexplored~\cite{Kahn:2021ttr, Zurek:2024, Essig:2024, Bringmann:2018cvk}. This motivates complementary probes that can exploit the high energies available in astrophysical environments. Galactic cosmic rays (CRs) extend beyond PeV energies and fill the Milky Way~\cite{Evoli:2018xif,Gabici:2019wzu,BeckerTjus:2020jla}, allowing even light, non-relativistic DM to participate in energetic interactions.

Inelastic CR--DM scattering can produce secondary mesons whose decays yield gamma rays and neutrinos~\cite{Cyburt:2002uw,TerolCalvo:2026xee}. Their energies are supplied by the incident CRs, while the DM mass sets the collision energy and production threshold. This photon signal is distinct from CR-boosted DM (CRDM), in which accelerated DM is detected through a subsequent collision~\cite{Bringmann:2018cvk, Ema_2019, Yin:2018yjn, Cappiello:2024acu,Super-Kamiokande:2022ncz,DeMarchi:2025uoo,DeMarchi:2025xag}. Both mechanisms probe interactions below the mass range accessible to conventional nuclear-recoil experiments, but their rates and detection strategies differ.

Diffuse gamma-ray observations now reach the energies relevant to this process. Following the Tibet AS$\gamma$ detection of sub-PeV diffuse emission~\cite{TibetASgamma:2021tpz}, LHAASO has measured the inner and outer Galactic plane from TeV energies towards the PeV scale~\cite{LHAASO:2023gne,LHAASO:2024lnz}. H.E.S.S.\ observations of the Galactic Centre and Ridge, and HAWC measurements of the Galactic plane, provide complementary sky coverage~\cite{HESS:2016pst,HESS:2017tce,HAWC:2023wdq}. These data permit a search for additional emission on top of the conventional component produced by CR interactions with interstellar gas~\cite{DeLaTorreLuque:2025zsv}.

Notably, the spatial distribution of this signal is readily distinguishable from the standard astrophysical emission. Conventional hadronic emission traces CRs interacting with the thin gas disk, whereas the CR--DM signal traces the product of the CR and DM densities along the line of sight. Since high-energy CRs occupy a halo several kpc thick, the latter signal extends beyond the gas disk. It is also broader than heavy-DM annihilation emission, which depends on the square of the DM density. Figure~\ref{fig:SkyMaps} illustrates these differences at $100$~TeV. Spectral information provides a second distinction: the threshold for scattering off the lightest DM selects the highest-energy CRs and shifts the signal towards the energies covered by LHAASO. Interestingly, the morphology is largely DM model-independent.

Earlier gamma-ray studies explored CR--DM interactions with a constant-cross-section prescription~\cite{Cyburt:2002uw,Hooper:2018bfw,Reis:2024wfy} or a heavy-mediator elastic cross section~\cite{Ambrosone:2022mvk}, while Ref.~\cite{TerolCalvo:2026xee} calculated the neutrino signal using a microscopic interaction model. Here we extend the latter approach to photons and confront it with diffuse measurements up to PeV energies. We simulate deep-inelastic scattering for four mediator types, propagate the resulting photons through the Galaxy, and assess how the constraints depend on the CR distribution, DM profile and conventional background. This also allows the gamma-ray and neutrino channels to be compared with the same physical inputs.

Our central question is how much these observations constrain specified DM--quark interactions, and how that reach compares with laboratory and cosmological probes of the same models. We retain the constant-cross-section prescription as a reference, but distinguish its phenomenological limits from the model-dependent conversion to elastic scattering. Sections~\ref{sec:DDetection}--\ref{sec:DMModels} define the observations and signal calculation, and Sec.~\ref{sec:Results} presents the limits and their uncertainties. Section~\ref{sec:Discussion} compares complementary bounds, and Sec.~\ref{sec:Conclusions} summarises the implications.

\section{TeV--PeV diffuse gamma-ray observations}
\label{sec:DDetection}

We consider complementary measurements of diffuse Galactic gamma-ray emission spanning the TeV--PeV range and different regions of the sky. The published LHAASO, H.E.S.S.\ and HAWC measurements define our baseline analysis. Below we describe their coverage and the source contributions relevant to interpreting them; the background prescriptions used to derive limits are specified in Sec.~\ref{sec:LimitProcedure}.

\begin{itemize}
\item \textbf{LHAASO:} The Large High Altitude Air Shower Observatory~\cite{LHAASO:2019qtb} combines the Water Cherenkov Detector Array (WCDA), covering approximately $1$--$25$~TeV~\cite{LHAASO:2024lnz}, with the Kilometer Square Array (KM2A), extending from about $10$~TeV towards the PeV scale~\cite{LHAASO:2023gne}. We use the diffuse measurements in the inner Galaxy, $15^\circ<l<125^\circ$, and outer Galaxy, $125^\circ<l<235^\circ$, both with $|b|<5^\circ$. Their high-energy coverage is particularly useful for probing the lightest DM.

The published measurements apply extensive masks around known sources, excluding a substantial part of the bright Galactic plane and suppressing much of the unresolved-source contribution~\cite{He2025UnresolvedGammaRaySources,vecchiotti}. The remaining emission is well described by CR interactions with interstellar gas~\cite{DeLaTorreLuque:2025zsv}. We use the full inner- and outer-Galaxy regions; the smaller spatial subdivisions available at WCDA energies do not extend the highest-energy coverage central to this search.

LHAASO has also presented preliminary spectra obtained by subtracting detected sources rather than masking their surroundings~\cite{Zhang2025LHAASO}. This retains more low-latitude emission and increases the sky coverage. We include the inner-Galaxy spectra for illustration, without using them to define our principal limits.\footnote{The spectra are taken from slide 11 of Ref.~\cite{LHAASO_unmasked_Slides}.} Their interpretation also requires attention to any unresolved-source component retained by the different selection.

\item \textbf{H.E.S.S.:} The High Energy Stereoscopic System~\cite{Hinton:2004eu} is an imaging atmospheric Cherenkov telescope array whose angular resolution allows detailed observations of the Galactic Centre. We use the Galactic Ridge measurements in approximately $|l|<1^\circ$, $|b|<0.3^\circ$, with point sources excluded~\cite{HESS:2006vje,HESS:2017tce}, and observations of the very inner region (VIR), within an angular radius of $0.1^\circ$ around the Centre~\cite{Aharonian:2009zk,HESS:2016pst}. These multi-TeV measurements probe the innermost Galaxy and complement LHAASO's larger regions.

The same regions contain the Central Molecular Zone (CMZ), with dense gas and intense source activity. Our large-scale CR model does not describe this environment in detail. Constraints that include a predicted gas background therefore carry larger modelling uncertainties than those from the masked LHAASO regions, and a dedicated treatment of the CMZ and its sources would be needed to refine them.

\item \textbf{HAWC:} The High-Altitude Water Cherenkov Observatory~\cite{historical:2023opo} is a wide-field air-shower detector. We use its measurement of diffuse TeV emission in $43^\circ<l<73^\circ$, $|b|<4^\circ$~\cite{HAWC:2023wdq}, providing an independent comparison away from the Galactic Centre. The inferred spectrum assumes a power law, and the source selection leaves a potentially sizeable unresolved contribution. CR interactions with gas together with unresolved sources can account for these data~\cite{DeLaTorreLuque:2025zsv}. We therefore show separately how including an unresolved-source estimate changes the HAWC limits.
\end{itemize}

Tibet AS$\gamma$ provides an additional comparison at sub-PeV energies~\cite{TibetASgamma:2021tpz}. Its smaller event sample and sensitivity to contamination from known and unresolved sources lead to weaker constraints than LHAASO~\cite{He2025UnresolvedGammaRaySources,vecchiotti,DeLaTorreLuque:2025zsv}. We show its points together with ARGO-YBJ data~\cite{ARGO-YBJ:2015cpa} in Fig.~\ref{fig:Observables}. Across these observations, the range of sky regions and source selections helps distinguish the role of high-energy coverage from that of the Galactic background.

\section{Cosmic-ray distribution}
\label{sec:CRFlux}

We adopt the $\gamma$-optimized Min model for the spatial and energy distributions of CR protons and helium~\cite{Luque:2022buq,DelaTorreLuque:2022ats,DeLaTorreLuque:2025zsv,Tovar-Pardo:2024cbp}. The signal calculation needs the CR flux at each scattering location, rather than only its value near the Solar System. This model supplies a spatially varying density and spectrum, constrained by local CR data and diffuse Galactic gamma-ray observations. We use the same CR population to predict the CR--DM signal and the conventional emission from interactions with gas.

The motivation for spatially varying transport comes from the hardening and enhanced intensity of the inner-Galaxy gamma-ray emission, observed by Fermi-LAT, relative to conventional homogeneous-diffusion predictions,~\cite{Gaggero:2015xza,Fermi-LAT:2016zaq,Yang2016prd,Lipari_2018}. The $\gamma$-optimized models extend the KRA$_\gamma$ framework~\cite{Gaggero:2014xla,Cerri2017jcap}: diffusion varies with Galactocentric position, producing a harder CR population towards the Galactic Centre. Fermi-LAT observations constrain this behaviour at GeV energies~\cite{Fermi-LAT:2012edv,Fermi-LAT:2016zaq}, while higher-energy gamma rays and neutrinos test its continuation into the regime relevant here. Recent Galactic-neutrino analyses support such scenario of inhomogeneous diffusion as well~\cite{abbasi2026highenergyneutrinoemissionmilky,neste2025measuringneutrinofluxsegments,Neste:2026}.

This evidence motivates the benchmark but does not uniquely determine CR transport. The contribution of unresolved sources and the limited coverage of the Galactic Centre prevent the present TeV--PeV data from selecting a propagation model unambiguously. Local measurements also leave uncertainty in the CR spectrum around the knee, precisely where the lightest DM requires the incident particles to be most energetic. We therefore vary both the transport description and the high-energy CR spectrum when assessing robustness.

The alternatives are the Base and $\gamma$-optimized Max models. Base assumes spatially uniform diffusion and lacks the inner-Galaxy hardening of the $\gamma$-optimized models. Max retains spatially varying diffusion but has a larger CR flux at PeV energies than Min. These variations probe different uncertainties: Base changes the relation between local and inner-Galaxy CRs, while Max tests the effect of a larger high-energy population. Their impact on the limits is summarised in Sec.~\ref{sec:Results_Robustness} and quantified in App.~\ref{sec:AppB}. The full model specifications and public distributions are given in Refs.~\cite{DeLaTorreLuque:2025zsv,TerolCalvo:2026xee,deLaTorre2024GalacticDiffuse}.

\section{Dark matter models and cross sections}
\label{sec:DMModels}

\subsection{Gamma-ray flux}
\label{sec:SignalFlux}

The photon flux follows the line-of-sight calculation used for neutrinos in Ref.~\cite{TerolCalvo:2026xee}, with an additional attenuation factor:
\begin{equation}
    \begin{split}
    \frac{d\phi_{\gamma}}{d\Omega\, dE_{\gamma}}(l,b,E_\gamma) = \int ds\; e^{-\tau_{\gamma\gamma}(E_\gamma,s)}\,\frac{\rho_{\DM}(r)}{m_{\chi}} &\int dE_{\rm CR}\, \frac{d\phi_{\rm CR}}{d\Omega\, dE_{\rm CR}}(E_{\rm CR}) \\
    &\times \sigma_{\rm CR\chi}(E_{\rm CR})\; \frac{dN_\gamma}{dE_\gamma}(E_\gamma;E_{\rm CR})\,.
    \end{split}
    \label{eq:GammaFlux}
\end{equation}
Here $s$ is the distance along direction $(l,b)$, $r$ is the corresponding Galactocentric radius, and the CR flux is evaluated at that location. We sum protons and helium and adopt a Navarro--Frenk--White (NFW) DM profile~\cite{Navarro:1995iw,Navarro:1996gj}, normalised to $\rho_\DM(8.3~{\rm kpc})=0.4\GeV\,{\rm cm}^{-3}$. Alternative profiles are specified in App.~\ref{sec:AppB}. We include photon absorption by pair production on the cosmic microwave background, relevant above about $100$~TeV over Galactic distances~\cite{Moskalenko:2005ng,Vernetto:2016alq}; infrared and starlight absorption are not relevant in this calculation~\cite{Rocamora:2025ddt}.

The interaction enters through $\sigma_{\rm CR\chi}\,dN_\gamma/dE_\gamma$, the cross section multiplied by the photon yield per collision. We first calculate both quantities in simplified particle models, then introduce a constant-cross-section prescription for comparison with earlier work.

\subsection{Simplified interactions}
\label{sec:SimplifiedModels}

We consider Dirac DM, $\chi$, coupled to quarks through a scalar $\phi$, pseudoscalar $a$, vector $V$ or axial vector $V'$, with interactions
\begin{align}
    \mathcal{L}_{\phi} &= g_{\chi\phi}\,\bar\chi\chi\,\phi + g_{q\phi}\sum_q\bar q q\,\phi\,, \label{eq:Lphi}\\
    \mathcal{L}_{a} &= i g_{\chi a}\,\bar\chi\gamma_5\chi\,a + i g_{qa}\sum_q\bar q\gamma_5 q\,a\,, \label{eq:La}\\
    \mathcal{L}_{V} &= g_{\chi V}\,\bar\chi\gamma^\mu\chi\,V_\mu + g_{qV}\sum_q\bar q\gamma^\mu q\,V_\mu\,, \label{eq:LV}\\
    \mathcal{L}_{V'} &= g_{\chi V'}\,\bar\chi\gamma^\mu\gamma_5\chi\,V'_\mu + g_{qV'}\sum_q\bar q\gamma^\mu\gamma_5 q\,V'_\mu\,, \label{eq:LA}
\end{align}
together with the mediator mass terms. The coupling $g_{qY}$ is common to all six quark flavours and does not scale with quark mass. These benchmarks have been used in CR--DM studies~\cite{DeMarchi:2025uoo,TerolCalvo:2026xee,Guo:2020oum,Dent:2019krz,Bondarenko:2019vrb,Ema:2020ulo,Wang:2025ztb}. For a specified mediator, the scattering signal depends on $m_\chi$, $m_Y$ and the product
\begin{equation}
    G_D^2 \equiv \left(g_{\chi Y}\,g_{qY}\right)^2.
    \label{eq:GD}
\end{equation}
Other observables, including on-shell production and mediator decay, can depend on the couplings individually. We study $1~{\rm keV}\leq m_\chi\leq100\MeV$ and mediator masses $100\MeV$ and $5\GeV$. These are benchmarks for the interaction's Lorentz structure; their consistency and complementary bounds are discussed in Sec.~\ref{sec:Discussion}~\cite{Knapen:2017xzo,DeMarchi:2025uoo}.

\subsection{Scattering and photon production}
\label{sec:SignalSimulation}

At the collision energies considered here, DM resolves the quarks in the nucleon. We calculate deep-inelastic scattering (DIS), $\chi q\to\chi q$, above $Q_0^2\simeq4\GeV^2$. The scalar and pseudoscalar signals coincide to better than $1\%$ over our DM mass range, as do the vector and axial-vector signals: the differences associated with $\gamma_5$ are suppressed by the small DM mass. Vector exchange favours small momentum transfer and a larger total rate; scalar exchange transfers more momentum per collision and produces more photons, but with a softer spectrum. Appendix~\ref{sec:AppTheory_DIS} gives the amplitudes and cross-section limits explaining these features.

For a CR nucleus of mass number $A$, we use
\begin{equation}
    \sigma_{\rm CR\chi}(E_{\rm CR}) = A^{2/3}\,G_D^2\,\sigma^{(0)}_{\chi p,Y}(E_{\rm CR};m_\chi,m_Y),
    \label{eq:sigmaCRchi}
\end{equation}
where $\sigma^{(0)}_{\chi p,Y}$ denotes the proton DIS cross section at unit couplings, distinct from a partonic cross section $\hat\sigma$. For helium, $E_{\rm CR}$ is the kinetic energy per nucleon. The adopted $A^{2/3}$ scaling is conservative: incoherent scattering with mild nuclear shadowing would increase the helium cross section by a factor of $1.4$--$1.5$ (App.~\ref{sec:AppTheory_Flavour}). 
CRs heavier than He can contribute substantially to secondary-particle production, given that they sum up to more than $10\%$ of the proton flux. Extending the CR composition to include nuclei up to $\mathrm{Fe}$, and assuming linear scaling with the nuclear mass number ($\alpha=1$), can enhance the photon production from CRDM interactions by more than a factor of two. However, such an enhancement is subject to significant uncertainties in the high-energy abundances of heavy nuclei and requires additional assumptions about the CR composition. 

We generate the hard interactions at leading order with \textsc{MadGraph5\_aMC@NLO}~\cite{Alwall:2014hca}, using the five-flavour scheme and NNPDF4.0 MC LO parton distributions~\cite{Cruz-Martinez:2024nnpdf}. \textsc{Pythia~8} showers and hadronizes the events and decays all unstable particles~\cite{Bierlich:2022pfr}. The DMsimp UFO implementations~\cite{Backovic:2015soa,Abdallah:2015ter} are used with the spin-0 Yukawa factors removed to obtain flavour-universal couplings. Their squared amplitudes and tree-level mediator widths agree with the analytic expressions.

The factorisation scale follows the momentum transfer, and events satisfy $Q^2\gtrsim Q_0^2$ and $\sqrt{s}\geq5\GeV$. The latter requirement leaves sufficient invariant mass for hadronization. Below it we neglect photon production; elastic and nucleon-resonance processes are also omitted. These omissions make the photon-yield calculation conservative. The simulated spectra are used in Eq.~\eqref{eq:GammaFlux}.

\subsection{Conversion to elastic scattering}
\label{sec:ElasticConversion}

The models differ at the low momentum transfers relevant to direct detection~\cite{DeMarchi:2025uoo}. Scalar and vector exchange give spin-independent (SI) scattering, axial-vector exchange gives spin-dependent (SD) scattering, and pseudoscalar exchange gives the operator $(\vec S_\chi\cdot\vec q)(\vec S_N\cdot\vec q)$, whose cross section is suppressed by four powers of momentum transfer. For $Y=\phi,V,V'$,
\begin{equation}
    \sigma^{\rm el}_{\chi p} = \frac{g_{\chi Y}^2\,g_{pY}^2\,\mu_{\chi p}^2}{\pi\,m_Y^4}\,,
    \label{eq:SigmaElY}
\end{equation}
where $\mu_{\chi p}$ is the reduced mass. The nucleon matrix elements give $g_{pV}=3g_{qV}$, $g_{p\phi}\simeq16.5g_{q\phi}$ and $g_{pV'}\simeq0.70g_{qV'}$ for our flavour-universal couplings (App.~\ref{sec:AppTheory_NR}). We quote pseudoscalar results in $G_D^2$ rather than assigning them a momentum-independent elastic cross section.

\subsection{Constant-cross-section reference}
\label{sec:ConstantXS}

To compare with phenomenological CRDM and gamma-ray limits~\cite{Bringmann:2018cvk,Cappiello:2018hsu,Wang:2021jic,Super-Kamiokande:2022ncz,Maity:2022exk,Hooper:2018bfw,Reis:2024wfy}, we also use a constant inelastic cross section and a fixed photon energy fraction:
\begin{equation}
    \sigma_{\rm CR\chi}(E_{\rm CR})\,\frac{dN_\gamma}{dE_\gamma}(E_\gamma;E_{\rm CR}) = \sigma_0\; \delta\!\left(E_\gamma - \kappa_\gamma E_{\rm CR}\right)\,\Theta\!\left(E_{\rm CR}-E^{\rm th}_{\rm CR}\right), \qquad \kappa_\gamma = 0.1\,.
    \label{eq:ConstXS}
\end{equation}
The choice $\kappa_\gamma=0.1$ approximates neutral-pion production in proton--proton collisions, where a pion receives about $17\%$ of the projectile kinetic energy and each decay photon about half of that~\cite{Kelner:2006tc}. The threshold for single-pion production on stationary DM is
\begin{equation}
    E^{\rm th}_{\rm CR} = \frac{m_{\pi^0}\,(2m_p + 2m_\chi + m_{\pi^0})}{2m_\chi} \simeq 136\GeV \left(\frac{1\MeV}{m_\chi}\right),
    \label{eq:Ethr}
\end{equation}
with $E_{\rm CR}$ expressed as kinetic energy, or kinetic energy per nucleon for helium. In this prescription the photons lie above $E_\gamma\simeq14\GeV(1\MeV/m_\chi)$, shifting the lightest-DM signal into the TeV--PeV range. The energy integral then gives
\begin{equation}
    \frac{d\phi_{\gamma}}{d\Omega\, dE_{\gamma}} = \frac{\sigma_0}{\kappa_\gamma\, m_\chi}\int ds\; e^{-\tau_{\gamma\gamma}(E_\gamma,s)}\, \rho_{\DM}(r)\, \frac{d\phi_{\rm CR}}{d\Omega\, dE_{\rm CR}}\bigg|_{E_{\rm CR}=E_\gamma/\kappa_\gamma}\,, \qquad E_\gamma > \kappa_\gamma E^{\rm th}_{\rm CR}\,.
    \label{eq:ConstXSFlux}
\end{equation}
Above threshold, the photon spectrum follows the CR spectrum shifted by $\kappa_\gamma$, and the flux scales as $\sigma_0/m_\chi$. We display the corresponding spectra in App.~\ref{sec:AppA} and adopt the conventional mapping $\sigma^{\rm el}_{\chi p}=3\sigma_0/8$ to compare limits~\cite{Reis:2024wfy,TerolCalvo:2026xee}. This mapping is not model independent: a physical mediator makes both the energy dependence and the inelastic-to-elastic ratio interaction dependent~\cite{DeMarchi:2025uoo}. Appendix~\ref{sec:AppTheory_ConstXS} quantifies the limitation.

\section{Results}\label{sec:Results}

\subsection{Limit-setting procedure}
\label{sec:LimitProcedure}

We constrain the signal by requiring it not to overproduce the measured flux in the selected regions. For a normalisation $N$, equal to $G_D^2$ in the particle models or $\sigma_0$ in the constant-cross-section reference, we use
\begin{equation}
    \chi^2(N) = \sum_i \left[\frac{\max\!\left(N\,\Phi^{\rm DM}_i + \Phi^{\rm bkg}_i - \Phi^{\rm obs}_i,\,0\right)}{\sigma_i}\right]^2,
    \label{eq:chi2lim}
\end{equation}
where $\Phi_i^{\rm obs}$ and $\sigma_i$ are the measured flux and uncertainty, $\Phi_i^{\rm DM}$ is the signal at unit normalisation, and $\Phi_i^{\rm bkg}$ is the conventional emission. Only excesses above the data contribute. We adopt $\chi^2(N)=4$ as the upper-limit criterion, corresponding nominally to approximately $95\%$ confidence for one parameter.

Each baseline result is calculated twice: with $\Phi_i^{\rm bkg}=0$, and with the CR--gas emission predicted by the $\gamma$-optimized Min model. The first is the more conservative bound; the second uses the same CR population for the signal and background. We hold the background fixed rather than fitting its normalisation. The resulting bands span these two assumptions and are not uncertainty intervals from a background fit. A joint template analysis would also need to model unresolved sources and their degeneracy with the diffuse components.

The treatment depends on the observing region. For masked LHAASO data we use the CR--gas prediction directly. For HAWC we also show a separate result including the unresolved-source estimate of Ref.~\cite{EcknerCalore2022ALP}; its lower bound is distinguished from the baseline band. For the H.E.S.S.\ Ridge we rescale the gas prediction by $0.2$, following the fitted normalisation in Ref.~\cite{Zuriaga_Puig_2023}, to illustrate the CMZ uncertainty. The predicted gas background in the VIR region is negligible, so its two limits nearly coincide. These prescriptions specify the comparison, without implying equal control of the backgrounds in every region.

\subsection{Comparison of observations}
\label{sec:Results_Observations}

\begin{figure*}[t!]
    \centering 
    \includegraphics[width=0.49\textwidth]{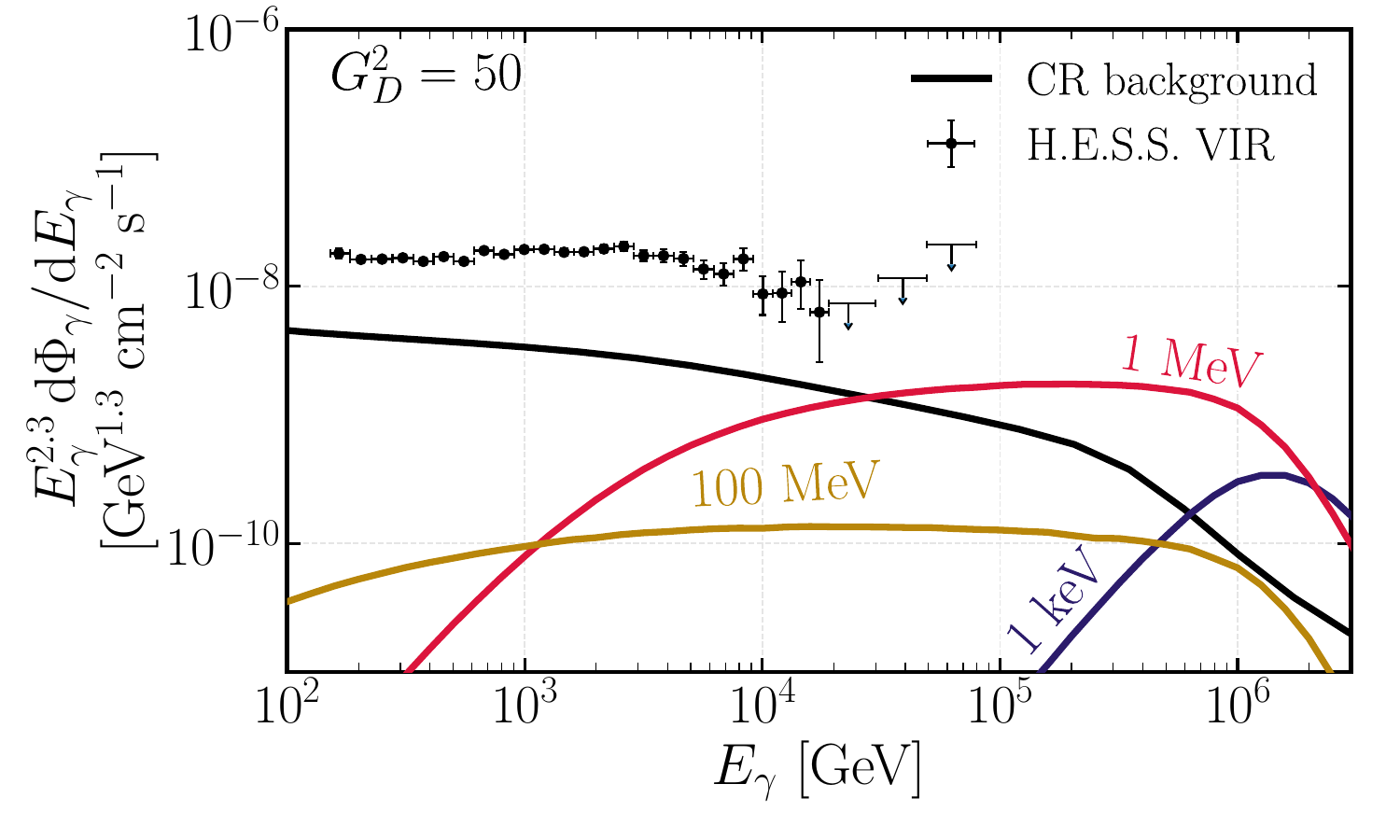}
    \includegraphics[width=0.49\textwidth]{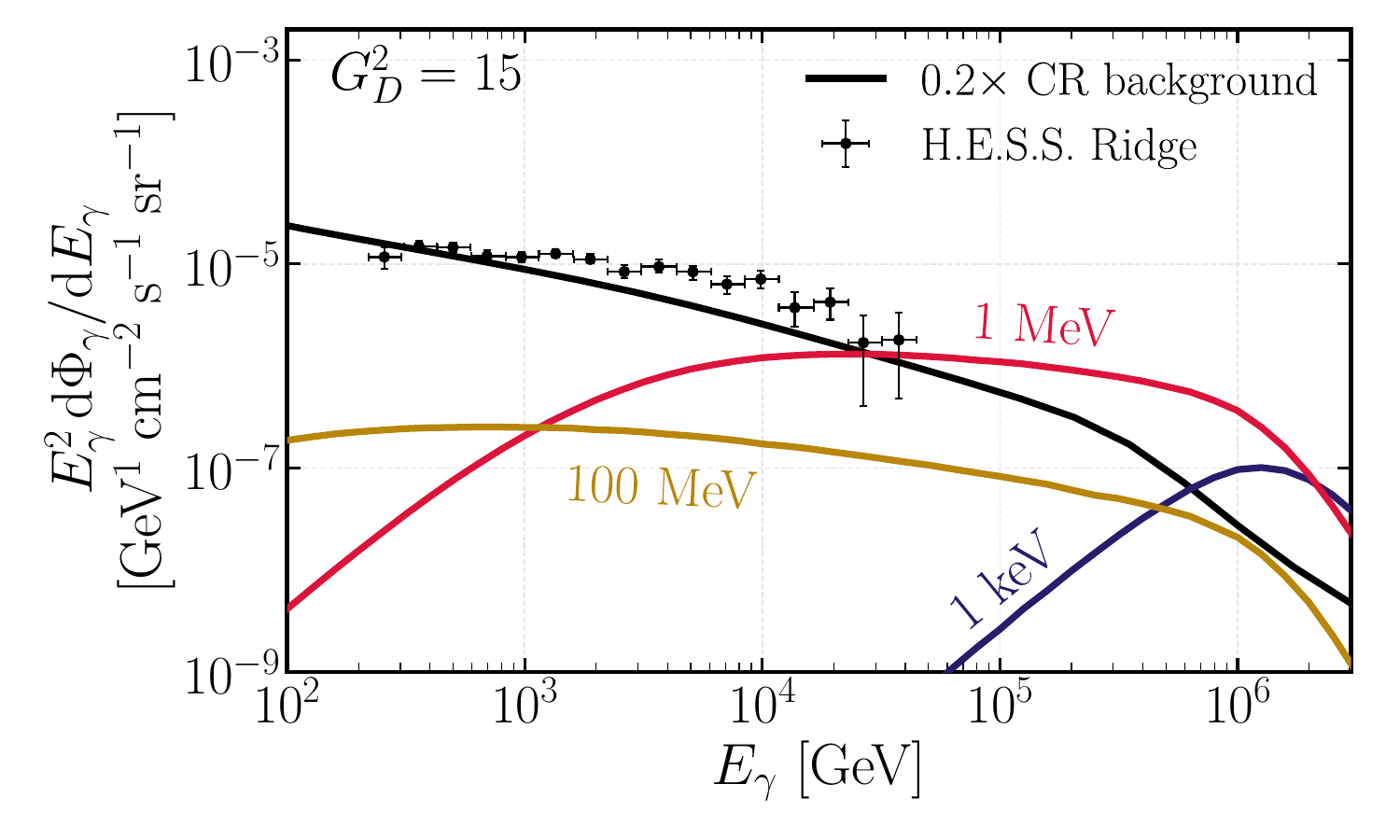}
    \includegraphics[width=0.49\textwidth]{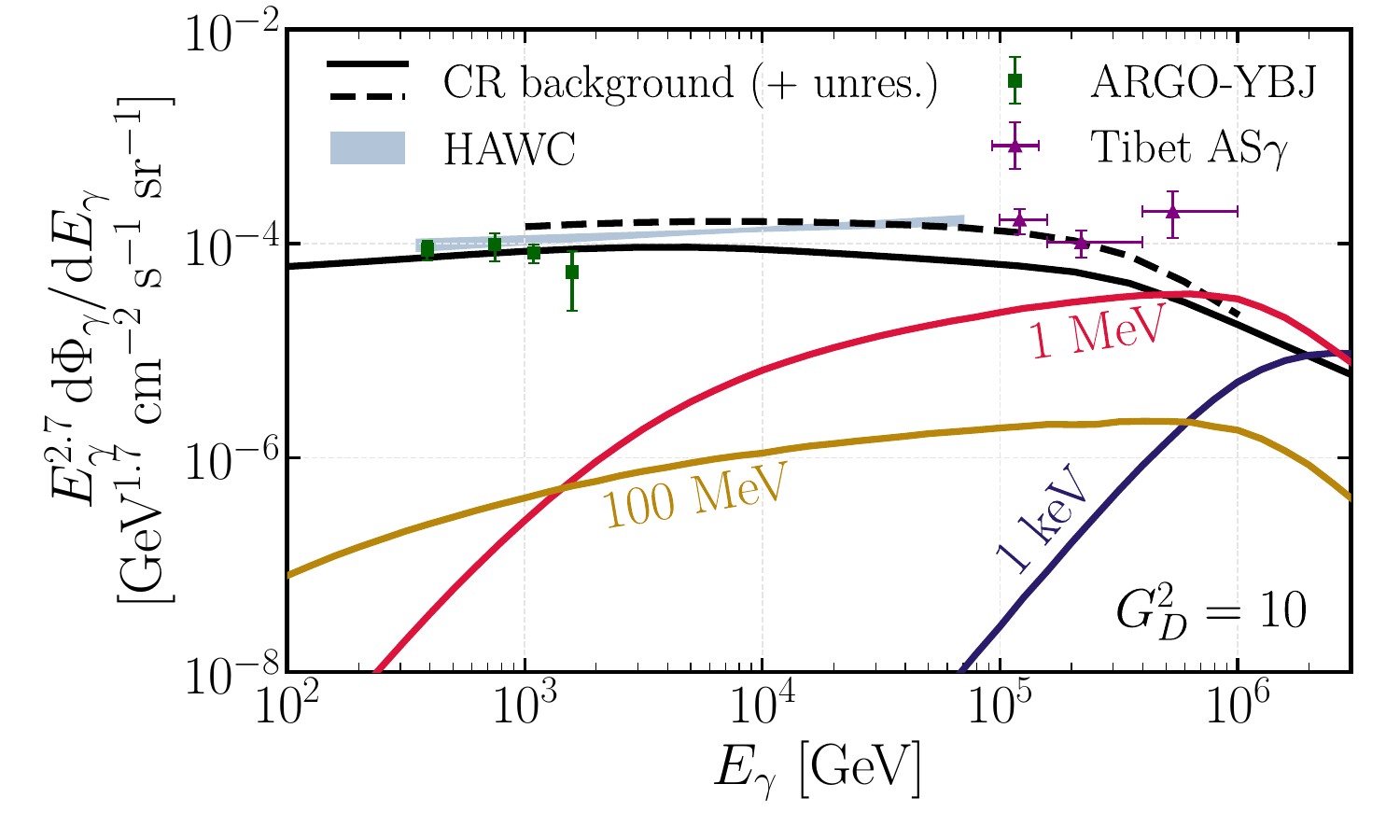}
    \includegraphics[width=0.49\textwidth]{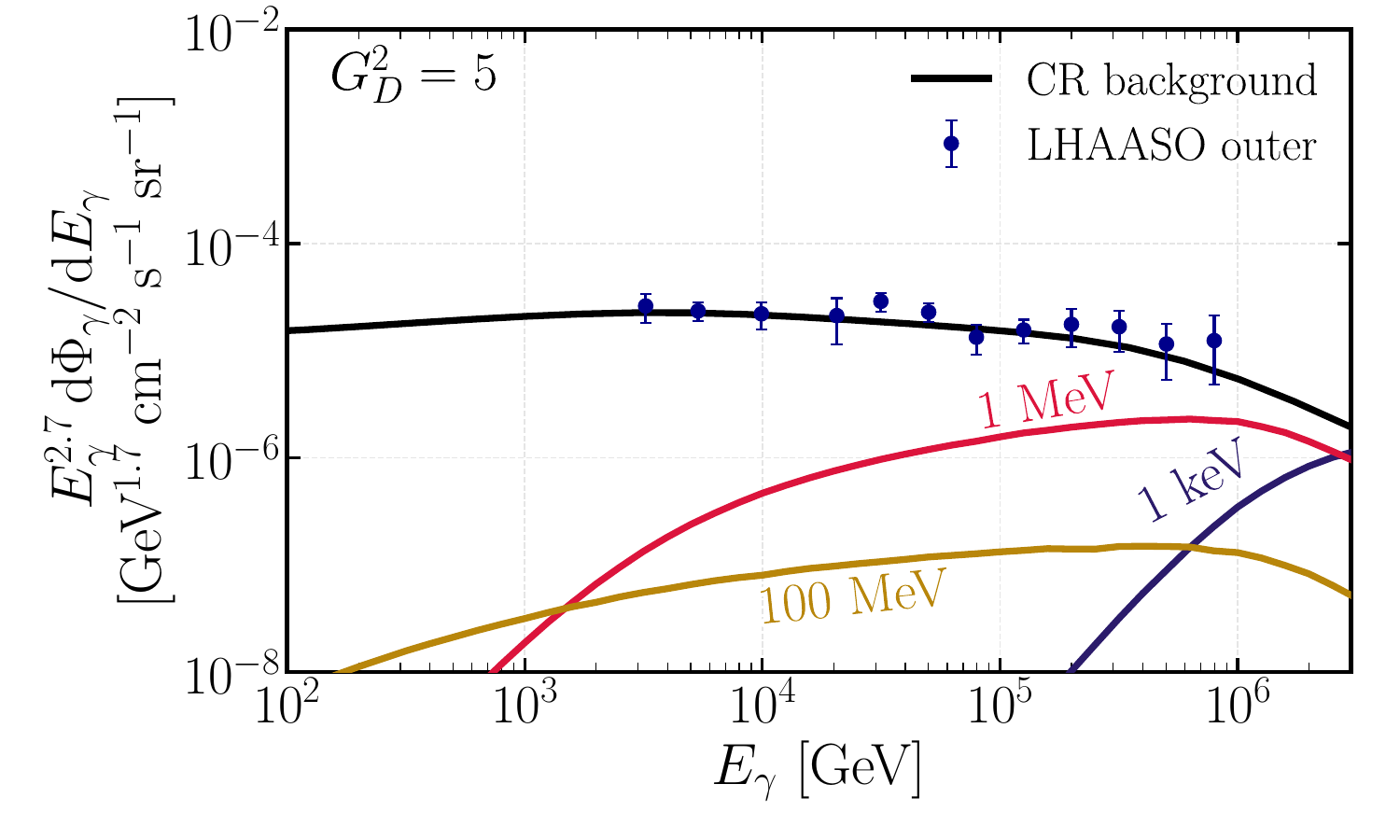}
    \includegraphics[width=0.49\textwidth]{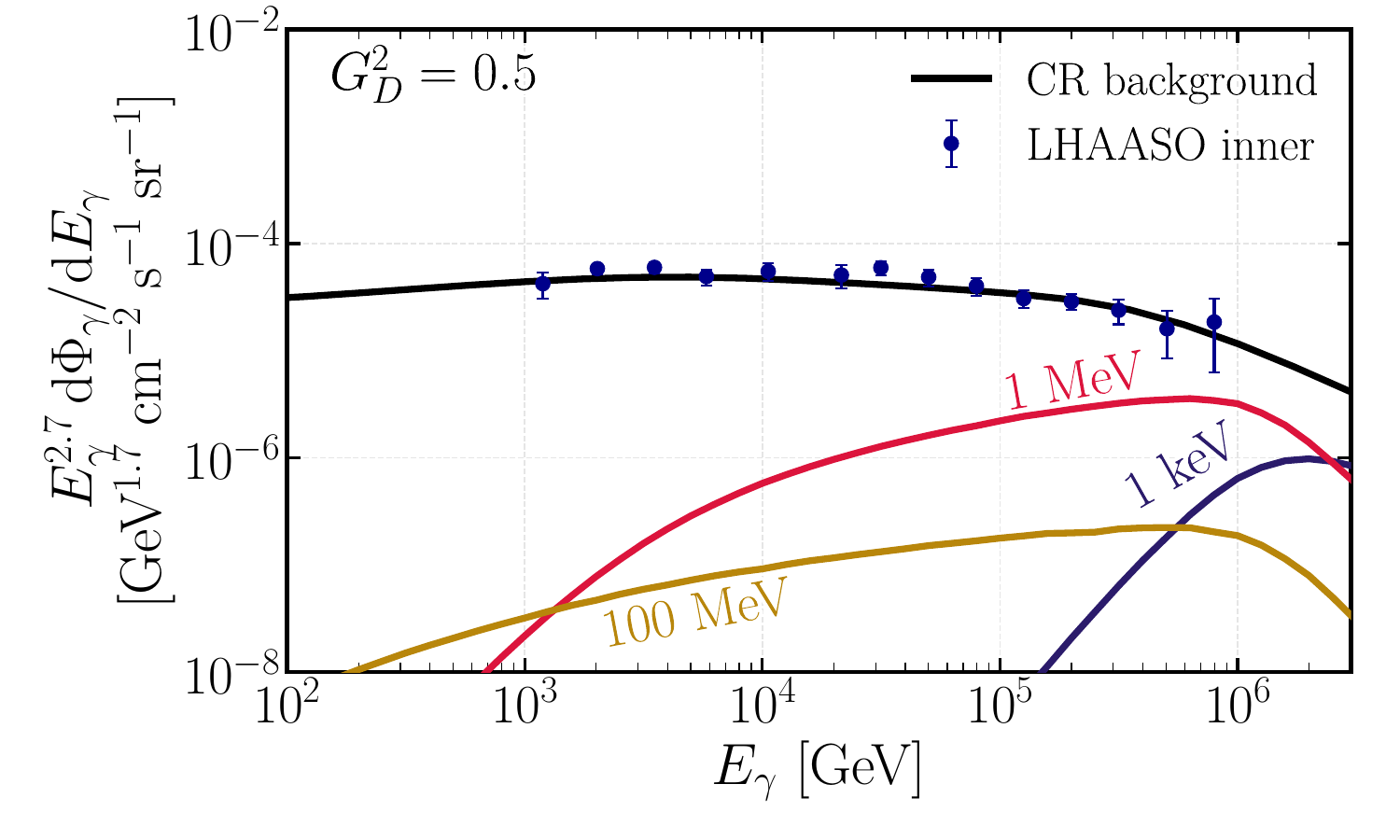}
    \includegraphics[width=0.49\textwidth]{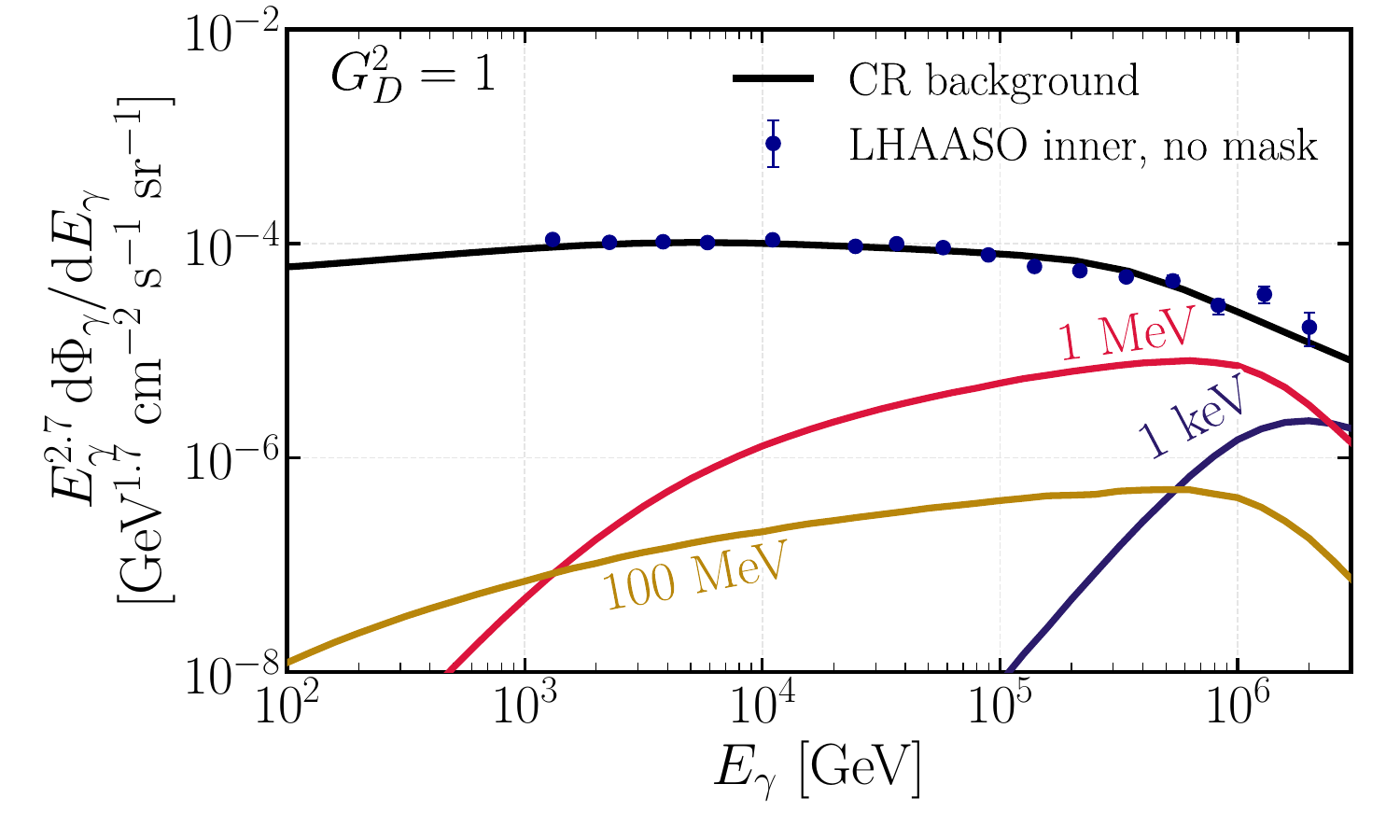}
    \caption{Gamma-ray spectra for the $5\GeV$ vector mediator, with $m_\chi=1$~keV--$100\MeV$ (colours) and the coupling specified in each panel. Regions are defined in Sec.~\ref{sec:DDetection}: H.E.S.S.\ VIR and Ridge (top), HAWC and LHAASO outer Galaxy (middle), and LHAASO inner Galaxy with the published mask and preliminary source subtraction (bottom left and right). The HAWC panel includes ARGO-YBJ and Tibet AS$\gamma$ points~\cite{ARGO-YBJ:2015cpa,TibetASgamma:2021tpz}. Black solid: Min CR--gas background, rescaled by $0.2$ for the Ridge; black dashed in HAWC: gas plus unresolved sources. The VIR flux is region-integrated and weighted by $E_\gamma^{2.3}$; the other panels show intensities weighted by $E_\gamma^2$ for the Ridge and $E_\gamma^{2.7}$ elsewhere.}
\label{fig:Observables} 
\end{figure*}

The LHAASO inner Galaxy gives the strongest baseline constraints across the mediator models. Its high-energy coverage tests the lightest DM, and its source mask leaves a background that is comparatively well described by CR--gas emission. HAWC can become competitive or stronger when an unresolved-source contribution is included, but that improvement depends on this less certain background component. H.E.S.S.\ and the LHAASO outer Galaxy provide complementary, but weaker bounds.

The spectral origin of this hierarchy is visible in Fig.~\ref{fig:Observables}, which uses a $5\GeV$ vector mediator, NFW DM and the Min CR model. The signal is generally harder than CR--gas emission. Above $m_\chi\sim1\MeV$ it extends across the TeV range, while lighter DM requires higher-energy CRs and produces spectra peaking above about $10^5\GeV$. KM2A therefore has a particular advantage at low masses. The angular morphology varies less strongly with mass, becoming somewhat more extended for the lighter DM. The constant-cross-section spectra in App.~\ref{sec:AppA} illustrate the corresponding threshold behaviour in the fixed-energy-fraction prescription.

\begin{figure*}[t!]
    \centering 
    \includegraphics[width=0.49\textwidth]{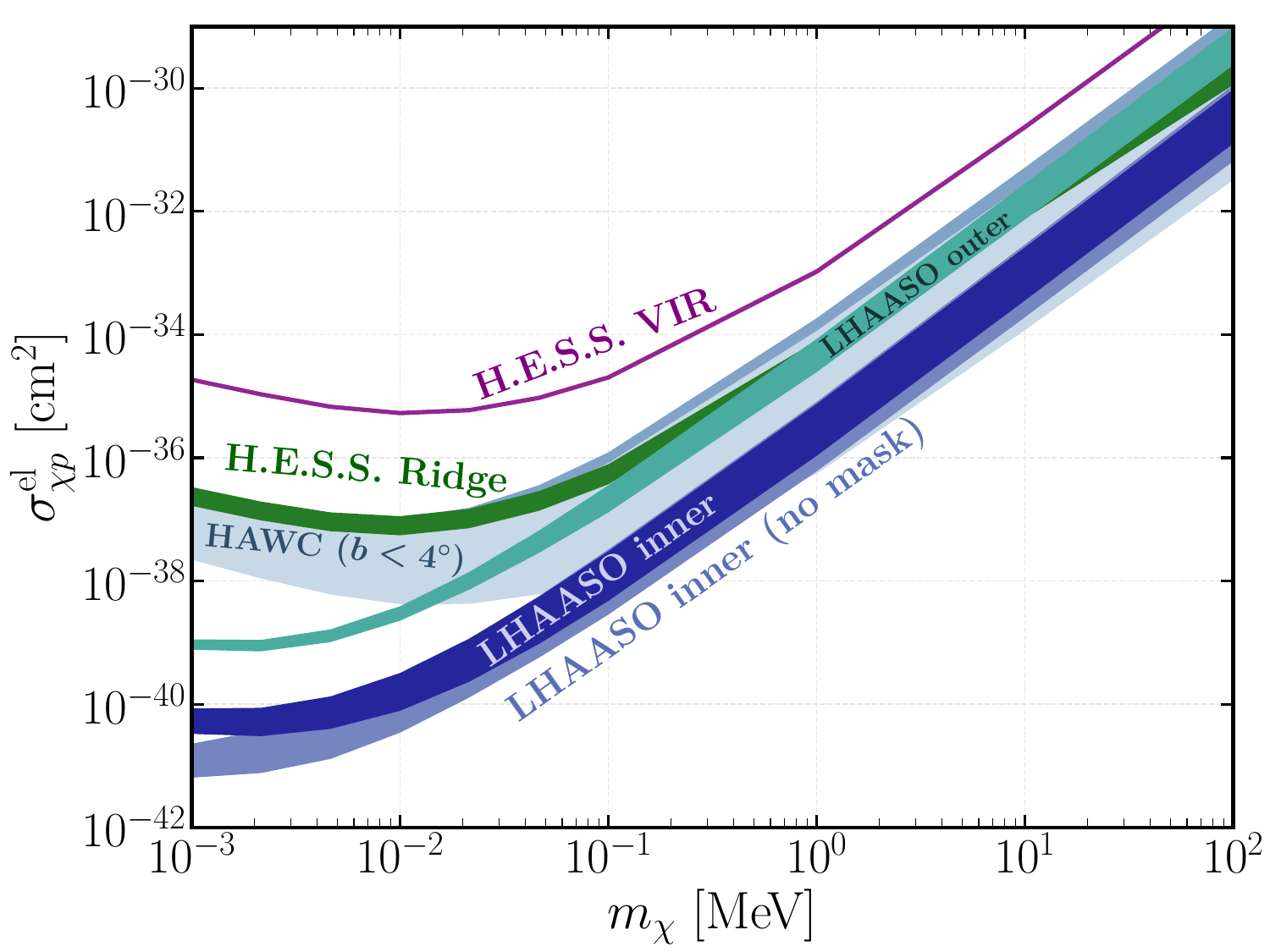}
    \includegraphics[width=0.49\textwidth]{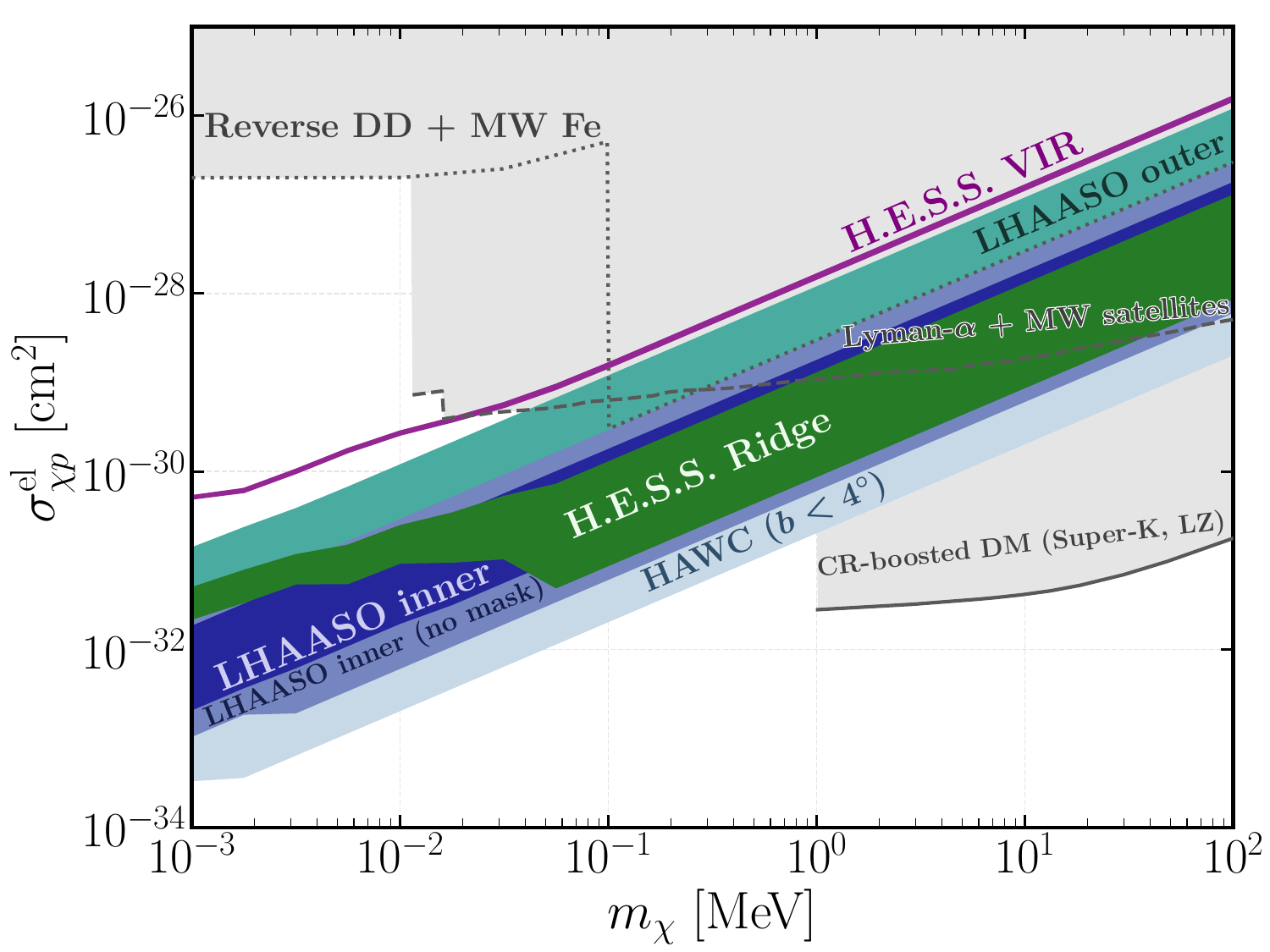}
    \caption{Upper limits on the elastic DM--proton cross section for the $5\GeV$ vector mediator (left) and constant-cross-section reference (right). Bands span the limits without background (upper edge) and with the Min CR--gas prediction (lower edge); the lighter HAWC band includes unresolved sources. Complementary curves are reverse direct detection~\cite{Cappiello:2018hsu}, Lyman-$\alpha$~\cite{Rogers:2021byl}, Milky Way satellites~\cite{DES:2020fxi}, LZ and Super-K CRDM~\cite{Maity:2022exk,Super-Kamiokande:2022ncz}, and CR iron spallation~\cite{Meighen-Berger:2025hrq}.}
\label{fig:Observables_Constraints} 
\end{figure*}

The limits in Fig.~\ref{fig:Observables_Constraints} compare the vector benchmark and constant-cross-section reference using the same background conventions. The two HAWC bands show how a model for unresolved sources can change the ranking of the observations. Even with the Ridge background rescaling, H.E.S.S.\ remains less constraining than the LHAASO inner Galaxy. The preliminary source-subtracted LHAASO comparison suggests improvements of a factor of $2$--$4$, but is illustrative and does not replace the published masked data in our quoted limits. Its larger coverage also changes the source-background requirements.

\subsection{Dependence on the mediator}
\label{sec:Results_Mediators}

\begin{figure}[t]
    \centering
    \includegraphics[width=0.8\textwidth]{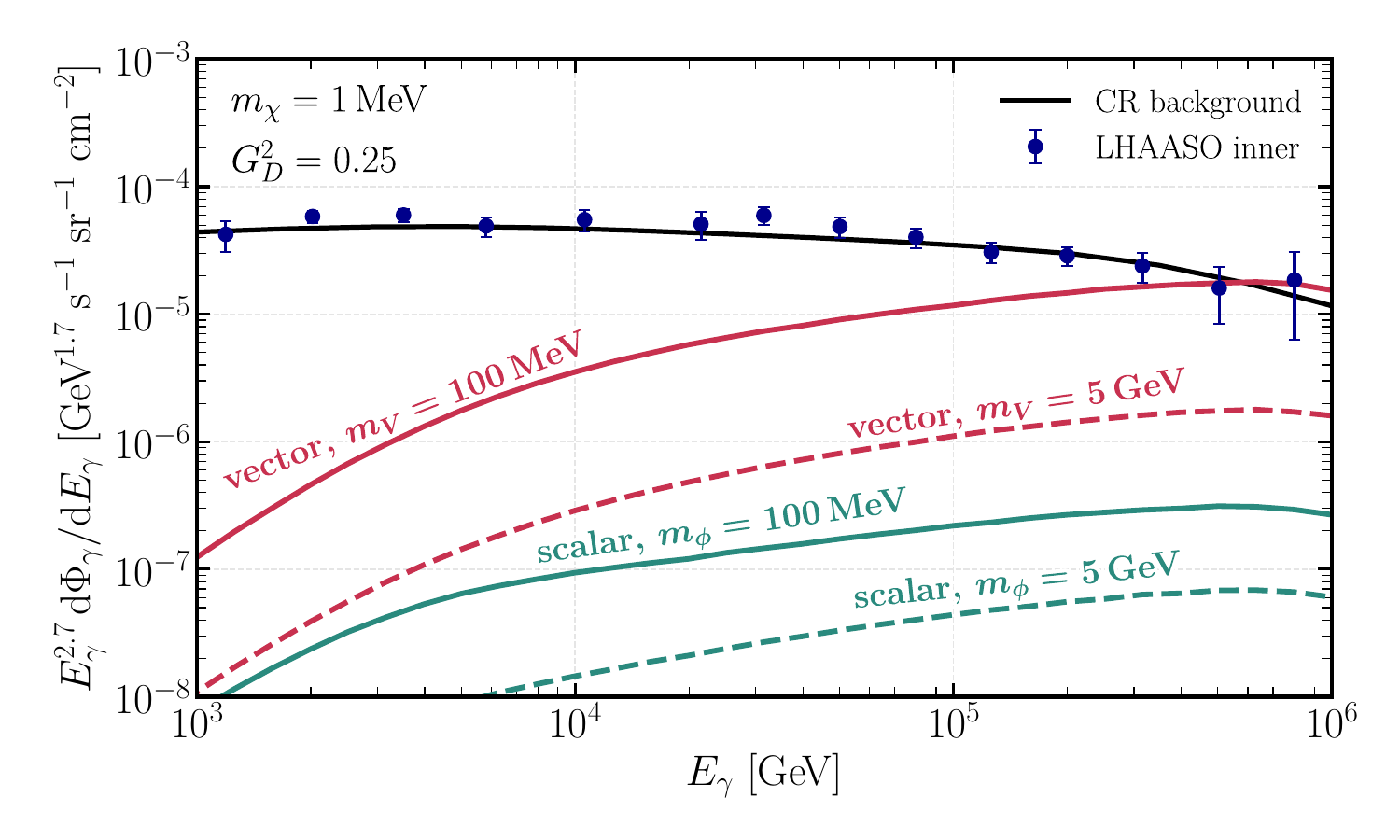}
    \caption{Diffuse gamma-ray flux from the inner Galaxy ($15^\circ<l<125^\circ$, $|b|<5^\circ$), weighted by $E_\gamma^{2.7}$. Points: LHAASO measurement~\cite{LHAASO:2023gne,LHAASO:2024lnz}. Black line: conventional emission in the $\gamma$-optimized (Min) CR model. Coloured lines: CR--DM signal for Dirac DM with $m_\chi=1\MeV$ and a vector (crimson) or scalar (teal) mediator $Y$ of mass $m_Y=100\MeV$ (solid) or $5\GeV$ (dashed), at the same coupling $G_D^2=0.25$.}
    \label{fig:FluxLHAASOInner}
\end{figure}

At fixed coupling product, vector exchange produces a larger gamma-ray signal than scalar exchange. Figure~\ref{fig:FluxLHAASOInner} compares the two at $m_\chi=1\MeV$ and $G_D^2=0.25$. The spectra peak near $0.6$--$0.8$~PeV, but the vector flux is larger by a factor of $20$--$60$, depending on mediator mass and photon energy. The scalar's smaller DIS cross section is only partly offset by its larger photon multiplicity, and its softer photons further reduce the high-energy flux. Table~\ref{tab:SimulationRatios} in App.~\ref{sec:AppTheory_DIS} gives the numerical decomposition.

Reducing the mediator mass from $5\GeV$ to $100\MeV$ raises the flux by roughly a factor of ten for the vector and five for the scalar, with relatively little change in shape at this DM mass. The larger vector response follows from the weight of low momentum transfers in its cross section; the scalar rate is less concentrated in that region (App.~\ref{sec:AppTheory_DIS}). The pseudoscalar and axial-vector signals follow the scalar and vector cases, respectively, within the accuracy discussed in Sec.~\ref{sec:SignalSimulation}.

\begin{figure}[t]
    \centering
    \includegraphics[width=0.8\textwidth]{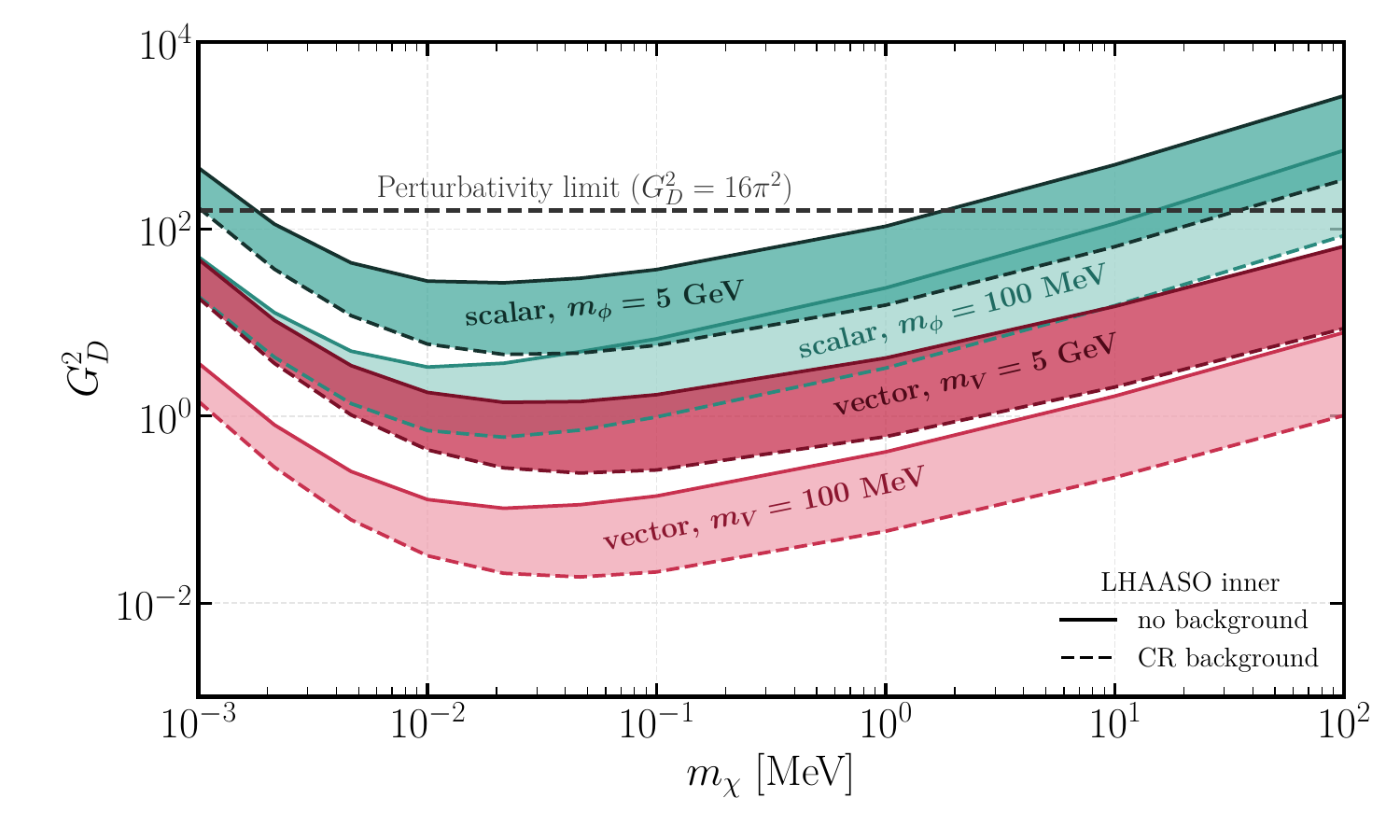}
    \caption{Upper limits on $G_D^2$ from the LHAASO inner-Galaxy data ($15^\circ<l<125^\circ$, $|b|<5^\circ$) for the vector and scalar mediators, with $m_Y=5\GeV$ and $100\MeV$. The upper edge of each band is obtained without astrophysical background (solid) and the lower edge including the CR background of the $\gamma$-optimized (Min) model (dashed). The axial-vector (pseudoscalar) limits coincide with the vector (scalar) ones. Dashed horizontal line: perturbativity limit, $G_D^2=16\pi^2$.}
    \label{fig:LimitsG2D}
\end{figure}

The corresponding coupling limits, Fig.~\ref{fig:LimitsG2D}, are strongest at $m_\chi\simeq10$--$50$~keV. Towards smaller masses the collision threshold removes all but the highest-energy CRs; towards larger masses the target number density $\rho_\DM/m_\chi$ falls. For the vector, the strongest limits including the CR background are $G_D^2\lesssim0.25$ at $m_V=5\GeV$ and $0.02$ at $100\MeV$, compared with $1.4$ and $0.10$ without background. Across the models, including the gas component strengthens the limits by factors of $3$--$8$. Scalar and pseudoscalar bounds are weaker because of their smaller photon flux, and parts of their sensitivity extend beyond the perturbativity boundary; Sec.~\ref{sec:Disc_Theory} specifies the resulting restriction.

\begin{figure}[t]
    \centering
    \includegraphics[width=0.495\textwidth]{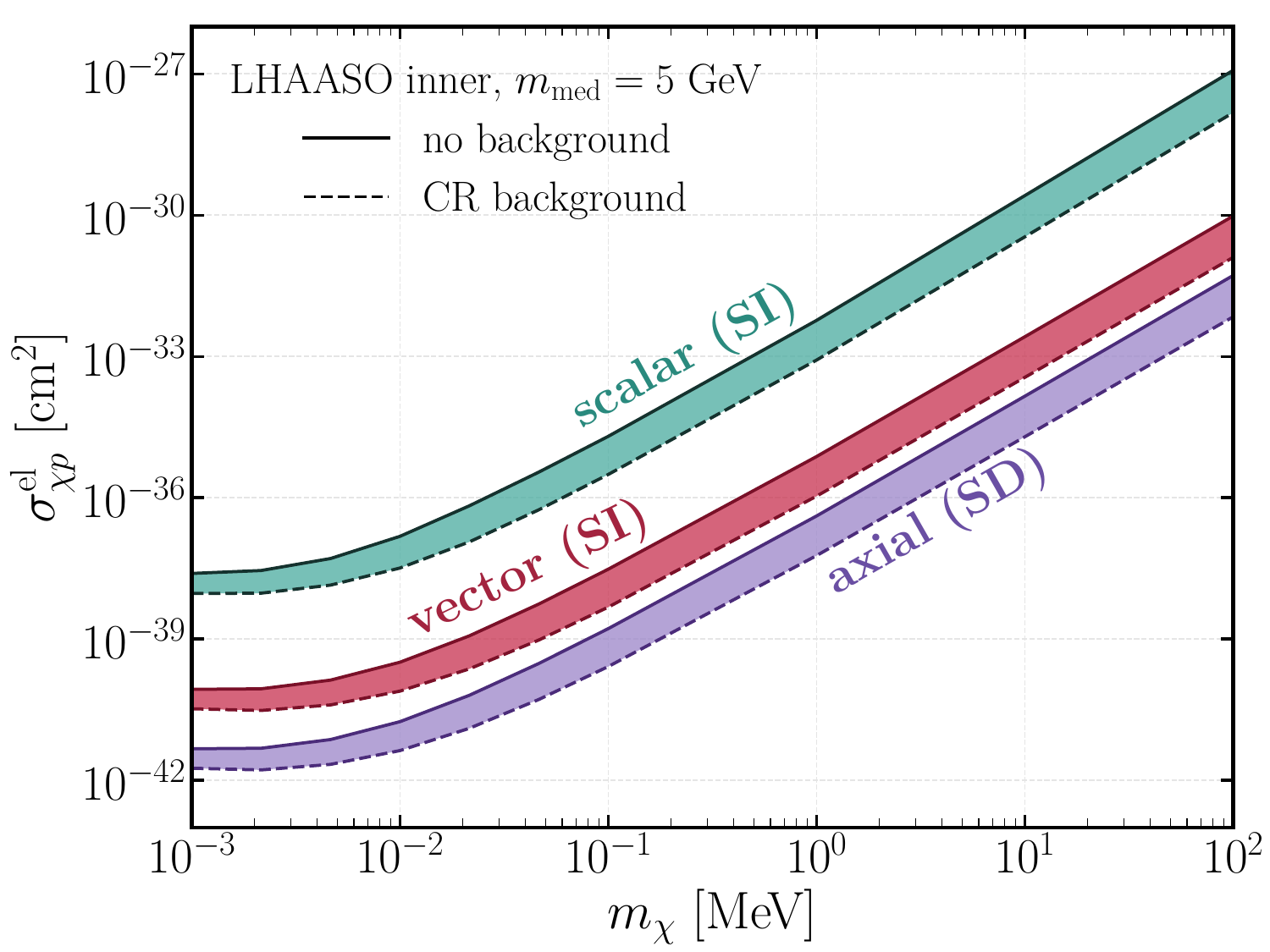}
    \includegraphics[width=0.495\textwidth]{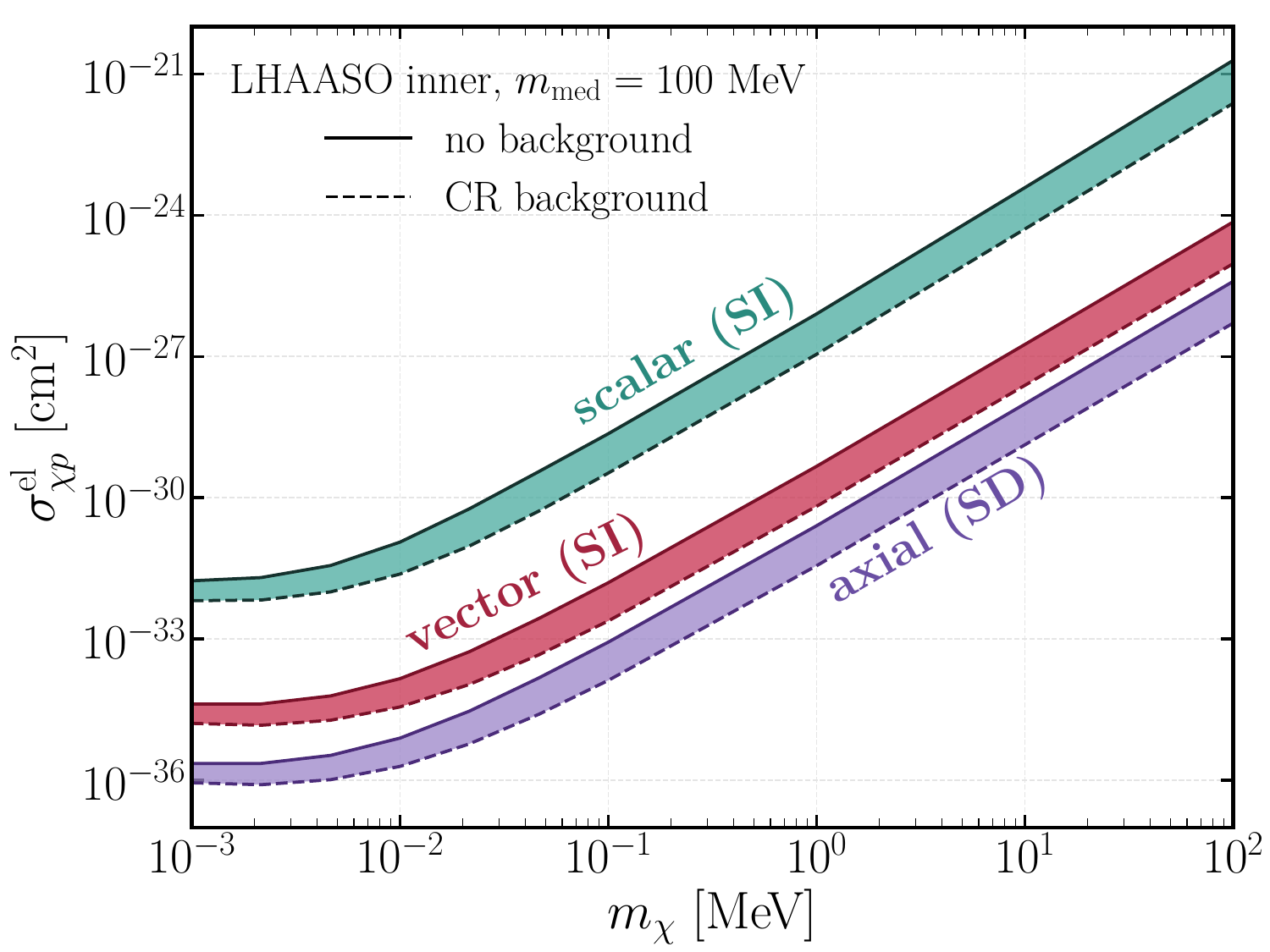}
    \caption{Upper limits on the elastic DM--proton cross section from the LHAASO inner-Galaxy data for the scalar and vector (spin-independent) and axial-vector (spin-dependent) mediators, with $m_Y=5\GeV$ (\textit{left}) and $100\MeV$ (\textit{right}), obtained from the $G_D^2$ limits of Fig.~\ref{fig:LimitsG2D} with Eq.~\eqref{eq:SigmaElY}. Solid (dashed) edges: without background (with the CR background). The pseudoscalar, whose tree-level elastic cross section is momentum suppressed, is not shown.}
    \label{fig:LimitsSigmaElMediators}
\end{figure}

Converting to elastic scattering changes the relative ordering, because Eq.~\eqref{eq:SigmaElY} also contains the proton coupling and the factor $m_Y^{-4}$. For $m_Y=5\GeV$ and $m_\chi=1$~keV, the background-included limits reach $3\times10^{-41}$, $9\times10^{-39}$ and $2\times10^{-42}$~cm$^2$ for the vector, scalar and axial vector, respectively (Fig.~\ref{fig:LimitsSigmaElMediators}). The scalar's larger proton coupling further weakens its bound in this plane, whereas the axial vector's smaller coupling makes its SD limit numerically lower than the vector SI limit. These compare different interactions, despite their similar high-energy signals.

The lighter mediator improves the limit on $G_D^2$ but gives a much weaker limit on the elastic cross section: the factor $(5\GeV/100\MeV)^4\simeq6\times10^6$ outweighs the gain in gamma-ray production. Likewise, the reduced-mass factor makes the elastic limits tighten towards light DM even where the coupling sensitivity no longer improves. The pseudoscalar is not included in this conversion because its tree-level elastic cross section depends on momentum transfer. Additional numerical ratios are given in App.~\ref{sec:AppTheory_DIS}.

\subsection{Astrophysical robustness}
\label{sec:Results_Robustness}

The main astrophysical uncertainties are the DM density profile and CR distribution. We vary them separately using the vector model at both mediator masses, with the same masks and attenuation as in the benchmark. These comparisons use background-free limits, so that changing the CR model tests the signal without simultaneously changing the adopted gas background. Appendix~\ref{sec:AppB} gives the profiles, propagation alternatives and full results.

Replacing NFW by Burkert, Moore or Einasto changes the LHAASO inner-Galaxy limits by less than a factor of two. The region avoids the Galactic Centre, where the profiles differ most. For the H.E.S.S.\ Ridge, the same variation can change the bound by about an order of magnitude. The weak profile dependence of the larger LHAASO and HAWC regions has also been noted in Ref.~\cite{Rocamora:2025ddt}.

Changing the Min CR model to Base or Max produces variations of a few at higher DM masses and up to an order of magnitude for the lightest DM, which depends on CRs near the knee. Thus the LHAASO bounds are relatively insensitive to the inner halo profile, but retain an appreciable dependence on the high-energy CR population. This distinction motivates improved CR measurements and background modelling as well as increased gamma-ray exposure.

\section{Discussion} \label{sec:Discussion}

The LHAASO inner-Galaxy limits substantially improve on the neutrino limits from the same CR--DM process and compete with direct detection searches, but accelerator searches generally can impose stronger constraints on the flavour-universal benchmarks. Their relative difference depends on the mediator, its quark couplings and relevant assumptions entering each search. The comparison therefore establishes both the parameter space already constrained by other probes and the complementary information carried by Galactic gamma rays.

\begin{figure}[t]
    \centering
    \includegraphics[width=0.49\textwidth]{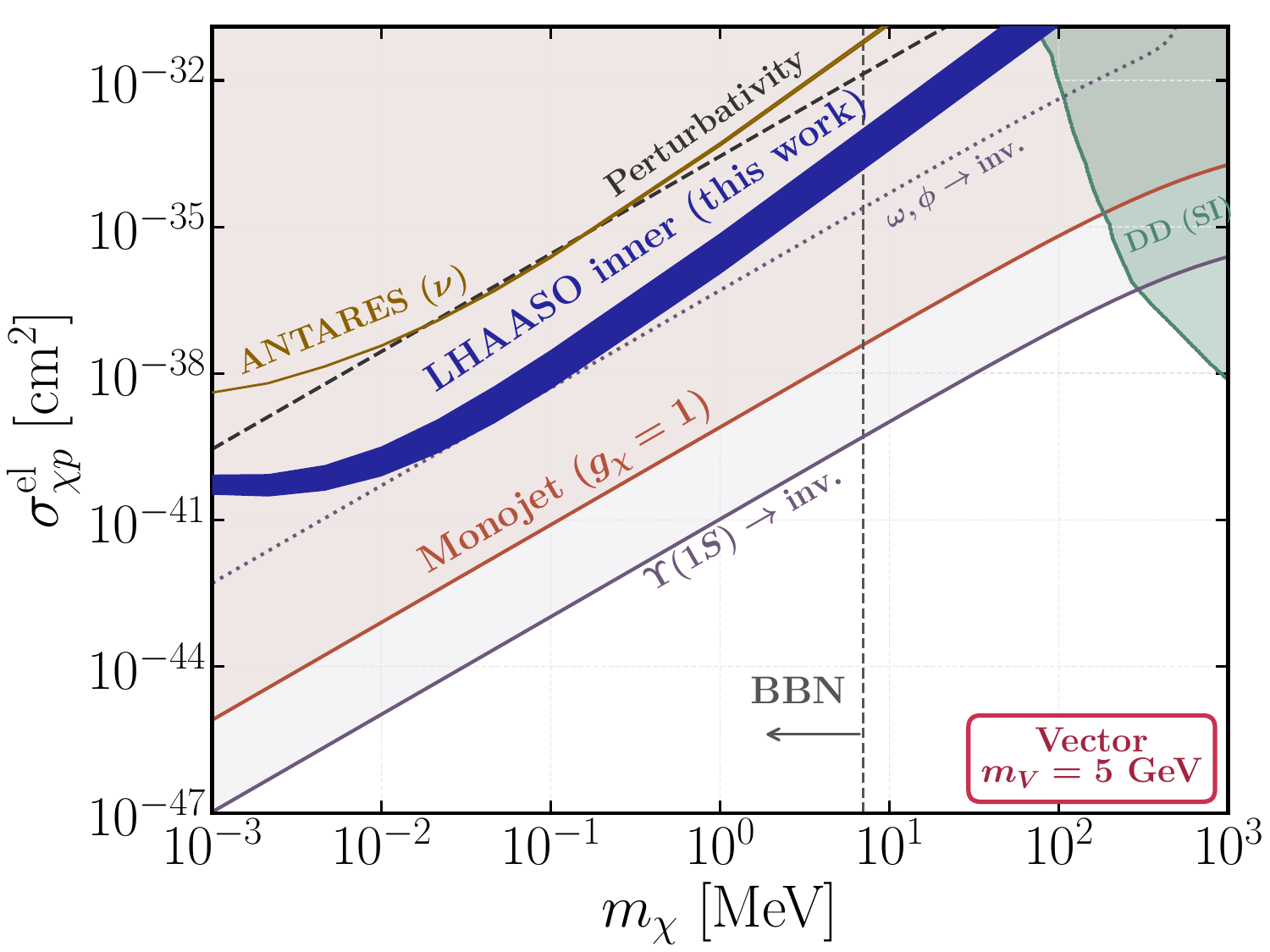}
    \includegraphics[width=0.49\textwidth]{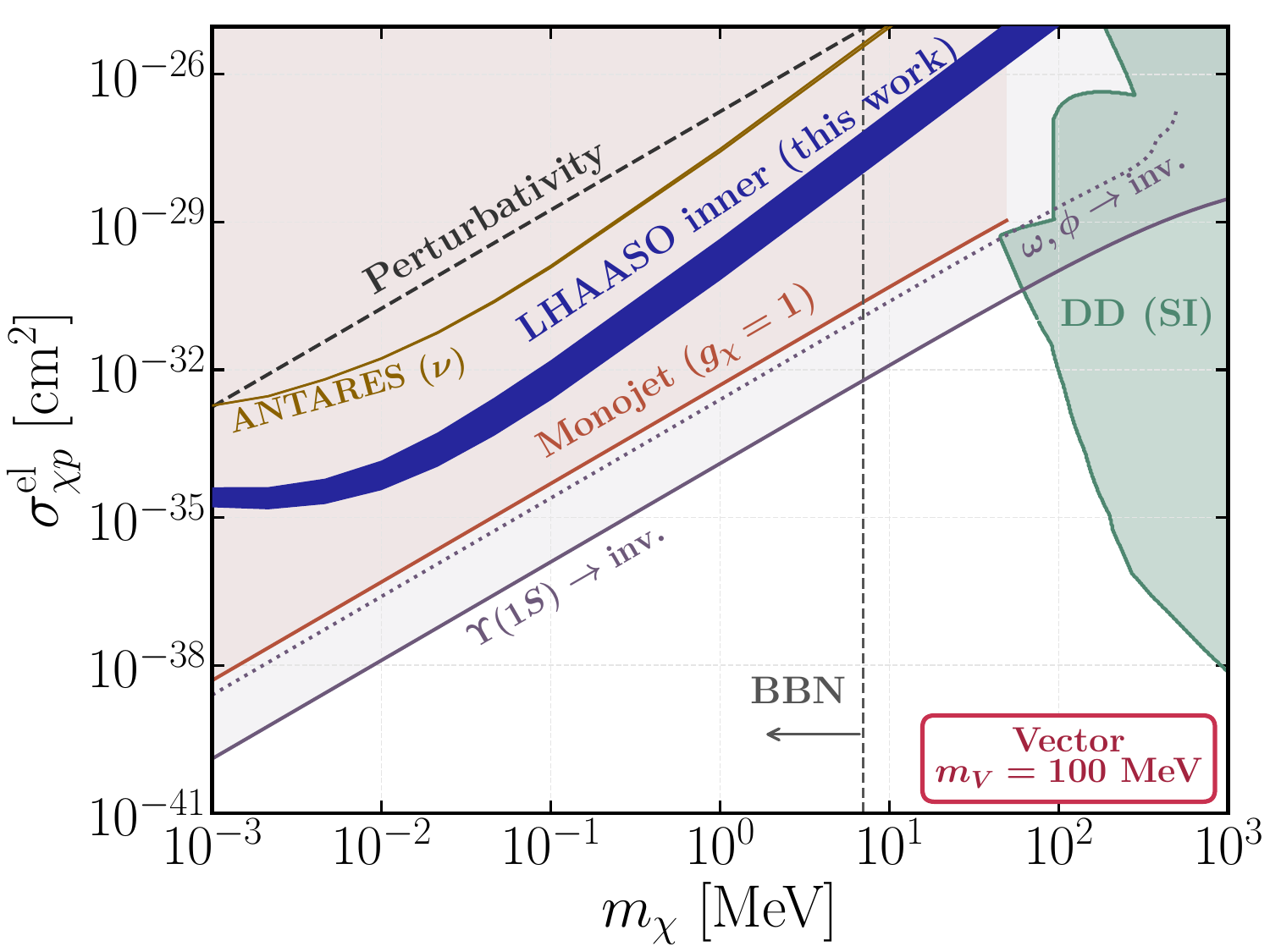}
    \caption{Upper limits on the SI DM--proton cross section for the vector mediator, with $m_V=5\GeV$ (\textit{left}) and $100\MeV$ (\textit{right}). The LHAASO band spans the limits without background (upper edge) and with the CR background (lower edge); the ANTARES band uses the analogous treatments (Sec.~\ref{sec:Disc_Neutrinos}). Complementary bounds are monojet searches for $g_{\chi V}=1$~\cite{Shoemaker:2011vi}, invisible $\Upsilon$ decays (shaded)~\cite{BaBar:2009gco}, invisible $\omega,\phi$ decays (dotted)~\cite{BESIII:2018bec}, DD (Sec.~\ref{sec:Disc_Scattering}). The BBN line excludes $m_\chi<7.0\MeV$ only if $\chi$ equilibrated with the SM plasma~\cite{Sabti:2019mhn}. Dashed: $G_D^2=16\pi^2$, the product boundary from individual-coupling perturbativity. Monojet bounds end at $m_\chi=m_V/2$; no off-shell continuation is included. DD ceilings use published attenuation boundaries where available; closure at $10^{-31}$~cm$^2$ elsewhere is only for display. Large-cross-section DD regions are indicative (App.~\ref{sec:AppBounds_DD}). CRDM regions requiring a mediator-mass recast are omitted (Sec.~\ref{sec:Disc_CRDM}).}
    \label{fig:LimitsVector}
\end{figure}

\begin{figure}[t]
    \centering
    \includegraphics[width=0.49\textwidth]{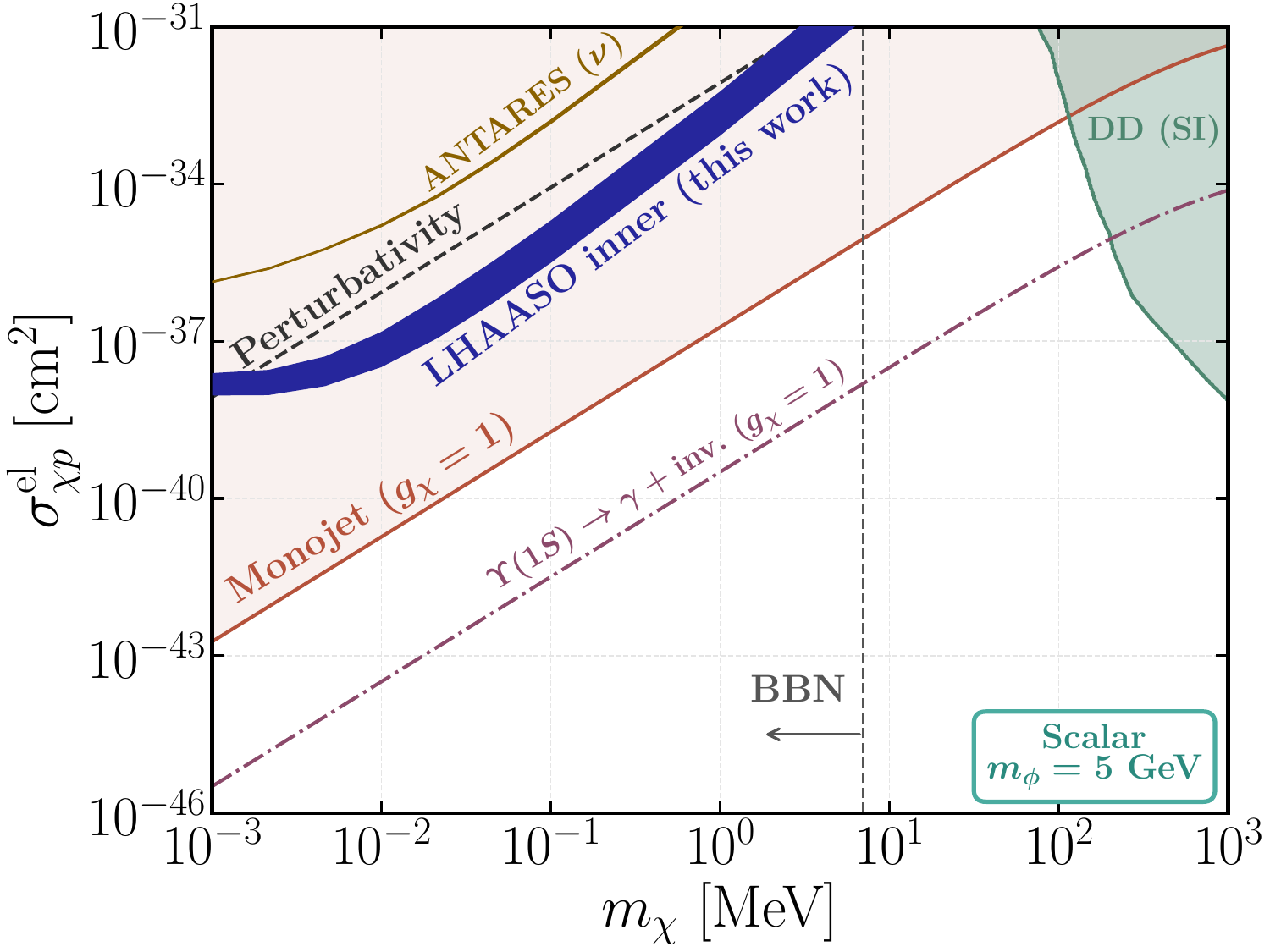}
    \includegraphics[width=0.49\textwidth]{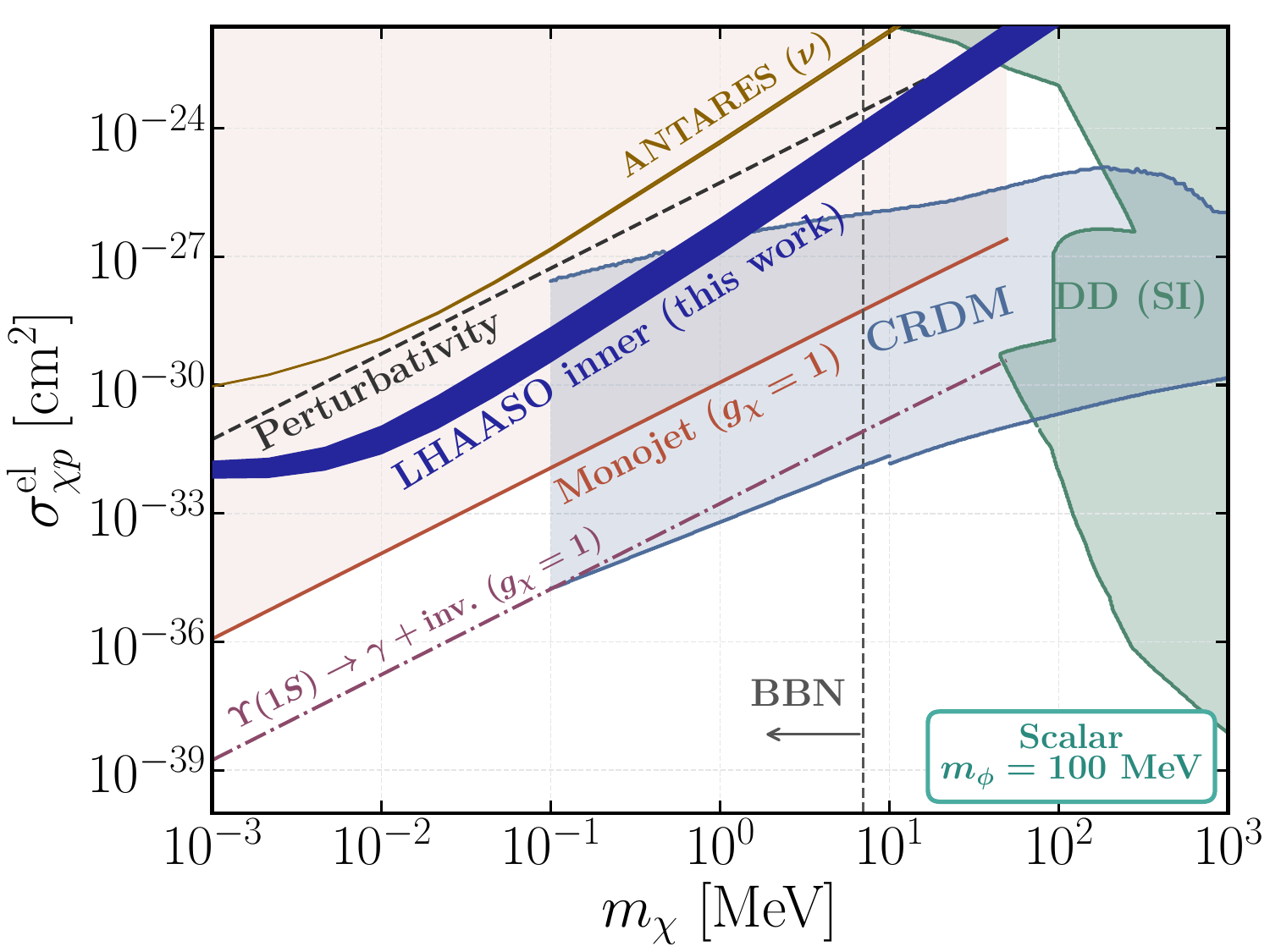}
    \caption{As in Fig.~\ref{fig:LimitsVector}, for the scalar mediator with $m_\phi=5\GeV$ (\textit{left}) and $100\MeV$ (\textit{right}). Monojet bounds~\cite{ATLAS:2021kxv,Ema:2020ulo} and radiative $\Upsilon(1S)\to\gamma\phi$, $\phi\to\chi\bar\chi$ bounds (dash-dotted)~\cite{BaBar:2010eww} use $g_{\chi\phi}=1$ and end at $m_\chi=m_\phi/2$. The SI DD region is shared with the vector case. At $100\MeV$, the CRDM region~\cite{Bondarenko:2019vrb,Alvey:2022pad} includes its attenuation ceiling; no CRDM region is drawn at $5\GeV$. The LHAASO/ANTARES bands, conditional BBN line and perturbativity boundary follow Fig.~\ref{fig:LimitsVector}. Kaon-decay bounds requiring a flavour-specific calculation are omitted (Sec.~\ref{sec:Disc_Mesons}).}
    \label{fig:LimitsScalar}
\end{figure}

\begin{figure}[t]
    \centering
    \includegraphics[width=0.49\textwidth]{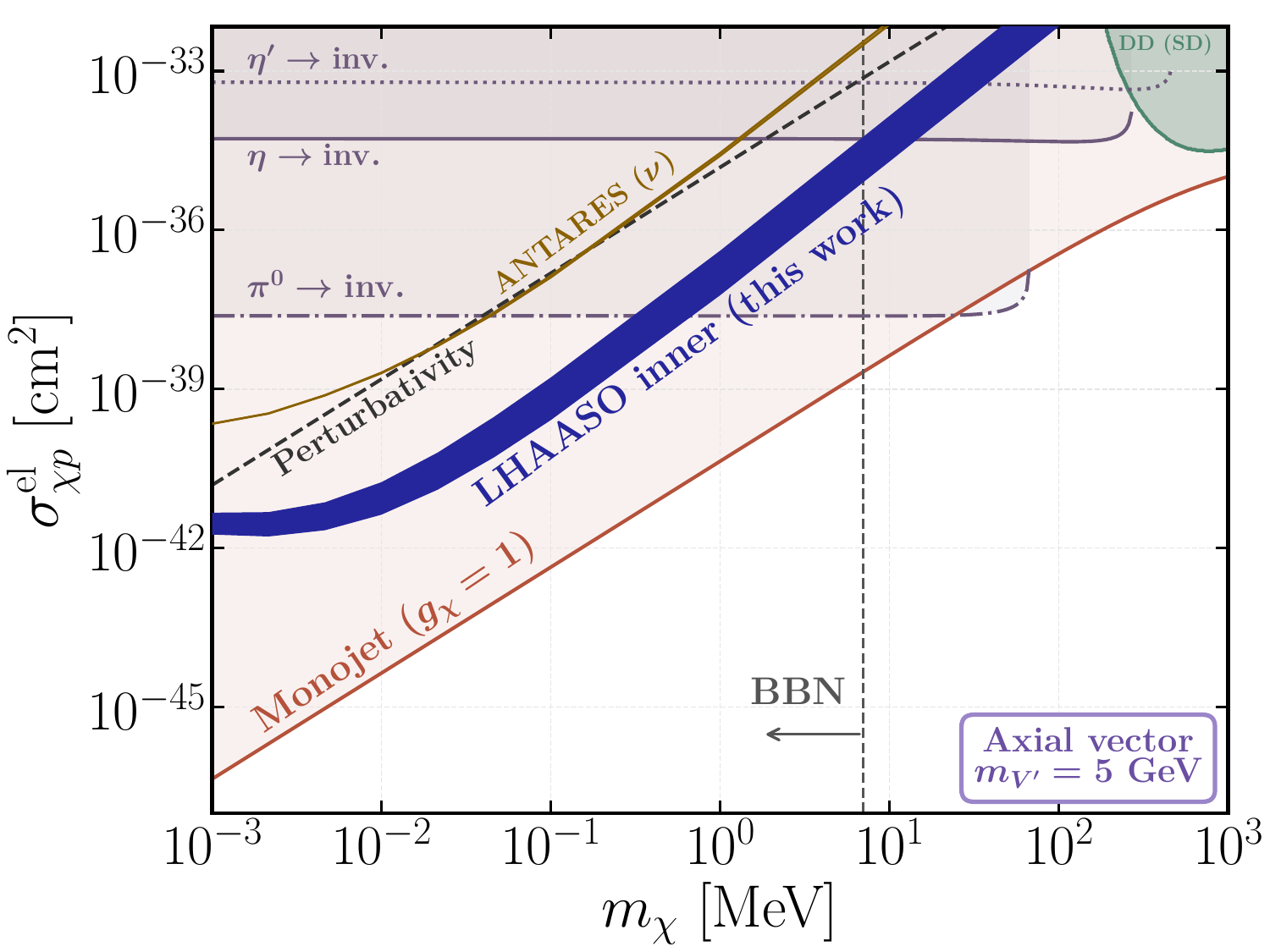}
    \includegraphics[width=0.49\textwidth]{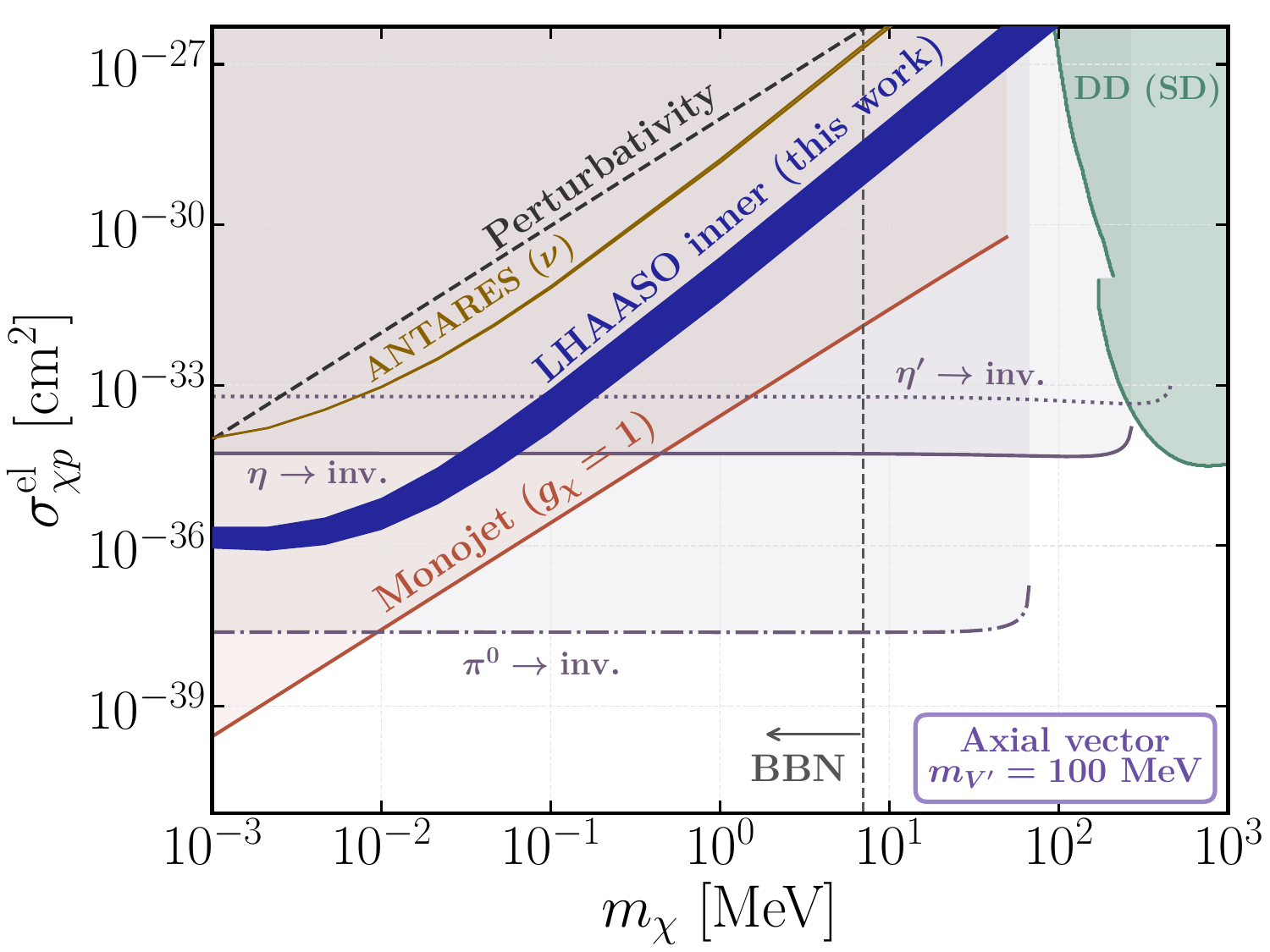}
    \caption{As in Fig.~\ref{fig:LimitsVector}, for the axial-vector mediator with $m_{V'}=5\GeV$ (\textit{left}) and $100\MeV$ (\textit{right}), in terms of the SD DM--proton cross section. Monojet bounds use $g_{\chi V'}=1$~\cite{Shoemaker:2011vi} and end at $m_\chi=m_{V'}/2$. Invisible $\pi^0,\eta,\eta'$ bounds~\cite{NA62:2020pwi,BESIII:2012nen} end at the respective meson thresholds; the $\pi^0$ bound uses leading-order isospin breaking. The SD DD response and attenuation qualifications are given in Sec.~\ref{sec:Disc_Scattering}. Bands and the conditional BBN and perturbativity lines follow Fig.~\ref{fig:LimitsVector}; additional axial unitarity restrictions are discussed in Sec.~\ref{sec:Disc_Theory}.}
    \label{fig:LimitsAxial}
\end{figure}

\begin{figure}[t]
    \centering
    \includegraphics[width=0.49\textwidth]{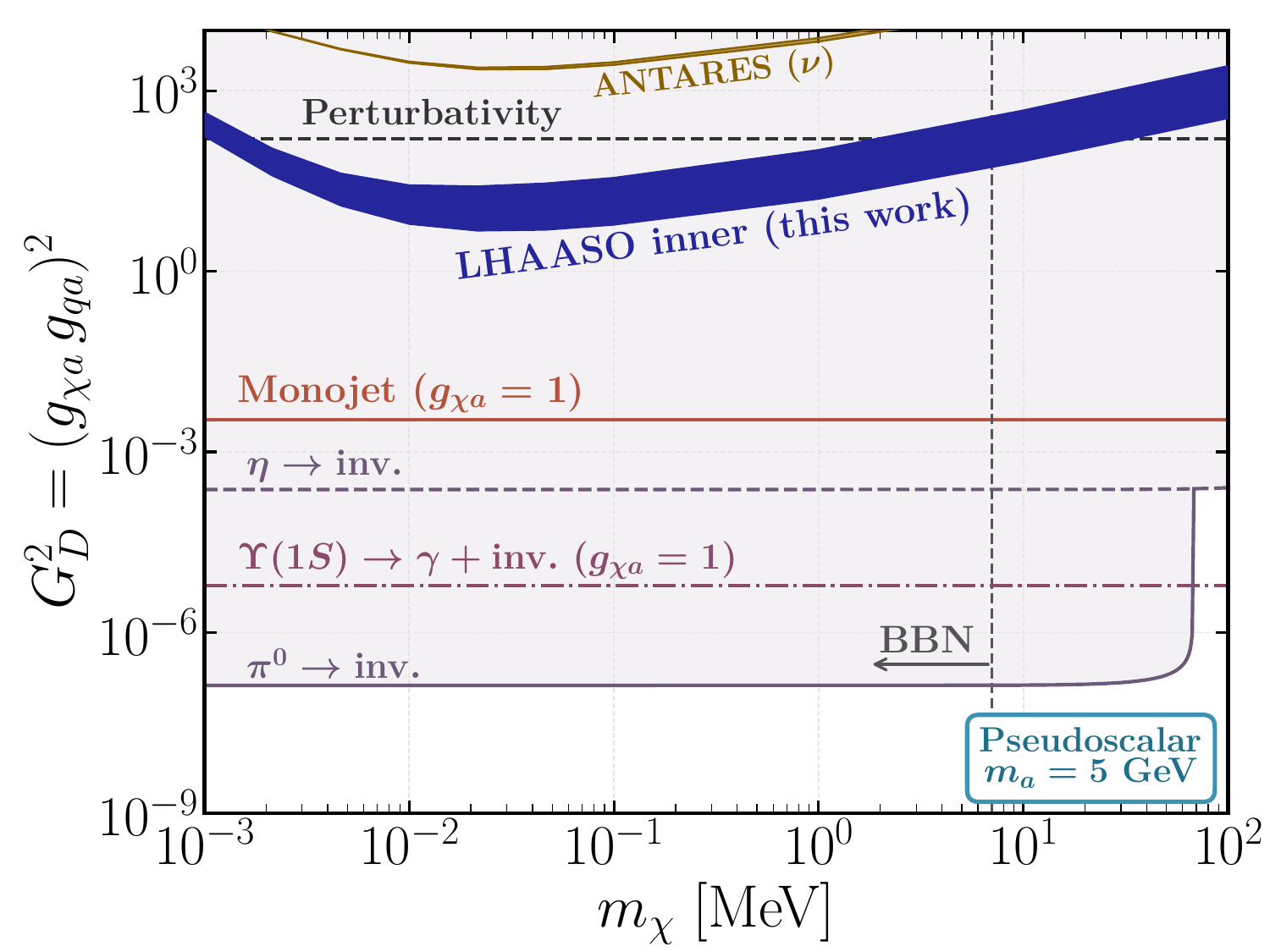}
    \includegraphics[width=0.49\textwidth]{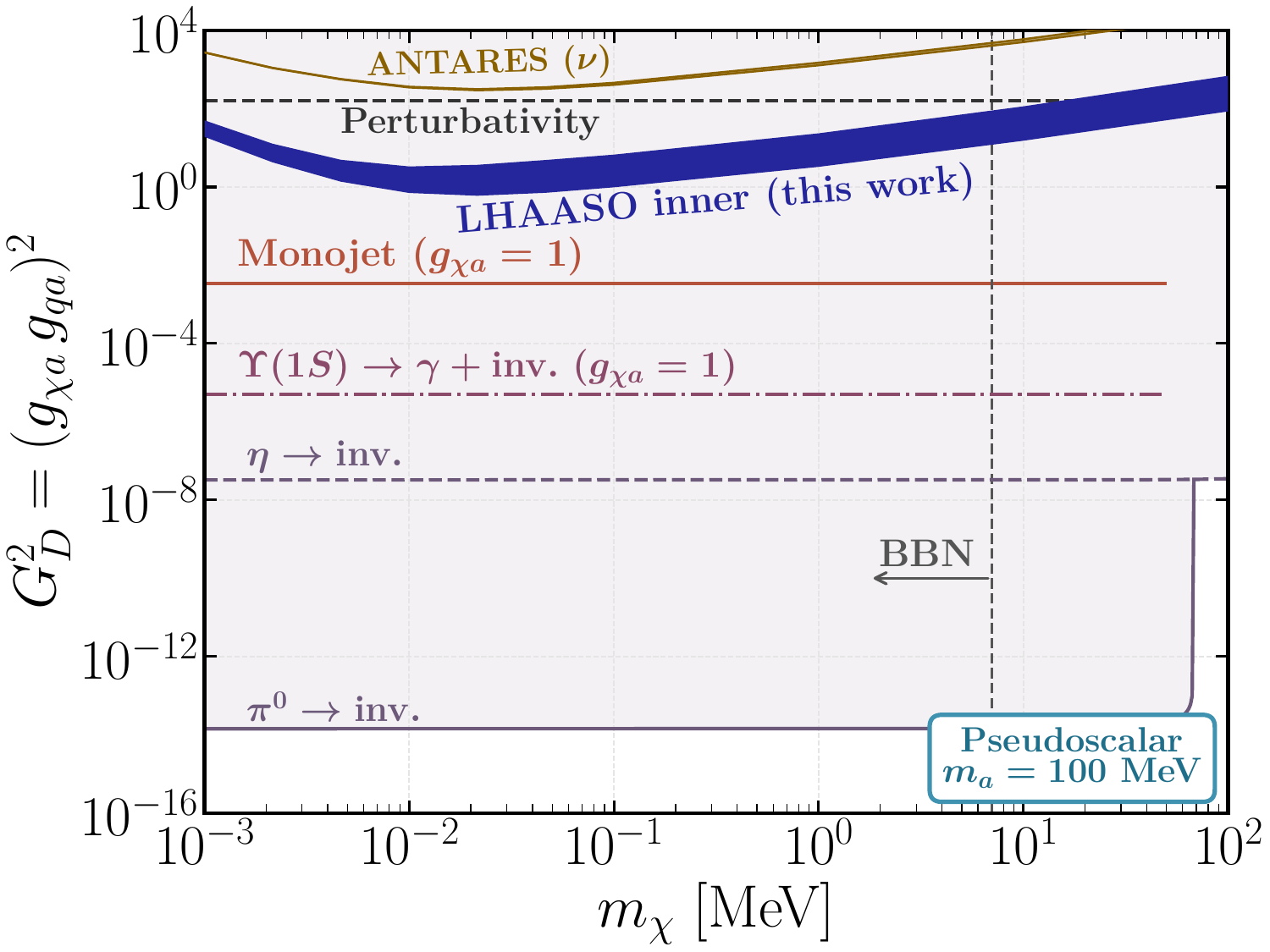}
    \caption{Upper limits on $G_D^2$ for the pseudoscalar mediator with $m_a=5\GeV$ (\textit{left}) and $100\MeV$ (\textit{right}). LHAASO and ANTARES bands, conditional BBN and the perturbativity boundary follow Fig.~\ref{fig:LimitsVector}. Shaded: invisible $\pi^0,\eta$ bounds~\cite{NA62:2020pwi,BESIII:2012nen}, recomputed with flavour-universal mediator--meson mixing~\cite{Ema:2020ulo,Feldmann:1998vh}; each curve ends at $m_\chi=m_a/2$. Radiative $\Upsilon$~\cite{BaBar:2010eww} and monojet~\cite{ATLAS:2021kxv,Ema:2020ulo} bounds use $g_{\chi a}=1$ and end at $m_\chi=m_a/2$.}
    \label{fig:LimitsPseudoscalar}
\end{figure}

Figures~\ref{fig:LimitsVector}--\ref{fig:LimitsPseudoscalar} compare the bounds for our two mediator masses in terms of the elastic DM--proton cross section, or $G_D^2$ for the pseudoscalar. Unless stated otherwise, numerical comparisons refer to our limits including the CR background; the figures retain both background treatments. Details of the conversions and additional numerical comparisons are collected in App.~\ref{sec:AppBounds}.

\subsection{Galactic neutrinos}
\label{sec:Disc_Neutrinos}

The closest comparison is with Galactic neutrinos: the charged pions produced in the same CR--DM collisions yield a diffuse neutrino flux~\cite{TerolCalvo:2026xee}. Using the same CR model, DM profile and DIS simulation, we obtain LHAASO limits on $G_D^2$ stronger than those from ANTARES observations of the Galactic ridge~\cite{ANTARES:2022izu} by a factor of $40$--$120$ without astrophysical backgrounds, and $120$--$520$ when conventional diffuse emission is included in both analyses. For ANTARES background we use the prediction of the $\gamma$-optimized Min model~\cite{Zuriaga-Puig:2026mfz}.

This improvement holds for all four mediators and both mediator masses. Its weak dependence on the interaction follows from the common hadronic origin of the photons and neutrinos. It is largest at $m_\chi\simeq10$--$50$~keV, and mainly reflects the larger statistics of LHAASO and its coverage up to PeV energies, where the signal of the lightest DM peaks. The conventional neutrino background changes the ANTARES limits by at most $20\%$.

\subsection{Cosmic-ray-boosted DM}
\label{sec:Disc_CRDM}

Elastic CR--DM collisions can accelerate DM to relativistic energies, allowing its detection through scattering in terrestrial detectors~\cite{Bringmann:2018cvk}. These CR-boosted DM (CRDM) bounds depend strongly on the Lorentz structure and mediator mass~\cite{Bondarenko:2019vrb,Alvey:2022pad,Yu:2025bwu}. We therefore compare only calculations with the appropriate interactions and equal proton and neutron couplings; bounds obtained with an energy-independent cross section cannot be applied directly (App.~\ref{sec:AppTheory_ConstXS}).

For the $5\GeV$ benchmark, the available CRDM curves were computed for lighter mediators. At fixed elastic cross section, they give conservative lower sensitivity edges, but their upper edges cannot be transferred: stronger scattering also increases the energy loss of DM before it reaches the detector. We consequently discuss these curves without plotting an excluded region for $m_Y=5\GeV$. For the vector, our limits improve on the published lower edges over an intermediate mass interval, by more than an order of magnitude around $m_\chi\simeq1$--$10\MeV$~\cite{Dent:2019krz,Diurba:2024dqo,Cappiello:2024acu}. For the scalar, the lower edges are stronger above $m_\chi\simeq0.4\MeV$, but the comparison becomes uncertain where our limits approach or exceed the attenuation ceilings~\cite{Bondarenko:2019vrb,Alvey:2022pad,Ema:2020ulo}. The Borexino axial-vector bound is weaker throughout our mass range~\cite{Dent:2019krz}. The numerical comparisons and the attenuation qualifications are given in App.~\ref{sec:AppBounds_CRDM}.

For $m_Y=100\MeV$, only bounds evaluated at that mediator mass can be used. The published scalar curves provide a full excluded region~\cite{Bondarenko:2019vrb,Alvey:2022pad}, which covers our limits for $0.1\lesssim m_\chi\lesssim1.6\MeV$. At higher masses our limits lie above the attenuation ceiling. A separate vector calculation with $m_V=3m_\chi$~\cite{Guo:2020oum} coincides with our benchmark only at $m_\chi\simeq33\MeV$; there, its XENON1T and MiniBooNE bounds cover our limits. Its Galactic gamma-ray analysis also provides a useful comparison, but uses different CR distributions and observations at lower photon energies (App.~\ref{sec:AppBounds_CRDM}).

CRs in blazar jets provide another source of boosted DM. The Super-K bounds for TXS~0506+056 and a stack of blazars~\cite{DeMarchi:2025uoo,Super-Kamiokande:2022ncz} are stronger than ours by one to four orders of magnitude for the vector, scalar and axial-vector models with $m_Y=5\GeV$. Their original restriction to $u,d$ couplings affects this comparison only modestly (App.~\ref{sec:AppBounds_CRDM}). Their principal uncertainty is the density of the DM spike around the central black hole, which can vary by orders of magnitude. 

\subsection{DM scattering in terrestrial detectors}
\label{sec:Disc_Scattering}

The nuclear-recoil direct-detection (DD) searches included here mainly overlap the upper end of our DM mass range. Apart from the large-cross-section XQC region, the combined SI and SD regions begin at $m_\chi\simeq45$ and $95\MeV$, respectively. For $m_Y=5\GeV$, DD reaches our vector and scalar limits only at masses of order $70$--$90\MeV$, and remains weaker for the axial vector. For $m_Y=100\MeV$, the vector limits fall in a gap between the terrestrial and XQC regions, while the scalar overlaps XQC where the nuclear-scattering treatment is uncertain; the axial-vector limits meet DD only near $100\MeV$.

The relevant searches use low-threshold calorimeters~\cite{TESSERACT:2025tfw,SuperCDMS:2020aus,CRESST:2019jnq,CRESST:2024cpr,EDELWEISS:2019vjv} and light targets~\cite{Collar:2018ydf,NEWS-G:2024jms}. Both mediator masses are far above the recoil momentum transfers, allowing the same contact limits to be used. Flavour-universal vector and scalar couplings give approximately isospin-conserving SI scattering. For the axial vector, we use the published proton- or neutron-only SD limits as conservative lower sensitivity edges for $a_p=a_n$, with the nuclear-response qualifications in App.~\ref{sec:AppBounds_DD}. The pseudoscalar's momentum-suppressed tree-level scattering is not covered by these limits; loop-induced SI bounds~\cite{Abe:2018emu} have not been computed for our flavour-universal benchmark.

At large cross sections, stopping in the atmosphere, Earth and shielding closes the DD regions. Consequently, these constraints are affected by the overburden problem, unlike gamma-ray observations.
We use published upper edges where available; the closure at $10^{-31}$~cm$^2$ elsewhere is a display convention, not a physical attenuation boundary. The SD ceilings require additional care for $a_p=a_n$, and the breakdown of coherent nuclear scaling makes the largest-cross-section surface and XQC regions indicative~\cite{Erickcek:2007jv,Mahdawi:2017cxz,Digman:2019wdm}.

We omit Migdal and bremsstrahlung limits, which rely on calculated atomic emission probabilities. The Migdal effect in nuclear recoils has been measured in a gas~\cite{Xu:2026acq}, but not in the liquid or solid targets relevant here, and the bremsstrahlung signal has not been measured. These searches could substantially extend the low-mass reach; their numerical impact and the DD region construction are detailed in App.~\ref{sec:AppBounds_DD}.

\subsection{Cosmological bounds}
\label{sec:Disc_Cosmo}

Cosmology constrains these interactions under assumptions distinct from those of present-day scattering searches. For a Dirac fermion in thermal equilibrium with photons and electrons, BBN with a CMB prior on the baryon density excludes $m_\chi<7.0\MeV$~\cite{Sabti:2019mhn,Krnjaic:2019dzc}, including the masses at which our $G_D^2$ limits are strongest. Such a particle changes the expansion rate and the relative heating of photons and neutrinos. Whether our quark-coupled benchmarks actually establish this equilibrium at temperatures of a few MeV requires a dedicated calculation, which we have not performed. The BBN line in the figures is therefore conditional on the thermal history. The Galactic gamma-ray limits require only the present halo population.

DM--baryon scattering also suppresses small-scale structure. The Lyman-$\alpha$ forest~\cite{Rogers:2021byl} gives stronger limits than Milky Way satellite counts~\cite{Maamari:2020aqz,Alvey:2022pad}. Both analyses include hydrogen scattering, so their velocity-independent bounds apply to the vector, scalar and axial-vector benchmarks at either mediator mass, but not to the momentum-suppressed pseudoscalar interaction. For $m_Y=5\GeV$, our vector and axial-vector limits are stronger, whereas the Lyman-$\alpha$ bound overtakes the scalar limits towards the upper end of the mass range. For $m_Y=100\MeV$, structure formation is more restrictive for the vector and scalar at masses around $10\MeV$ and above, and becomes competitive with the axial vector there. These bounds do not require thermal equilibrium, but assume that all DM interacts and depend on modelling the intergalactic thermal history or the relation between galaxies and subhalos. They are not plotted; numerical comparisons are given in App.~\ref{sec:AppBounds_Cosmo}.

CMB anisotropies~\cite{Lopez-Honorez:2013cua,Slatyer:2015jla,Planck:2018vyg} and, prospectively, the dark-ages 21-cm signal~\cite{Mohapatra:2025qpz} constrain energy injection from light DM annihilation. These bounds require comparable $\chi$ and $\bar\chi$ densities, whereas our signal requires only the present halo population and also applies to asymmetric DM. With quark-only couplings, tree-level annihilation into visible states opens only near the top of our mass range, e.g.\ $\chi\bar\chi\to\pi^0\gamma$ for the vector above $m_{\pi^0}/2$. At lower masses it proceeds through loop-induced couplings that depend on the completion, such as vector--photon mixing, which gives $e^+e^-$ for $m_\chi>m_e$. A rough estimate indicates that, for symmetric DM, the CMB bound then exceeds our vector limits by orders of magnitude for $m_V=100\MeV$, and becomes competitive above a few MeV for $m_V=5\GeV$. Below $m_e$, where our limits are strongest, this channel is closed.

\subsection{Monojet searches}
\label{sec:Disc_Colliders}

At hadron colliders, an invisibly decaying mediator produced with a jet gives missing transverse momentum. For on-shell production, the rate scales as $g_{qY}^2\,{\rm BR}(Y\to\chi\bar\chi)$. When $g_{\chi Y}\gg g_{qY}$ and $m_\chi<m_Y/2$, the invisible branching ratio is close to one and the search constrains the quark coupling, whereas our signal constrains the product $g_{\chi Y}g_{qY}$.

For $g_{\chi Y}=1$, the recasts used here give $g_{qY}<0.021$ for vector and axial-vector mediators~\cite{Shoemaker:2011vi}, and $g_{qY}<0.059$ for scalar and pseudoscalar mediators, after rescaling the ATLAS recast to flavour-universal couplings~\cite{ATLAS:2021kxv,Ema:2020ulo}. They are stronger than our limits by factors of $40$--$1300$. The rescaling, its uncertainties and the extrapolation to our mediator masses are described in App.~\ref{sec:AppBounds_Monojet}. 

No experimental collaboration reports monojet bounds for sub-GeV DM; the bounds used here are phenomenological recasts. Reference~\cite{Shoemaker:2011vi} finds a negligible dependence on the DM mass for on-shell decays of mediators down to $0.1\GeV$, while Ref.~\cite{Ema:2020ulo} recasts the ATLAS search for $m_Y=1$ and $3\GeV$, which we extend to our mediator masses. These bounds should therefore be taken as indicative; a dedicated analysis is beyond the scope of this work.

Changing the division of the coupling product weakens this comparison: increasing $g_{\chi Y}$ raises the bound on $G_D^2$, while increasing $g_{qY}/g_{\chi Y}$ reduces the invisible branching ratio. Within the on-shell recasts and requiring each coupling to remain below $\sqrt{4\pi}$, neither effect makes our limits stronger (App.~\ref{sec:AppBounds_Monojet}). For $m_Y=100\MeV$, however, the plotted bounds end at $m_\chi=m_Y/2$; an off-shell recast above this threshold has not been performed. Restricting the mediator to $u,d$ couplings changes the monojet limits only slightly, while weakening our $G_D^2$ limits by at most $25\%$ (App.~\ref{sec:AppTheory_Flavour}).

\subsection{Meson decays}
\label{sec:Disc_Mesons}

Meson decays probe the mediator's flavour couplings and current quantum numbers. Invisible decays through an off-shell mediator constrain $G_D^2$ directly, independently of its division between the DM and quark couplings. Production of an on-shell mediator followed by an invisible decay also depends on its branching ratio, as in monojet searches. We distinguish these cases below; numerical conversions are collected in App.~\ref{sec:AppBounds_Mesons}.

For the vector, the strongest meson bound comes from $\Upsilon(1S)\to\chi\bar\chi$ through an off-shell mediator. Normalising to the dimuon width, the BaBar invisible-decay limit~\cite{BaBar:2009gco} gives $G_D^2<5.9\times10^{-6}$ for $m_V=5\GeV$ and $1.1\times10^{-5}$ for $100\MeV$, about four and three orders of magnitude below our limits. The $J/\psi$ bound~\cite{BES:2007sxr} is weaker but remains restrictive. Both disappear if the mediator couples only to $u,d$ quarks.

The light vector mesons $\omega$ and $\phi$ provide analogous invisible-decay bounds~\cite{BESIII:2018bec}. For $m_V=5\GeV$ they are comparable to our limits, which are stronger around $m_\chi\simeq20$--$120$~keV. For $m_V=100\MeV$ both are substantially stronger. The $\omega$ bound remains relevant for a mediator coupled only to $u,d$, even when the quarkonium bounds are absent.

For scalar and pseudoscalar mediators, the relevant quarkonium channel is instead $\Upsilon(1S)\to\gamma Y$, followed by $Y\to\chi\bar\chi$. The BaBar search for a photon plus invisible particles~\cite{BaBar:2010eww} gives $G_D^2\lesssim6\times10^{-6}$ for $g_{\chi Y}=1$ and an approximately unit invisible branching ratio. In this regime the bound weakens as $g_{\chi Y}^2$, reaching our limits only for non-perturbative couplings. For the light mediator it ends at the invisible-decay threshold. The axial-vector current cannot mediate the direct invisible decay of the $1^{--}$ $\Upsilon$, although $\Upsilon\to\gamma V'$ is allowed. We do not apply the BaBar radiative bound to this case because the search assumed a spin-0 invisible state.

For axial-vector and pseudoscalar mediators, invisible decays of $\pi^0$, $\eta$ and $\eta'$ provide complementary bounds. With $g_{uY}=g_{dY}$ the $\pi^0$ contribution vanishes in the isospin limit, but is restored by isospin-breaking meson mixing. For the axial vector, the longitudinal mediator gives a helicity-suppressed rate. Despite this suppression, the NA62 $\pi^0$ limit~\cite{NA62:2020pwi} is the strongest of these meson bounds, subject to the leading-order treatment of isospin breaking. At $m_{V'}=5\GeV$, our limits improve on it below $m_\chi\simeq0.7\MeV$, and on the weaker $\eta,\eta'$ bounds~\cite{BESIII:2012nen} over a wider range. At $m_{V'}=100\MeV$, the $\pi^0$ bound is stronger up to its kinematic endpoint, with $\eta,\eta'$ decays constraining the higher masses.

For the pseudoscalar, we recompute mediator--meson mixing for flavour-universal couplings, including the strange quark, physical $\eta$--$\eta'$ mixing and $m_u\neq m_d$~\cite{Ema:2020ulo,Feldmann:1998vh}. The resulting invisible-decay bounds lie three to thirteen orders of magnitude below our limits at the two mediator masses. The uncertainties of the leading-order mixing calculation do not change this hierarchy.

Scalar-mediated $K^+\to\pi^+\chi\bar\chi$ could also be strongly constrained~\cite{NA62:2021zjw,Cox:2024rew,DeMarchi:2025uoo}. Published conversions use different flavour structures; our benchmark requires both the corresponding light-quark matrix elements and the top-loop contribution to $s\to d$. We therefore omit this bound pending a dedicated calculation. For $m_Y=100\MeV$, on-shell production in rare meson decays and beam dumps offers further constraints, also not included here. Their interpretation can depend on the UV completion, as discussed below.

\subsection{Theoretical consistency}
\label{sec:Disc_Theory}

We require each coupling to be perturbative, $g_{\chi Y},g_{qY}<\sqrt{4\pi}$, implying the dashed boundary $G_D^2<16\pi^2$ in the figures. A point below this line still requires a perturbative choice of both couplings. Our vector and axial-vector limits satisfy the product bound throughout. For the $5\GeV$ scalar and pseudoscalar, they exceed it above $m_\chi\simeq35\MeV$; the restrictions without the CR background are more severe.

The axial-vector benchmark has an additional restriction: perturbative unitarity of longitudinal-mediator scattering constrains every coupled fermion mass~\cite{Kahlhoefer:2015bea}. With flavour-universal couplings, the top quark restricts $G_D^2$ below the values probed here at both mediator masses. Setting $g_{tV'}=0$ leaves our five-flavour signal calculation and the plotted complementary limits unchanged. The remaining bottom-quark condition then permits the $5\GeV$ benchmark at our limits including the CR background, but still restricts the $100\MeV$ benchmark below our sensitivity. This flavour qualification is essential when interpreting the axial-vector comparison; the individual bounds are given in App.~\ref{sec:AppBounds_Theory}.

Additional constraints depend on the ultraviolet completion of the simplified interactions, as discussed in App.~\ref{sec:AppBounds_Theory}. Our gamma-ray limits apply provided that the assumed interactions describe the scattering over the kinematic range contributing to the signal.

\subsection{Summary by mediator}
\label{sec:Disc_Summary}

\textit{Vector.} Invisible quarkonium decays dominate for flavour-universal couplings. With only $u,d$ couplings, our $5\GeV$ benchmark is more competitive: our limits rival light-meson bounds and improve on several scattering probes at low masses. Monojet searches remain stronger, and blazar bounds are stronger under their DM-spike assumptions.

\textit{Scalar.} Monojet and radiative $\Upsilon$ searches strongly constrain the on-shell invisible regime. CRDM and structure formation also compete, but attenuation limits the CRDM comparison, particularly when the published mediator mass differs from ours. At high DM masses, perturbativity further limits the interpretation of the $5\GeV$ benchmark.

\textit{Axial vector.} With the top coupling removed, the $5\GeV$ model offers the strongest comparison with DD, Galactic CRDM and structure formation, and improves on light-meson bounds at low masses. Monojet and spike-dependent blazar bounds remain stronger. The flavour-universal model and the light-mediator benchmark face the additional unitarity restrictions above.

\textit{Pseudoscalar.} Invisible light-meson decays are much more restrictive at both mediator masses. The momentum suppression of elastic scattering prevents direct use of the standard DD and velocity-independent structure-formation bounds.

The common complementarity is the Galactic origin of the signal. Laboratory missing-energy searches constrain invisible particles irrespective of their cosmological abundance; our limits scale inversely with the fraction of halo DM in $\chi$. Within the scattering interpretation, an observed gamma-ray signal would therefore connect the interaction to a Galactic DM population and probe its spatial distribution, complementing a laboratory discovery.

\section{Conclusions} \label{sec:Conclusions}

We have calculated the diffuse TeV--PeV gamma-ray emission from inelastic collisions of Galactic CRs with DM of mass $1$~keV to $100\MeV$. The calculation combines spatially varying proton and helium distributions with simulations of deep-inelastic scattering, hadronization and decays for four mediator types. It predicts both the photon spectrum and its relation to the underlying interaction, which a constant-cross-section prescription cannot capture. The near equality of the scalar/pseudoscalar signals and of the vector/axial-vector signals reduces the high-energy calculation to two classes, while their low-energy constraints remain distinct.

The LHAASO inner-Galaxy data provide the strongest baseline limits among the observations considered. For a $5\GeV$ vector mediator, including conventional CR--gas emission gives $G_D^2\lesssim0.25$ near the most sensitive DM masses, $10$--$50$~keV, and an elastic cross-section limit of $3\times10^{-41}$~cm$^2$ at $1$~keV. The gamma-ray limits improve on the corresponding ANTARES limits by factors of $40$--$120$ without backgrounds and $120$--$520$ with them. Their dependence on the Galactic DM profile is very modest in the LHAASO region, whereas the high-energy CR distribution introduces uncertainties up to an order of magnitude for the lightest DM. HAWC can become more constraining under an additional unresolved-source prescription, illustrating the importance of background control.

Generally, these constraints are stronger than other astrophysical and cosmological scattering probes, although annihilation bounds can be stronger in part of the parameter space for symmetric DM. However, laboratory bounds  constrain the flavour-universal benchmarks more strongly in most cases. Quarkonium and light-meson decays dominate several comparisons, while monojet limits remain stronger within the on-shell, perturbative regime considered here. Relative to other scattering probes, the gamma-ray limits are more competitive for a $5\GeV$ vector with only $u,d$ couplings and for a $5\GeV$ axial vector with its top coupling removed. The latter qualification is required by the unitarity analysis. The complementary information is the signal's dependence on a Galactic DM population: within the scattering interpretation, an observation would connect the interaction to halo particles and their spatial distribution, beyond the evidence supplied by a laboratory invisible final state.

A dedicated spectral and spatial template analysis is the next step, treating the gas emission, unresolved sources and CR--DM signal together. Meanwhile, the preliminary source-subtracted LHAASO comparison already indicates a  potential gain of a factor 2-4. Extending LHAASO's observation period by another year would triple the accumulated statistics with respect to the published data, offering substantial gains in sensitivity as well. CTAO's angular resolution and SWGO's wide-field southern coverage offer complementary opportunities to study the inner Galaxy~\cite{Acharyya:2020,SWGO:2025}, provided their diffuse-emission analyses account for exposure and background systematics.

\section*{Acknowledgments}
\noindent
The authors want to thank Ernesto Arganda, Christopher Cappiello, Víctor Miralles, Paolo Panci, Farinaldo Queiroz and Filippo Sala for useful discussions.
PDL has been supported by the Juan de la Cierva JDC2022-048916-I grant, funded by MCIU/AEI/10.13039/501100011033 European Union "NextGenerationEU"/PRTR, and is currently supported by Ramón y Cajal RYC2024-048445-I grant, which is funded by MCIU/AEI/10.13039/501100011033 and FSE+. The work of PDL is also supported by the grants PID2021-125331NB-I00 and CEX2020-001007-S, both funded by MCIN/AEI/10.13039/ 501100011033 and by ``ERDF A way of making Europe''. PDL also acknowledges the MultiDark Network, ref. RED2022-134411-T. JTC acknowledges support from the Generalitat Valenciana
through PROMETEO PROJECT CIPROM/2022/13 and from MCIN/AEI (10.13039/501100011033) through Project PID2025-171322NB-C21, the  Research grant TAsP (Theoretical Astroparticle Physics) funded by \textsc{infn}, from the Italian Ministry of University and Research (MIUR) via the PRIN 2022 Project No. 2022F2843 “Addressing systematic uncertainties in searches for dark matter” and from the European Research Area (ERA) via the UNDARK project (project number 101159929). This project used computing resources from the National Academic Infrastructure for Supercomputing in Sweden (NAISS) under projects NAISS 2025-5-729 and NAISS 2024/5-666 and from MATCON cluster of the physics department of Universitat d'Alacant.

\bibliographystyle{bibi}
\bibliography{biblio_gamma.bib}

\appendix
\addtocontents{toc}{\protect\setcounter{tocdepth}{1}} 

\section{Particle-physics details of the CR--DM interaction}
\label{sec:AppTheory}

In this appendix, we collect the particle-physics details of the two descriptions of the CR--DM interaction introduced in Sec.~\ref{sec:DMModels}.

\subsection{Limitations of the constant cross section}
\label{sec:AppTheory_ConstXS}

The constant cross section is a convenient phenomenological device, but it is not a particle-physics model. As emphasized in Ref.~\cite{DeMarchi:2025uoo}, no known model of particle DM yields the same cross section over the range of energies spanned by CR--DM collisions. In our case the centre-of-mass energy ranges from a few GeV near threshold to a few TeV, and the momentum transfer ranges from the keV scale of non-relativistic elastic scattering to $Q^2 \gtrsim 1\GeV^2$, where the DM resolves the quarks inside the proton rather than the proton as a whole. Any interaction mediated by a particle of mass $m_Y$ responds to this hierarchy through the mediator propagator and the parton content of the proton. Two consequences matter here. First, the photon yield is not universal: how much of $E_{\rm CR}$ ends up in photons, and with what spectrum, depends on the momentum transferred in the hard process, and hence on the Lorentz structure of the interaction (App.~\ref{sec:AppTheory_DIS}). Second, the ratio between inelastic and elastic cross sections is not a constant. For the vector-mediator model of Ref.~\cite{TerolCalvo:2026xee}, for instance, at $\hat s\gg m_V^2,\,Q_0^2$
\begin{equation}
    \frac{\sigma^{\rm inel}_{\chi p}}{\sigma^{\rm el}_{\chi p}} \simeq \frac{N_{\rm eff}(s)}{36}\,\frac{m_V^4}{\mu_{\chi p}^2\,(Q_0^2+m_V^2)}\,,
    \label{eq:InelElRatio}
\end{equation}
where $\mu_{\chi p}$ is the DM--proton reduced mass and $N_{\rm eff}(s)$ the number of quarks and antiquarks resolved at momentum transfers above $Q_0^2\simeq4\GeV^2$, which grows from $\simeq7$ to $\simeq50$ between $s\simeq10^4$ and $10^7\GeV^2$. The elastic cross section is suppressed by $\mu_{\chi p}^2\simeq m_\chi^2$ for light DM, while the inelastic one is not, so this ratio exceeds the value $8/3$ implied by the constant-cross-section mapping by a factor of a few for $m_\chi=1\MeV$ and $m_V=100\MeV$, and by $\sim10^6$--$10^7$ for $m_\chi=1\MeV$ and $m_V=5\GeV$.

\subsection{Elastic cross sections and proton couplings}
\label{sec:AppTheory_NR}

The proton couplings $g_{pY}$ entering Eq.~\eqref{eq:SigmaElY} follow from the quark couplings through nucleon matrix elements. For the vector only the valence quarks carry net charge, so $g_{pV}=2g_{uV}+g_{dV}=3g_{qV}$ exactly, and Eq.~\eqref{eq:SigmaElY} reduces to $\sigma^{\rm el}_{\chi p}=9\,G_D^2\,\mu_{\chi p}^2/(\pi m_V^4)$, as in Ref.~\cite{TerolCalvo:2026xee}. For the scalar the proton coupling is set by the quark scalar densities,
\begin{equation}
    g_{p\phi} = m_p\left[\sum_{q=u,d,s}\frac{f^p_{Tq}}{m_q}\,g_{q\phi} + \frac{2}{27}\,f^p_{TG}\sum_{Q=c,b,t}\frac{g_{Q\phi}}{m_Q}\right],
    \label{eq:gpphi}
\end{equation}
with $f^p_{TG}=1-\sum_{q=u,d,s}f^p_{Tq}$. The $u$ and $d$ quarks contribute $g_{p\phi}\simeq 8.4\,g_{u\phi}+7.6\,g_{d\phi}$~\cite{DeMarchi:2025uoo}, the heavier flavours add a few percent, and the result carries an uncertainty of order $40\%$ from the pion--nucleon sigma term. For the axial vector, $g_{pV'}=\sqrt{3}\sum_q\Delta^p_q\,g_{qV'}$ is set by the quark spin fractions $\Delta_q^p$, so that Eq.~\eqref{eq:SigmaElY} is the standard SD cross section~\cite{DeMarchi:2025uoo}. Ref.~\cite{DeMarchi:2025uoo} includes only the $u$ and $d$ quarks, with $\Delta^p_u=0.845$ and $\Delta^p_d=-0.405$. For our flavour-universal couplings we add the strange-quark contribution $\Delta^p_s=-0.035$ from lattice QCD~\cite{Liang:2018pis} and neglect the heavy quarks, which gives $g_{pV'}\simeq0.70\,g_{qV'}$. For the pseudoscalar the tree-level elastic cross section is spin-dependent and momentum suppressed, and direct-detection rates are dominated by loop-induced SI scattering~\cite{Abe:2018emu}.

\subsection{Deep-inelastic scattering}
\label{sec:AppTheory_DIS}

For massless quarks, the spin-averaged squared amplitudes for $\chi q\to\chi q$ through $t$-channel exchange are
\begin{align}
    \overline{|\mathcal{M}|^2}_{\phi,\,a} &= G_D^2\,\frac{Q^2\left(Q^2+4\,\xi_Y\, m_\chi^2\right)}{\left(Q^2+m_Y^2\right)^2}\,, \qquad \xi_\phi=1,\;\; \xi_a=0\,, \label{eq:M2spin0}\\
    \overline{|\mathcal{M}|^2}_{V,\,V'} &= 2\,G_D^2\,\frac{(\hat s-m_\chi^2)^2+(\hat u-m_\chi^2)^2 \mp 2\,m_\chi^2\,Q^2}{\left(Q^2+m_Y^2\right)^2}\,, \label{eq:M2spin1}
\end{align}
where $Q^2=-\hat t$, and the upper (lower) sign refers to $V$ ($V'$). The scalar/pseudoscalar difference is a relative correction $4m_\chi^2/Q^2$, while the vector/axial-vector difference is of order $m_\chi^2/\hat s$. Both remain below $1\%$ for $m_\chi\leq100\MeV$ and $\hat s\geq Q^2\gtrsim Q_0^2\simeq4\GeV^2$. The same pairwise degeneracy appears in the cross sections of Ref.~\cite{DeMarchi:2025uoo}; our simulations confirm it for the photon yields as well as the rates.

Vector exchange is forward peaked, with
\begin{equation}
    \frac{d\hat\sigma_V}{dQ^2}\propto\frac{1+(1-y)^2}{\left(Q^2+m_V^2\right)^2}\,, \qquad y=\frac{Q^2}{\hat s-m_\chi^2}\,.
    \label{eq:dsigmaV}
\end{equation}
The integral is dominated by $Q^2\sim\max(Q_0^2,m_V^2)$. For scalar exchange the amplitude vanishes in the forward direction, since the scalar current flips the helicity of a massless quark. When $m_\phi^2\ll Q^2$, its momentum-transfer distribution is approximately flat up to $\hat s$. The corresponding partonic cross sections are~\cite{TerolCalvo:2026xee}
\begin{align}
    \hat\sigma_V &\simeq\frac{G_D^2}{4\pi}\left[\frac{1}{Q_0^2+m_V^2}-\frac{1}{\hat s+m_V^2}\right], \label{eq:sigmaV}\\
    \hat\sigma_\phi &\simeq\frac{G_D^2}{16\pi\hat s}\qquad (m_\phi^2\ll Q^2)\,. \label{eq:sigmaPhi}
\end{align}
For $m_V^2\ll Q_0^2\ll\hat s$, their ratio is approximately $Q_0^2/(4\hat s)$. Scalar collisions transfer more energy to DM: in the laboratory frame $y$ is the fraction of the incident quark energy carried away by DM, approximately uniform for the scalar and small for the vector. Less energy remains for the hadronic final state, giving softer photons despite a somewhat larger multiplicity.

Table~\ref{tab:SimulationRatios} separates these effects in the numerical comparison of Sec.~\ref{sec:Results_Mediators}. The production-cross-section ratios refer to $\sqrt{s}\simeq60$--$1200\GeV$, and the gamma-ray flux ratios to $m_\chi=1\MeV$ between $10\TeV$ and $1$~PeV. They describe different stages of the calculation and need not coincide.

\begin{table}[t]
\centering
\small
\renewcommand{\arraystretch}{1.15}
\begin{tabular}{lcc}
\hline
Quantity at equal $G_D^2$ & $m_Y=5\GeV$ & $m_Y=100\MeV$ \\
\hline
Vector/scalar proton DIS cross section & $25$--$55$ & $60$--$120$ \\
Vector/scalar gamma-ray flux & $20$--$30$ & $40$--$60$ \\
Scalar/vector photon multiplicity & \multicolumn{2}{c}{$1.20$--$1.35$ across the samples} \\
Scalar/vector mean photon energy & \multicolumn{2}{c}{$0.70$--$0.85$ across the samples} \\
\hline
\end{tabular}
\caption{Numerical origin of the scalar--vector signal hierarchy. The larger scalar photon yield only partly compensates its lower scattering rate, and its softer spectrum increases the flux difference at higher energies.}
\label{tab:SimulationRatios}
\end{table}

Changing $m_Y$ from $5\GeV$ to $100\MeV$ strengthens the $G_D^2$ limits by factors of $8$--$14$ for the vector and $4$--$9$ for the scalar. The scalar limits are weaker than the vector ones by factors of $9$--$40$ at $5\GeV$ and $13$--$90$ at $100\MeV$, with the ratio growing with $m_\chi$. In the elastic plane, the scalar proton-coupling coefficient gives a cross section about $30$ times larger at fixed $G_D^2$, making its limits roughly $300$--$3000$ times weaker than the vector's. The axial coefficient gives an SD cross section about $18$ times smaller than the vector SI cross section at the same product. For the lighter mediator the elastic limits are weaker by roughly $(5$--$7)\times10^5$, because the $m_Y^{-4}$ conversion outweighs the improvement in coupling sensitivity.

\subsection{Quark flavours and nuclear targets}
\label{sec:AppTheory_Flavour}

Ref.~\cite{DeMarchi:2025uoo} couples the mediators to $u$ and $d$ quarks only. Switching on the heavier flavours adds $s$, $c$ and $b$ initial states and raises the DIS cross section by $10$--$30\%$ in our simulations, with no visible change in the shape of the secondary spectra. With $u$ and $d$ couplings only, the resulting limits on $G_D^2$ are weaker by $3$--$5\%$ for $m_\chi\lesssim50$~keV and by up to $25\%$ at $m_\chi=100\MeV$, for all mediators and data sets.

For helium we use the factor $A^{2/3}$ of Eq.~\eqref{eq:sigmaCRchi}, which is conservative. At $Q^2\gtrsim Q_0^2$ the nucleons scatter incoherently, $\sigma_{A\chi}\simeq R_A\,A\,\sigma_{N\chi}$ with $\sigma_n=\sigma_p$ for flavour-universal couplings, as in ultra-high-energy neutrino DIS~\cite{Gandhi:1995tf,Garcia:2020jwr}. Shadowing gives $R_A\leq1$~\cite{Armesto:2006ph,Frankfurt:2011cs,Kovarik:2015cma,Eskola:2021nhw,AbdulKhalek:2022fyi}, e.g.\ $R_{\rm He}\simeq0.94$ at $x\simeq3\times10^{-3}$~\cite{NewMuon:1995cua,E665:1995xur,NewMuon:1996yuf}, and an effective scaling $A^{\alpha}$ with $\alpha\simeq0.90$--$0.95$ down to $x\sim10^{-4}$~\cite{Frankfurt:2011cs}. For helium, $A^{2/3}$ thus underestimates the cross section by $4^{\alpha-2/3}\simeq1.4$--$1.5$.


\section{Details of the complementary-bound comparisons}
\label{sec:AppBounds}

This appendix gives the conversions, numerical comparisons and applicability qualifications supporting Sec.~\ref{sec:Discussion}. Unless explicitly stated otherwise, comparisons use the LHAASO inner-Galaxy limits including the CR background. Parenthetical alternatives labelled ``without background'' refer to the same analysis without that contribution.

\subsection{Direct-detection responses and excluded regions}
\label{sec:AppBounds_DD}

For halo DM, $q\lesssim2\mu_{\chi N}v$ is at most a few MeV for $m_\chi\lesssim1\GeV$, below both mediator masses. Flavour-universal couplings give $g_n=g_p$ exactly for the vector and up to small isospin-breaking corrections for the scalar. For the axial vector, $a_p=a_n$, and the nuclear response is proportional to $(a_p\langle S_p\rangle+a_n\langle S_n\rangle)^2$. We use SD-neutron limits for germanium and oxygen and SD-proton limits for hydrogen and aluminium. The subdominant spin contribution has the same sign or is negligible, making the corresponding lower sensitivity edges conservative.

At large cross sections, this argument does not fix the attenuation ceiling: stopping in the overburden is stronger for $a_p=a_n$ than for proton- or neutron-only couplings, lowering the SD upper edges by a factor of a few. Surface runs~\cite{TESSERACT:2025tfw,SuperCDMS:2020aus,EDELWEISS:2019vjv} reach larger cross sections than underground detectors. XQC~\cite{Erickcek:2007jv,Mahdawi:2017cxz} reaches $10^{-19}$~cm$^2$ for $m_\chi\gtrsim10\MeV$, but coherent $A^2$ scaling breaks down around $10^{-28}$--$10^{-27}$~cm$^2$ in the cases studied in Ref.~\cite{Digman:2019wdm}. The largest-cross-section regions are consequently indicative. Where no upper edge is published, we close the displayed region at $10^{-31}$~cm$^2$, following Ref.~\cite{DeMarchi:2025uoo}; this is not a stopping calculation.

For $m_Y=5\GeV$, the DD regions cover our limits above $m_\chi\simeq65$--$80\MeV$ for the vector and $90\MeV$ for the scalar, depending on the background treatment, but not for the axial vector. For $m_Y=100\MeV$, the vector limits lie below XQC and above the upper edges of terrestrial searches. XQC covers the scalar limits above $30$--$50\MeV$, subject to the coherent-scaling caveat, and SD searches reach the axial-vector limits only near $100\MeV$.

The omitted Migdal searches~\cite{SENSEI:2023zdf,SuperCDMS:2023sql,XENON:2019zpr,PandaX:2023xgl,DarkSide:2022dhx,EDELWEISS:2019vjv,EDELWEISS:2022ktt,CDEX:2021cll} use ionisation accompanying a nuclear recoil; bremsstrahlung searches~\cite{XENON:2019zpr,SuperCDMS:2023sql} use photon emission. Subject to validating their atomic-emission predictions in the relevant targets, Migdal limits would extend the SI region to $30\MeV$~\cite{PandaX:2023xgl}, or $4\MeV$ at cross sections above $10^{-31}$~cm$^2$~\cite{SENSEI:2023zdf}, lower it by one to two orders of magnitude up to $1\GeV$, and extend the SD region to $32\MeV$~\cite{CDEX:2021cll}. The experimental qualification is discussed in Sec.~\ref{sec:Disc_Scattering}.

\subsection{Structure-formation comparisons}
\label{sec:AppBounds_Cosmo}

The Lyman-$\alpha$ bound is $\sigma_{\chi p}<6\times10^{-30}$~cm$^2$ at $m_\chi=100$~keV and $1.7\times10^{-29}$~cm$^2$ at $10\MeV$~\cite{Rogers:2021byl}; the satellite bound is weaker by about an order of magnitude~\cite{Maamari:2020aqz,Alvey:2022pad}. For $m_Y=5\GeV$, our vector and axial-vector limits are stronger by more than an order of magnitude. The scalar crossover occurs at $m_\chi\simeq25$--$60\MeV$, depending on the CR background. For $m_Y=100\MeV$ and $m_\chi=100$~keV, Lyman-$\alpha$ is weaker than our vector limit and comparable to the scalar one. At $10\MeV$ it is stronger by factors of $15$--$100$ for the vector and four to five orders of magnitude for the scalar, and comparable to the axial-vector limits. Its applicability assumptions are stated in Sec.~\ref{sec:Disc_Cosmo}.

\subsection{CRDM mediator masses, attenuation and earlier gamma-ray results}
\label{sec:AppBounds_CRDM}

Several searches~\cite{PandaX-II:2021kai,Super-Kamiokande:2022ncz,CDEX:2022fig,LZ:2025iaw} and recasts~\cite{Bringmann:2018cvk,Xia:2021vbz,Maity:2022exk} assume an energy-independent cross section. We instead use interaction-specific calculations: XENON1T for the vector~\cite{Dent:2019krz,Diurba:2024dqo} and scalar~\cite{Bondarenko:2019vrb,Alvey:2022pad}, Super-K and KamLAND for the scalar~\cite{Ema:2020ulo}, IceCube for the vector~\cite{Cappiello:2024acu}, and Borexino for the axial vector~\cite{Dent:2019krz}. Reference~\cite{Yu:2025bwu} provides LZ and Borexino results at fixed $m_\chi=1$, $10$, $100\MeV$ and $1\GeV$ as functions of mediator mass; we do not plot these as continuous DM-mass exclusion curves.

\subsubsection*{Mediator conversion and attenuation}

For comparison with $m_Y=5\GeV$, the published curves either use $m_Y=1$--$3\GeV$~\cite{Diurba:2024dqo,Ema:2020ulo,Bondarenko:2019vrb,Alvey:2022pad}, or vary $m_Y$ at fixed couplings~\cite{Dent:2019krz,Cappiello:2024acu}; in the latter case we consider only the portions with $m_Y\leq5\GeV$. At fixed $\sigma_{\chi p}\propto(g_{\chi Y}g_{pY})^2/m_Y^4$, the propagator gives each differential production or detection cross section a factor
\begin{equation}
    \left[\frac{m_Y^2}{Q^2+m_Y^2}\right]^2
\end{equation}
relative to the contact limit. This grows with $m_Y$, making the lighter-mediator lower sensitivity edges conservative in the regime of negligible attenuation. The same increase in scattering strengthens energy loss in the Earth, so the published upper edges cannot be used as exclusions for $5\GeV$ without a transport calculation. For the $100\MeV$ scalar we use the full matched-mass regions~\cite{Bondarenko:2019vrb,Alvey:2022pad}, converting the effective cross section of Ref.~\cite{Alvey:2022pad} with its Eq.~(5.6).

\subsubsection*{Numerical comparisons}

For the $5\GeV$ vector, the IceCube and XENON1T lower edges lie below our limits for $m_\chi\lesssim10$~keV ($0.3\MeV$ without background) and $m_\chi\gtrsim75\MeV$ ($30\MeV$ without background), respectively. In the latter range our limits lie two to three orders of magnitude below the XENON1T ceilings computed for $m_V\simeq1$--$2\GeV$, suggesting that an exclusion would persist at $5\GeV$, although it has not been recomputed. The IceCube ceilings use mediators lighter than about $0.3\GeV$, and require a dedicated attenuation calculation.

For the $5\GeV$ scalar, the CRDM lower edges are stronger above $m_\chi\simeq0.4\MeV$ ($0.1\MeV$ without background), by two to three orders of magnitude for Super-K above $1\MeV$. Our limits nevertheless exceed the published Super-K and KamLAND ceilings for $m_\chi\gtrsim2$ and $6\MeV$, respectively, and at all masses without background. XENON1T lower edges are up to two orders of magnitude stronger, while our limits remain more than an order of magnitude below its published ceilings for $m_\chi\lesssim30\MeV$ ($10\MeV$ without background). At larger masses the comparison is particularly sensitive to attenuation. These ceilings were computed at $m_\phi=3\GeV$ for Super-K/KamLAND and $1$--$3\GeV$ for XENON1T. For the axial vector the Borexino bound is weaker by one to three orders of magnitude.

\subsubsection*{Blazars and earlier gamma-ray results}

For blazar-boosted DM~\cite{DeMarchi:2025uoo}, the bounds are stronger than ours by two to three orders of magnitude for the vector, two to four for the scalar, and one to three for the axial vector. Production and detection each contribute one power of the elastic cross section, so its bound depends only weakly on the nucleon-coupling conversion. Adding the heavier quarks to the original $u,d$ benchmark increases the DIS contribution by $10$--$30\%$, making the published curves slightly conservative. The larger uncertainty is the DM spike, as discussed in Sec.~\ref{sec:Disc_CRDM}.

At the matching point $m_\chi\simeq33\MeV$, $m_V=100\MeV$, Ref.~\cite{Guo:2020oum} excludes $2.9\times10^{-6}<G_D^2<47$ with XENON1T and MiniBooNE, covering our limits $G_D^2<0.49$ ($3.7$ without background). Its H.E.S.S. GC limit is $G_D^2<2.5$, and its Fermi-LAT GC-excess limit is $G_D^2<0.075$. It assumes CR protons with the local spectrum throughout a cylinder of radius $10$~kpc and half-height $1$~kpc. Our model includes helium, a spatially varying CR spectrum and density, and a diffusion halo several kpc thick. Its Fermi-LAT limit permits a signal up to $1.28$ times the upper systematic edge of the GeV excess, whereas our LHAASO analysis uses TeV--PeV photons and correspondingly higher-energy CRs.

\subsection{Monojet rescaling and coupling dependence}
\label{sec:AppBounds_Monojet}

We start from quark-coupling limits for an on-shell mediator with approximately unit invisible branching ratio. For spin-1 mediators we use the flavour-universal Tevatron/early-LHC recast of Ref.~\cite{Shoemaker:2011vi}, $g_{qY}<0.021$. For spin-0 mediators, Ref.~\cite{Ema:2020ulo} recasts the $139$~fb$^{-1}$ ATLAS search~\cite{ATLAS:2021kxv} as $g_u=g_d<0.06$ at both $m_Y=1$ and $3\GeV$. We apply it to our mediator masses, also well below the recoil thresholds. For $u,d$ couplings only, the published quark-coupling bounds are $0.022$ for spin 1 and $0.06$ for spin 0~\cite{Shoemaker:2011vi,Ema:2020ulo}.

For spin 0, we convert the $u,d$ bound to flavour-universal couplings using the production ratio $R\equiv\sigma_{\rm universal}/\sigma_{u,d}=1.05$, evaluated at equal quark coupling with \textsc{MadGraph5\_aMC@NLO}. Since the production rate is proportional to $g_{qY}^2$, the rescaled bound is $g_{qY}<0.06/\sqrt{1.05}\simeq0.059$. At leading order, $R$ varies from $1.0$ to $1.1$ across recoil thresholds of $200$--$900\GeV$, corresponding to about $2\%$ uncertainty on the rescaled coupling. This estimates only the flavour-rescaling uncertainty. The underlying recast carries about $10\%$ merging-scale uncertainty on the coupling and assumes that the simulation reproduces the experimental acceptance and efficiency~\cite{Ema:2020ulo}; these estimates do not quantify the extrapolation to our mediator masses. Loop-induced $gg\to Yg$ is omitted and is expected to be subleading for couplings independent of quark mass.

In the on-shell, narrow-width approximation, the monojet rate scales as
\begin{equation*}
    \sigma(pp\to j+\mathrm{invisible})\propto g_{qY}^2\,{\rm BR}(Y\to\chi\bar\chi).
\end{equation*}
For $g_{\chi Y}=1$ and approximately unit invisible branching ratio, the quark-coupling bounds therefore give $G_D^2=(g_{\chi Y}g_{qY})^2<4.5\times10^{-4}$ and $3.4\times10^{-3}$ for spin 1 and spin 0, respectively. As long as the invisible branching ratio remains close to one, the bound on $G_D^2$ scales as $g_{\chi Y}^2$. It would reach our vector limits only at $g_{\chi Y}\gtrsim7$ and $23$ for $m_V=100\MeV$ and $5\GeV$, and the scalar limits at $g_{\chi Y}\gtrsim13$ and $37$, beyond $\sqrt{4\pi}$.

When the quark coupling is comparable to or larger than the DM coupling, visible mediator decays can reduce the invisible branching ratio. In a partonic-width approximation, neglecting final-state mass effects and additional decay channels, ${\rm BR}(Y\to\chi\bar\chi)\simeq g_{\chi Y}^2/(g_{\chi Y}^2+3N_f g_{qY}^2)$, so the on-shell rate scales as
\begin{equation}
    \sigma(pp\to j+\mathrm{invisible})\propto\frac{G_D^2}{g_{\chi Y}^2+3N_f g_{qY}^2},
\end{equation}
where $N_f$ counts open quark flavours and the factor of three counts colour. Requiring both couplings below $\sqrt{4\pi}$ bounds the denominator by $4\pi(1+3N_f)$, which limits the weakening relative to the $g_{\chi Y}=1$, unit-branching-ratio conversion. With $N_f=4$ at $5\GeV$, the corresponding bounds are $G_D^2\lesssim0.07$ and $0.6$ for spin 1 and spin 0. Below hadronic thresholds at $100\MeV$, taking $N_f=0$ gives $6\times10^{-3}$ and $0.04$. These remain below our limits within this on-shell approximation. The plotted curves end at $m_\chi=m_Y/2$; above this threshold an off-shell recast is required.

\subsection{Meson-decay conversions}
\label{sec:AppBounds_Mesons}

For off-shell vector exchange, normalising the invisible width to the electromagnetic dilepton width removes the vector-meson decay constant. The experimental inputs are ${\rm BR}(\Upsilon(1S)\to{\rm invisible})<3.0\times10^{-4}$~\cite{BaBar:2009gco}, ${\rm BR}(\omega\to{\rm invisible})<7.3\times10^{-5}$ and ${\rm BR}(\phi\to{\rm invisible})<1.7\times10^{-4}$~\cite{BESIII:2018bec}. The conversions quoted in the Discussion are
\begin{center}
\begin{tabular}{lcc}
\hline
Decay & $G_D^2$ upper bound, $m_V=5\GeV$ & $m_V=100\MeV$ \\
\hline
$\Upsilon(1S)\to\chi\bar\chi$ & $5.9\times10^{-6}$ & $1.1\times10^{-5}$ \\
$\omega\to\chi\bar\chi$ & $0.37$ & $2.2\times10^{-4}$ \\
$\phi\to\chi\bar\chi$ & $0.28$ & $5.2\times10^{-4}$ \\
\hline
\end{tabular}
\end{center}
The $J/\psi$ bound~\cite{BES:2007sxr} is weaker than the $\Upsilon$ one by a factor of $4$--$20$. For the light mediator, the $\omega,\phi$ bounds improve on our limits by factors of roughly $40$--$6000$. Radiative scalar/pseudoscalar $\Upsilon$ bounds~\cite{BaBar:2010eww}, with unit invisible branching ratio, would reach our limits only at $g_{\chi Y}\sim10^2$--$10^3$.

For axial-vector exchange, the pseudoscalar-meson decay rate scales as $G_D^2m_\chi^2/m_{V'}^4$. Away from the endpoint and for $m_\chi\ll m_p$, its bound on $\sigma^{\rm SD}_{\chi p}$ is therefore independent of $m_\chi$ and $m_{V'}$. The NA62 input ${\rm BR}(\pi^0\to{\rm invisible})<4.4\times10^{-9}$~\cite{NA62:2020pwi} gives $\sigma^{\rm SD}_{\chi p}<2.4\times10^{-38}$~cm$^2$, with an endpoint at $67.5\MeV$. The BESIII $\eta$ and $\eta'$ limits~\cite{BESIII:2012nen} give $5\times10^{-35}$ and $6\times10^{-34}$~cm$^2$. At $m_{V'}=5\GeV$, our limits are stronger below $m_\chi\simeq0.7$, $15$ and $35\MeV$, respectively. At $100\MeV$, only the $\eta,\eta'$ comparisons favour our limits, below $0.07$ and $0.2\MeV$, respectively.

For pseudoscalar exchange with flavour-universal couplings and physical meson mixing~\cite{Ema:2020ulo,Feldmann:1998vh}, the $\pi^0,\eta,\eta'$ bounds at $m_a=5\GeV$ are $G_D^2<1.3\times10^{-7}$, $2.4\times10^{-4}$ and $3.5\times10^{-3}$, respectively. At $m_a=100\MeV$, the proximity to the pion mass enhances mixing and the $\pi^0$ bound reaches $G_D^2<1.4\times10^{-14}$.

For the scalar, Ref.~\cite{Cox:2024rew} computes $K^+\to\pi^+\chi\bar\chi$ for contact interactions with gluons or mass-weighted quark scalar densities. Reference~\cite{DeMarchi:2025uoo} applies the gluon-coupled result to a $u,d$-coupled scalar, finding a bound six to seven orders of magnitude below ours at $5\GeV$. The decay amplitude and nucleon scattering involve different flavour combinations; a particular $u/d$ coupling ratio can evade that conversion. Our universal benchmark also includes a top-loop $s\to d$ contribution, known to dominate for mass-proportional couplings~\cite{Cox:2024rew}. Both the off-shell decay and on-shell $K^+\to\pi^+\phi$ at $m_\phi=100\MeV$ therefore require a dedicated calculation before applying the NA62 measurements~\cite{NA62:2021zjw}.

\subsection{Perturbativity, unitarity and completion dependence}
\label{sec:AppBounds_Theory}

The scalar/pseudoscalar limits exceed $G_D^2=16\pi^2$ for $m_Y=5\GeV$ above $m_\chi\simeq35\MeV$ with background and $2\MeV$ without it. The $5\GeV$ background-free limits also exceed the boundary below approximately $2$~keV. At $m_Y=100\MeV$ they exceed this boundary only without background, above $15\MeV$.

Axial-vector perturbative unitarity requires $m_f\lesssim\sqrt{\pi/2}\,m_{V'}/g_{fV'}$ for every coupled fermion~\cite{Kahlhoefer:2015bea}. The DM condition is satisfied throughout our mass range at $5\GeV$ unless $g_{\chi V'}\gtrsim60$; at $100\MeV$ and $m_\chi\simeq100\MeV$ it requires $g_{\chi V'}\lesssim1.3$. With universal quark couplings, the top condition gives $g_{qV'}\lesssim0.036$ and $7\times10^{-4}$ at mediator masses $5\GeV$ and $100\MeV$. Combining these with $g_{\chi V'}<\sqrt{4\pi}$ gives $G_D^2\lesssim0.017$ and $7\times10^{-6}$. Removing the top coupling leaves bottom-quark bounds $g_{bV'}\lesssim1.5$ and $0.03$, or $G_D^2\lesssim28$ and $0.01$. The former permits all our background-included limits, and the background-free ones for $1.3~{\rm keV}\lesssim m_\chi\lesssim27\MeV$; the latter remains below our sensitivity.

Further constraints depend on how the simplified interactions are embedded in a complete theory~\cite{Knapen:2017xzo,DeMarchi:2025uoo}. Our flavour-universal vector couples to the SM baryon-number current, $J_B^\mu=\frac13\sum_q\bar q\gamma^\mu q$, with coupling $g_B=3g_{qV}$. In gauge completions where the electroweak anomalies are cancelled by heavy fermions whose masses preserve the SM gauge symmetry, integrating out those fermions leaves effective Wess--Zumino interactions. These induce longitudinal-vector production in rare $Z$ and meson decays with rates enhanced by $(E/m_V)^2$~\cite{Dror:2017ehi,Dror:2017nsg}. The resulting bounds depend on the mediator decay modes and can change if the heavy-fermion masses receive substantial electroweak-breaking contributions.

Axial-vector completions can generate additional interactions. Mixing with vector-like quarks generically produces vector and axial couplings together~\cite{Kahn:2016vjr,DeMarchi:2025uoo}. If axial couplings instead arise from chiral gauge charges and renormalizable quark Yukawa interactions, the Higgs doublets generating the quark masses must carry appropriate charges under the new symmetry. Their vacuum expectation values generally induce mediator--$Z$ mass mixing, modifying the physical couplings and introducing electroweak precision constraints~\cite{Kahlhoefer:2015bea}. A single Higgs doublet with ordinary Yukawa interactions cannot reproduce our equal, same-sign axial couplings to up- and down-type quarks; a larger Higgs sector or a different origin of the effective couplings is required~\cite{Kahn:2016vjr}. The induced vector couplings and mixing therefore require a specified completion to quantify.

For SM-singlet spin-0 mediators, the quark bilinears in Eqs.~\eqref{eq:Lphi} and~\eqref{eq:La} connect electroweak doublets to singlets and require additional Higgs-dependent structure. Dimension-five operators such as $\phi\bar Q_L\widetilde H u_R/\Lambda$ and $\phi\bar Q_L H d_R/\Lambda$, together with their Hermitian conjugates, give $g_{q\phi}=c_qv/(\sqrt{2}\Lambda)$ after electroweak symmetry breaking; analogous operators generate pseudoscalar couplings. Here $c_q$ is a dimensionless coefficient and $\Lambda$ is the scale of the additional physics. These operators also contain interactions involving the physical Higgs. Simple Higgs mixing instead gives mass-proportional scalar couplings and does not reproduce our flavour-universal benchmark. Naturalness criteria and rare-meson bounds depend on the heavy states and flavour structure, so numerical limits derived for flavour-specific scalar models cannot be transferred directly to our benchmark~\cite{Batell:2017kty,Batell:2018fqo,Cox:2024rew}.

\section{Additional signal and limit plots}
\label{sec:AppA}

\begin{figure*}[h!] 
    \centering 
    \includegraphics[width=0.49\textwidth]{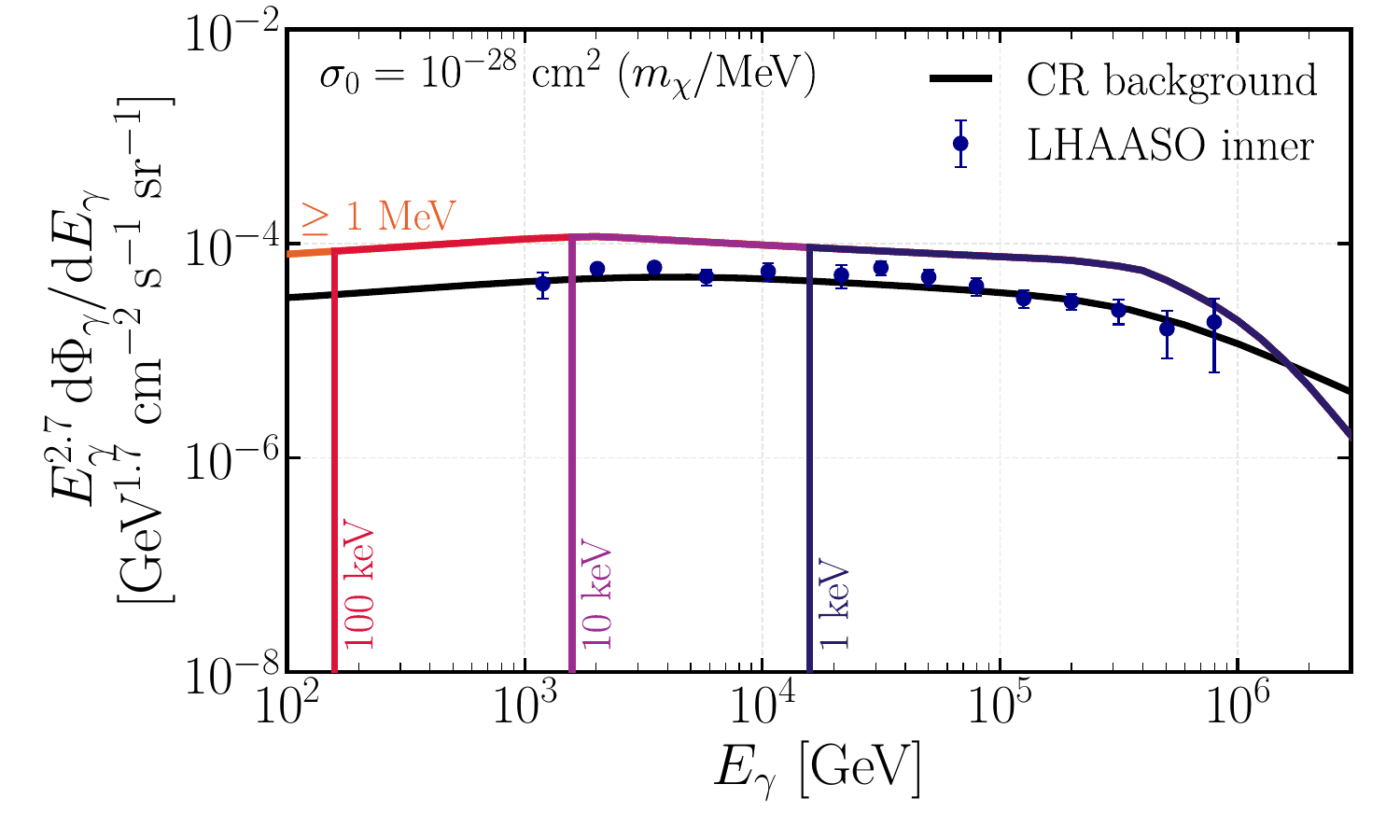}
    \includegraphics[width=0.49\textwidth]{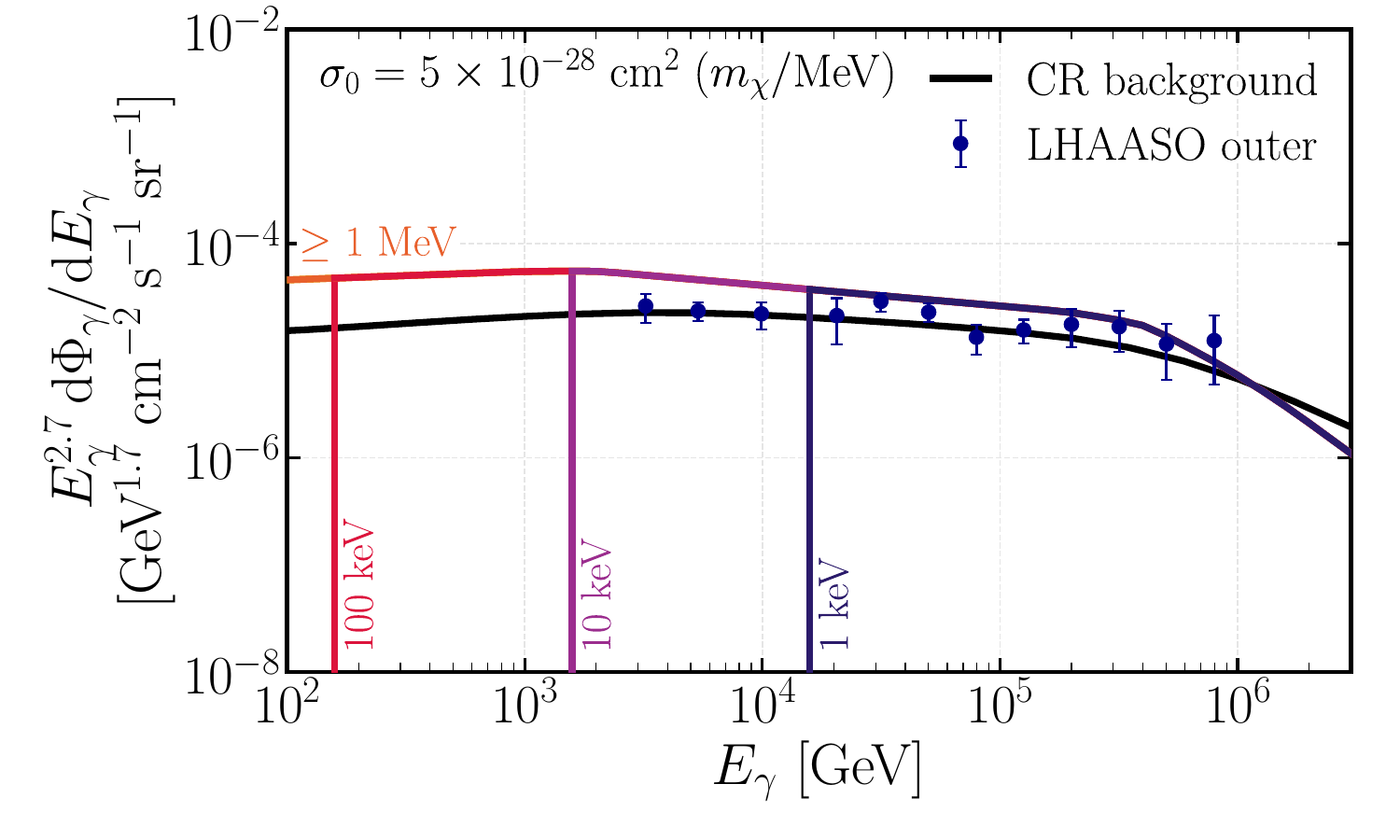}
    \includegraphics[width=0.49\textwidth]{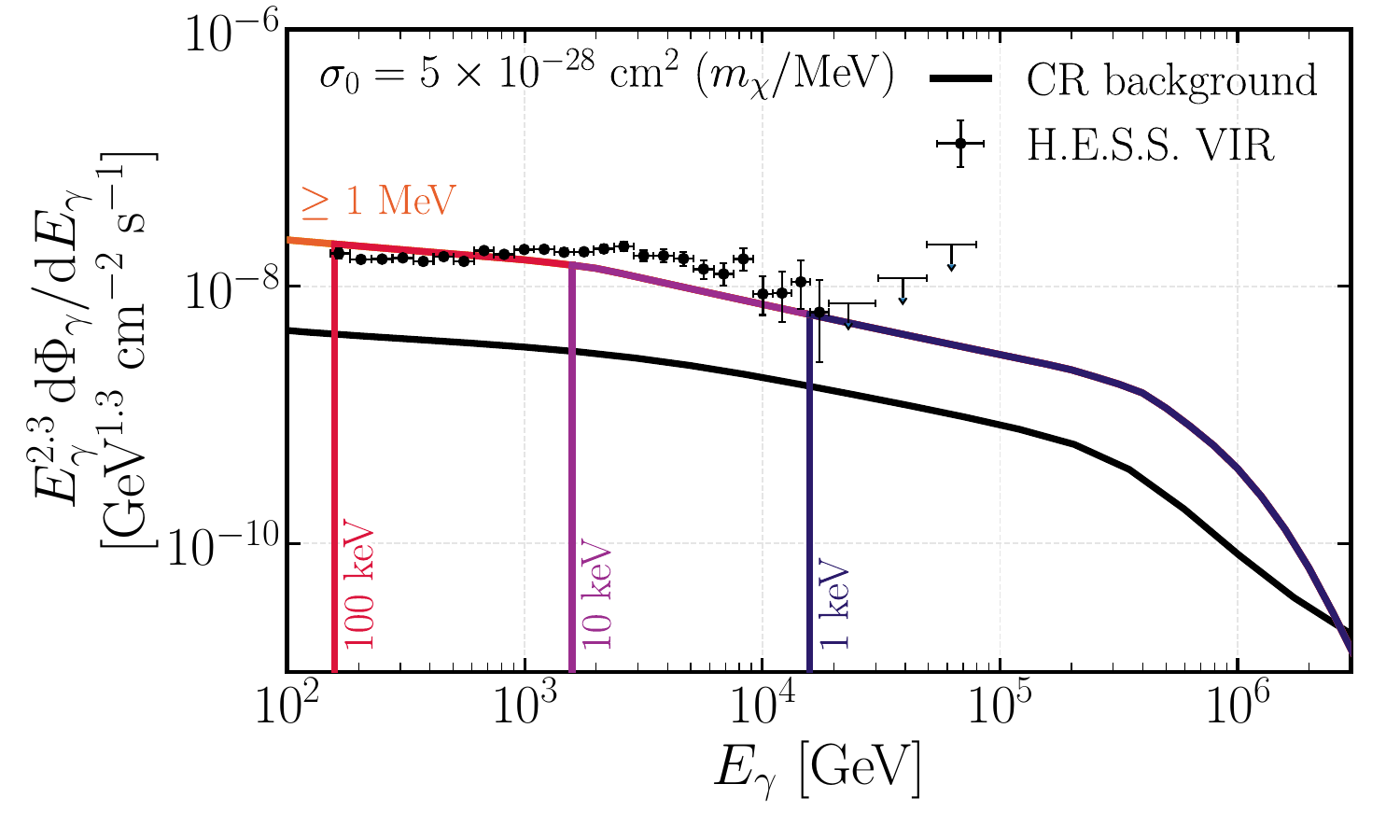}
    \includegraphics[width=0.49\textwidth]{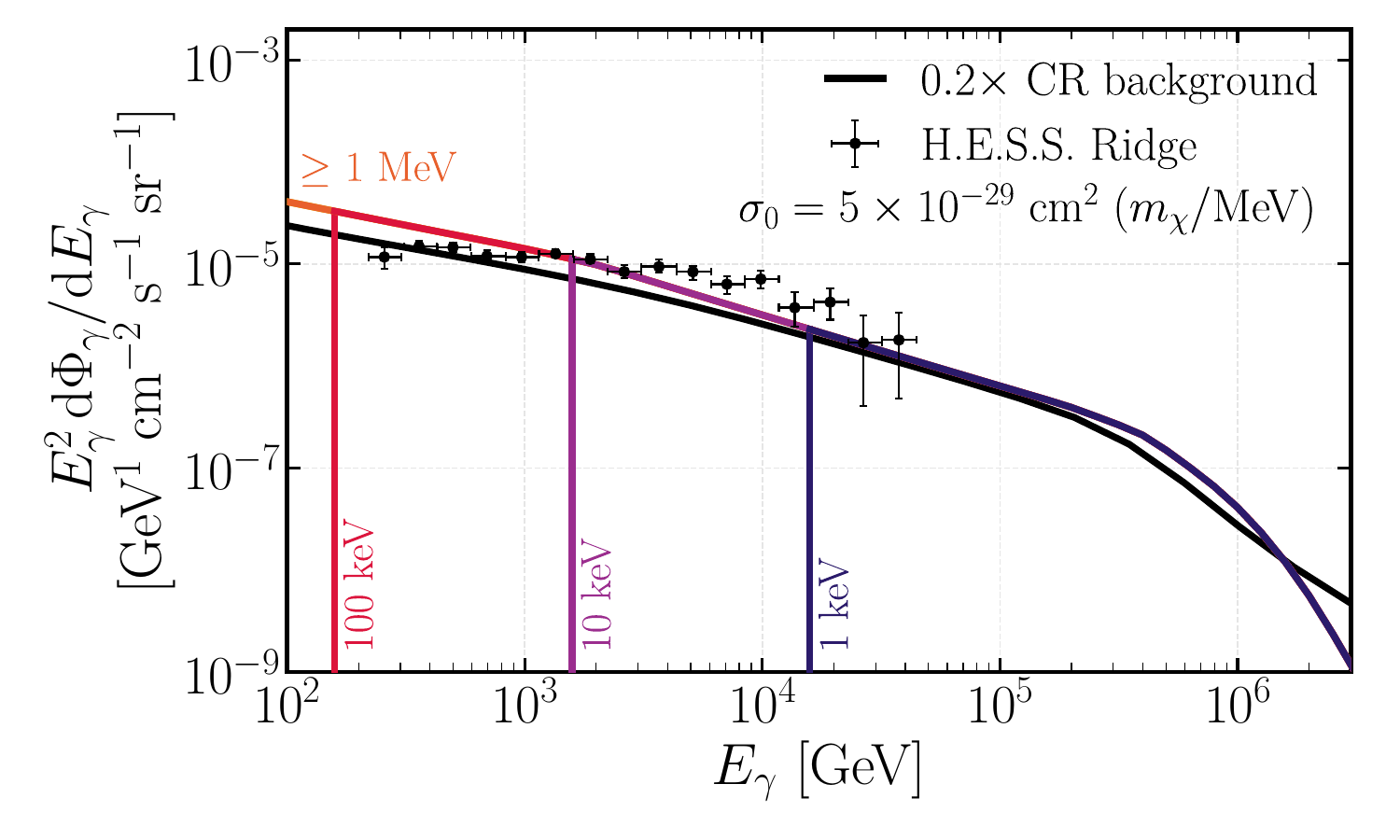}
    \includegraphics[width=0.49\textwidth]{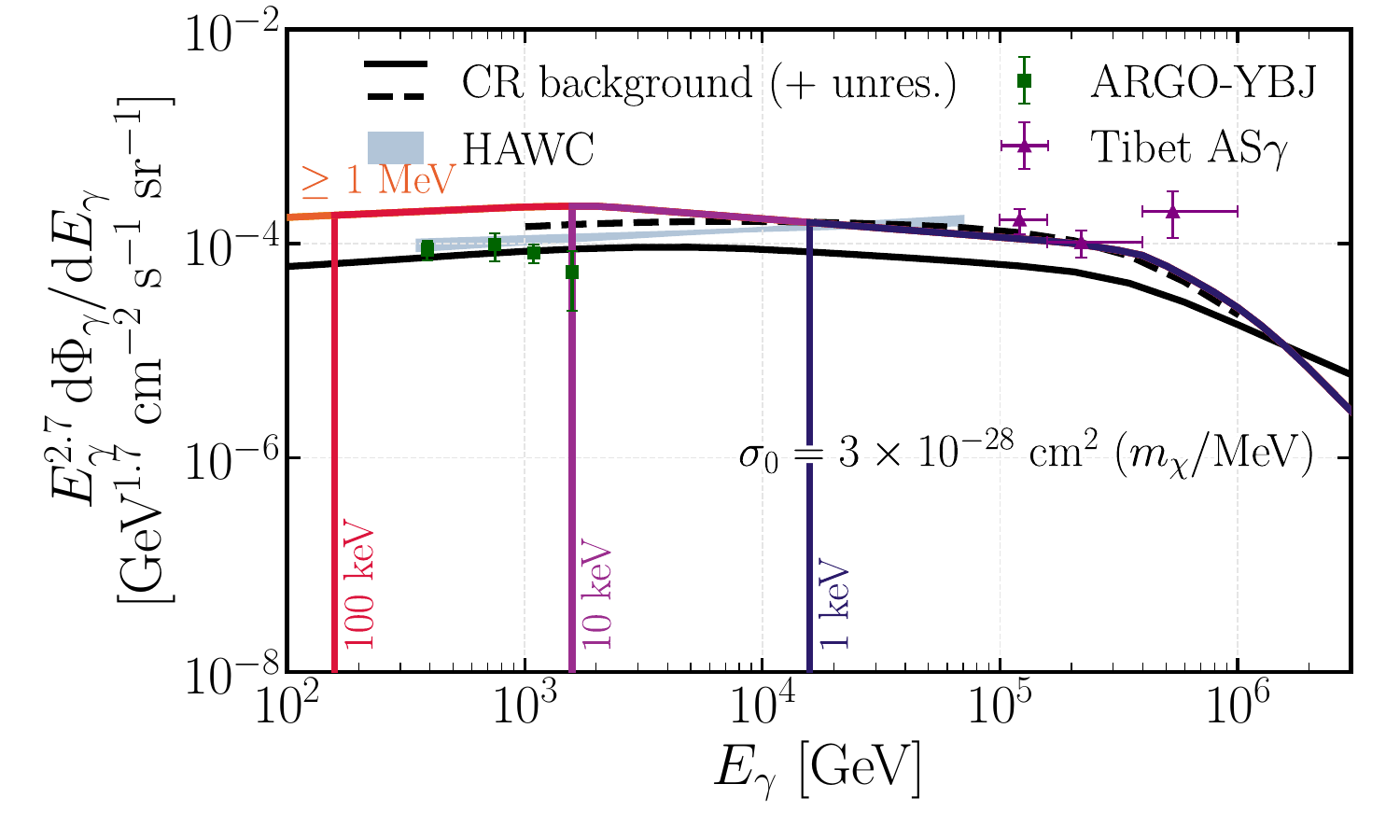}
    \caption{Gamma-ray spectra from CR--DM scattering in the constant-cross-section prescription, compared with LHAASO inner/outer Galaxy (top), H.E.S.S.\ VIR/Ridge (middle) and HAWC (bottom) data. The cross-section normalisation is specified in each panel. The fixed-energy-fraction threshold and elastic-scattering convention are defined in Sec.~\ref{sec:ConstantXS}.}
\label{fig:Observables_ConstantXS} 
\end{figure*}

\begin{figure*}[h!] 
    \centering 
    \includegraphics[width=0.49\textwidth]{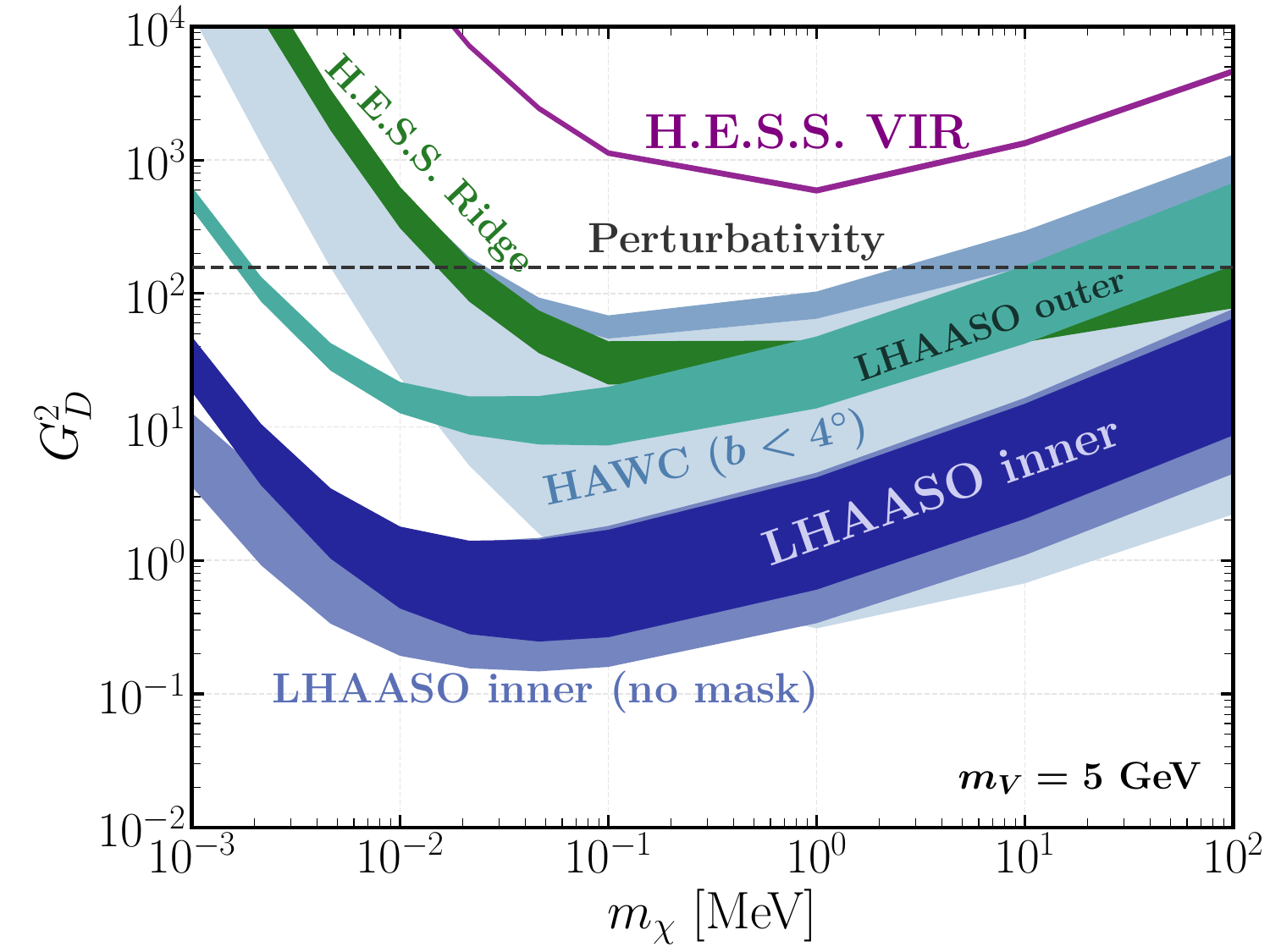}
    \includegraphics[width=0.49\textwidth]{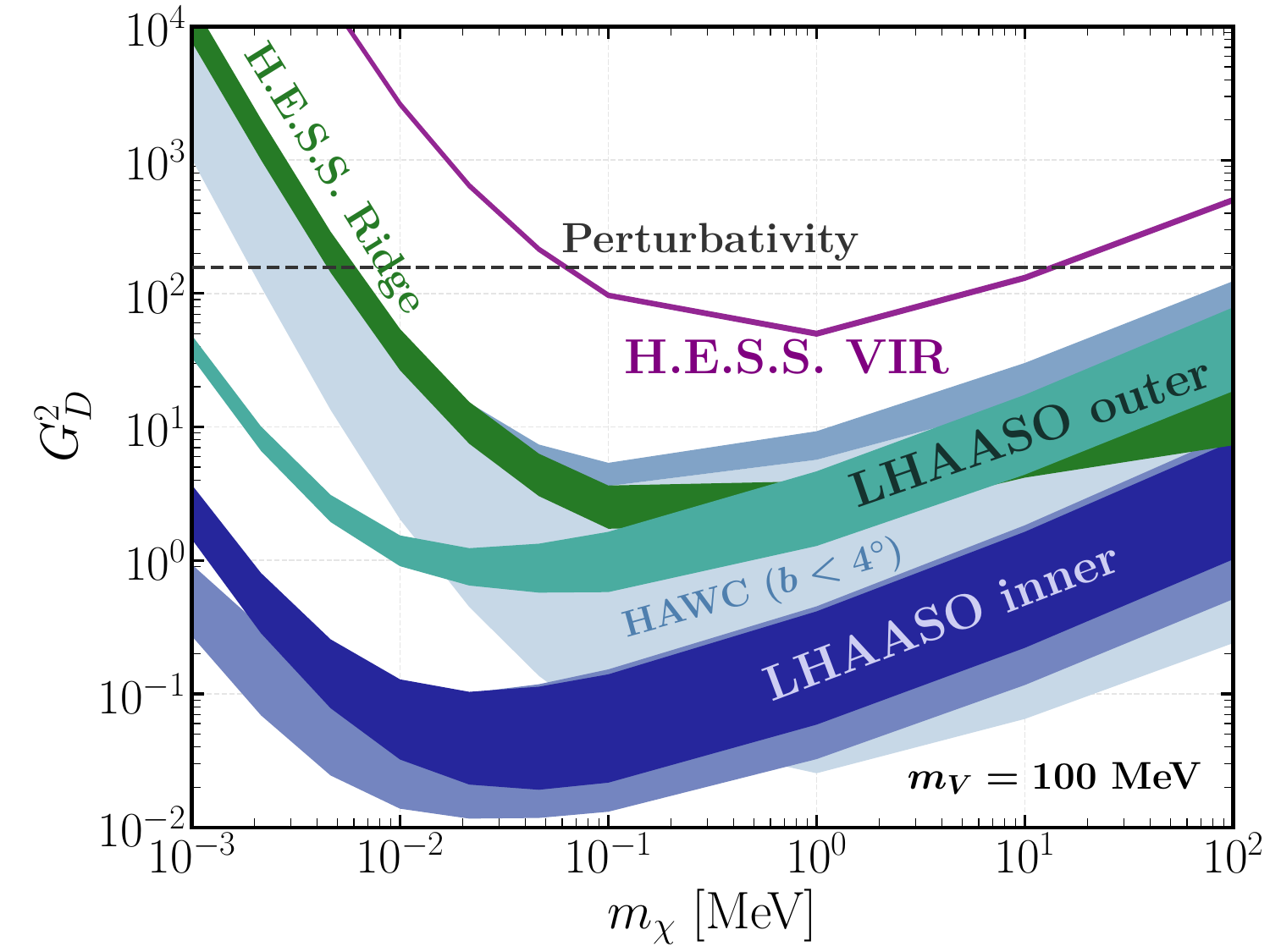}
    \caption{Upper limits on the coupling parameter $G_D^2$ for the vector-mediator DM model with $m_V=5\GeV$ (\textit{left}) and $m_V=100\MeV$ (\textit{right}), derived from the different diffuse Galactic gamma-ray observations considered in this work. The constraints are shown as a function of the DM mass and illustrate the relative constraining power of LHAASO, H.E.S.S. and HAWC. The upper edge of each band is obtained without astrophysical background and the lower edge including the CR background of the $\gamma$-optimized (Min) model; for HAWC, the light band extends down to the limit with the CR background plus unresolved sources. Dashed line: perturbativity limit, $G_D^2=16\pi^2$.}
\label{fig:Observables_G2D} 
\end{figure*}

Figure~\ref{fig:Observables_ConstantXS} gives the constant-cross-section spectra corresponding to the observational comparison in Sec.~\ref{sec:Results_Observations}. Figure~\ref{fig:Observables_G2D} shows the vector limits from each data set directly in $G_D^2$ for both mediator masses, before the elastic-scattering conversion. The background conventions are those of Sec.~\ref{sec:LimitProcedure}.

\section{Uncertainties from astrophysical assumptions}
\label{sec:AppB}

The predicted gamma-ray emission from CR--DM interactions depends on the spatial distributions of both Galactic CRs and DM. In this appendix, we quantify the impact of these astrophysical ingredients by recomputing the limits on the vector-mediator model, for $m_V=5\GeV$ and $100\MeV$, with alternative DM density profiles and CR propagation scenarios. We consider the two regions that bracket the sensitivity to the inner Galaxy: the LHAASO inner Galaxy, the region giving our strongest limits, and the H.E.S.S. Ridge, centred on the GC. The line-of-sight integration, the source masks and the gamma-ray absorption are the same as for the benchmark. Since the conventional diffuse emission used as background is itself a prediction of the CR model, we compare background-free limits throughout.

\subsection{DM density profiles}

We compare NFW with the Burkert~\cite{Burkert:1995yz}, Moore~\cite{Moore:1999nt} and Einasto~\cite{Einasto:1965czb} profiles,
\begin{equation}
\begin{aligned}
    \rho_{\rm Bur}(r)&\propto\frac{1}{(1+r/r_s)\left[1+(r/r_s)^2\right]},\qquad
    \rho_{\rm Moo}(r)\propto\frac{(r_s/r)^{1.16}}{(1+r/r_s)^{1.84}},\\
    \rho_{\rm Ein}(r)&\propto\exp\left\{-\frac{2}{\alpha}\left[\left(\frac{r}{r_s}\right)^{\alpha}-1\right]\right\},
\end{aligned}
\end{equation}
with the shapes of Fig.~1 of Ref.~\cite{CirelliPPPC}: $r_s=12.67$, $30.28$ and $28.44$~kpc, respectively, and $\alpha=0.17$. All profiles are normalised to the local density of our benchmark, $\rho(8.3~{\rm kpc})=0.4\GeV\,{\rm cm}^{-3}$, so that the comparison isolates the effect of the inner halo shape. Burkert, with a constant-density core, and Moore, with a cusp steeper than NFW, bracket the plausible inner behaviour.

\begin{figure*}[h!]
    \centering
    \includegraphics[width=0.49\textwidth]{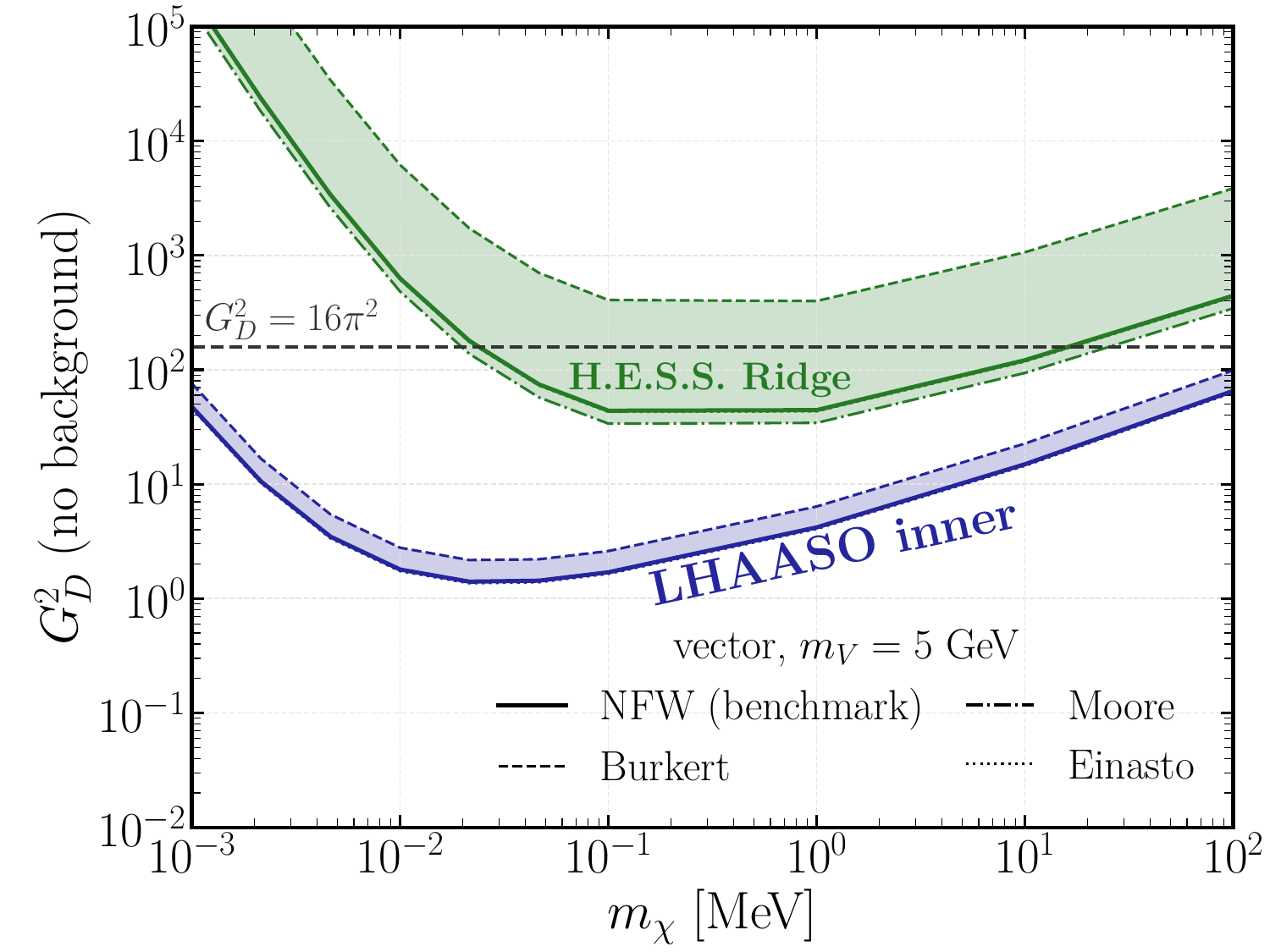}
    \includegraphics[width=0.49\textwidth]{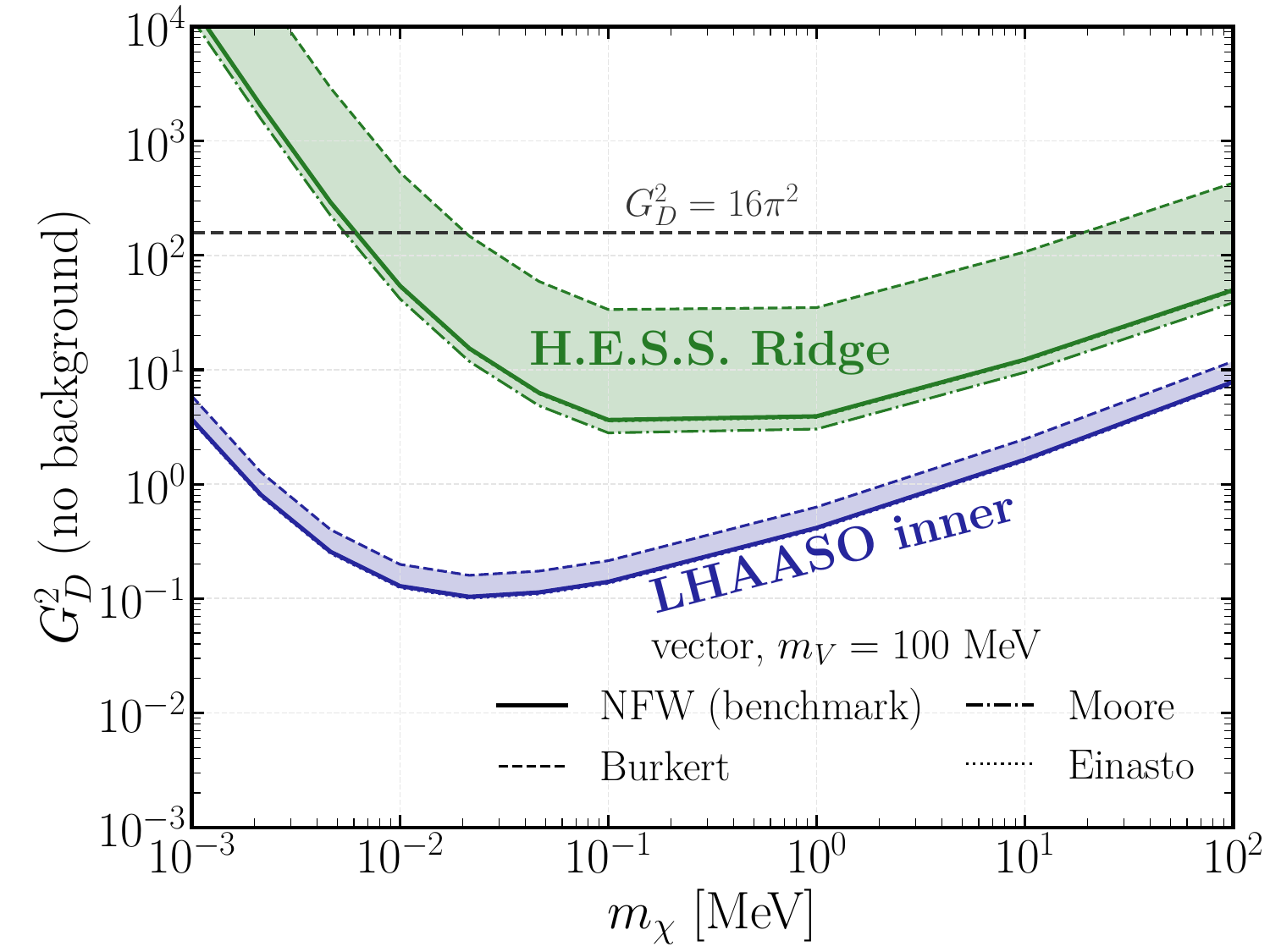}
    \caption{Background-free upper limits on $G_D^2$ for the vector-mediator model with $m_V=5\GeV$ (\textit{left}) and $m_V=100\MeV$ (\textit{right}) from the LHAASO inner Galaxy (blue) and the H.E.S.S. Ridge (green), for different DM density profiles and the $\gamma$-optimized Min CR model. Thick solid lines: NFW benchmark; dashed: Burkert; dash-dotted: Moore; dotted: Einasto (indistinguishable from NFW). The shaded bands span all profiles. Horizontal dashed line: perturbativity limit, $G_D^2=16\pi^2$.}
\label{fig:DM_uncerts}
\end{figure*}

Figure~\ref{fig:DM_uncerts} shows the resulting limits, with the CR distribution fixed to the $\gamma$-optimized Min model. For the LHAASO inner Galaxy, the Burkert profile weakens the limits by a factor of 1.5--1.6, while the Moore and Einasto profiles change them by less than $3\%$. Since the region starts at $|l|=15^\circ$, it probes Galactocentric radii $r\gtrsim2$~kpc, where profiles with the same local density differ little; the steeper Moore cusp only matters within $\sim1$~kpc of the GC. The H.E.S.S. Ridge, instead, is dominated by the innermost hundreds of parsecs: the Burkert core weakens its limits by a factor of 9--11, and the Moore profile strengthens them by a factor of 1.3. The Einasto profile stays within $3\%$ of NFW in both regions. These ratios depend only weakly on the DM mass and are the same for both mediator masses, since the DM profile mostly rescales the line-of-sight integral of the signal.

\subsection{CR propagation models}

\begin{figure*}[h!]
    \centering
    \includegraphics[width=0.49\textwidth]{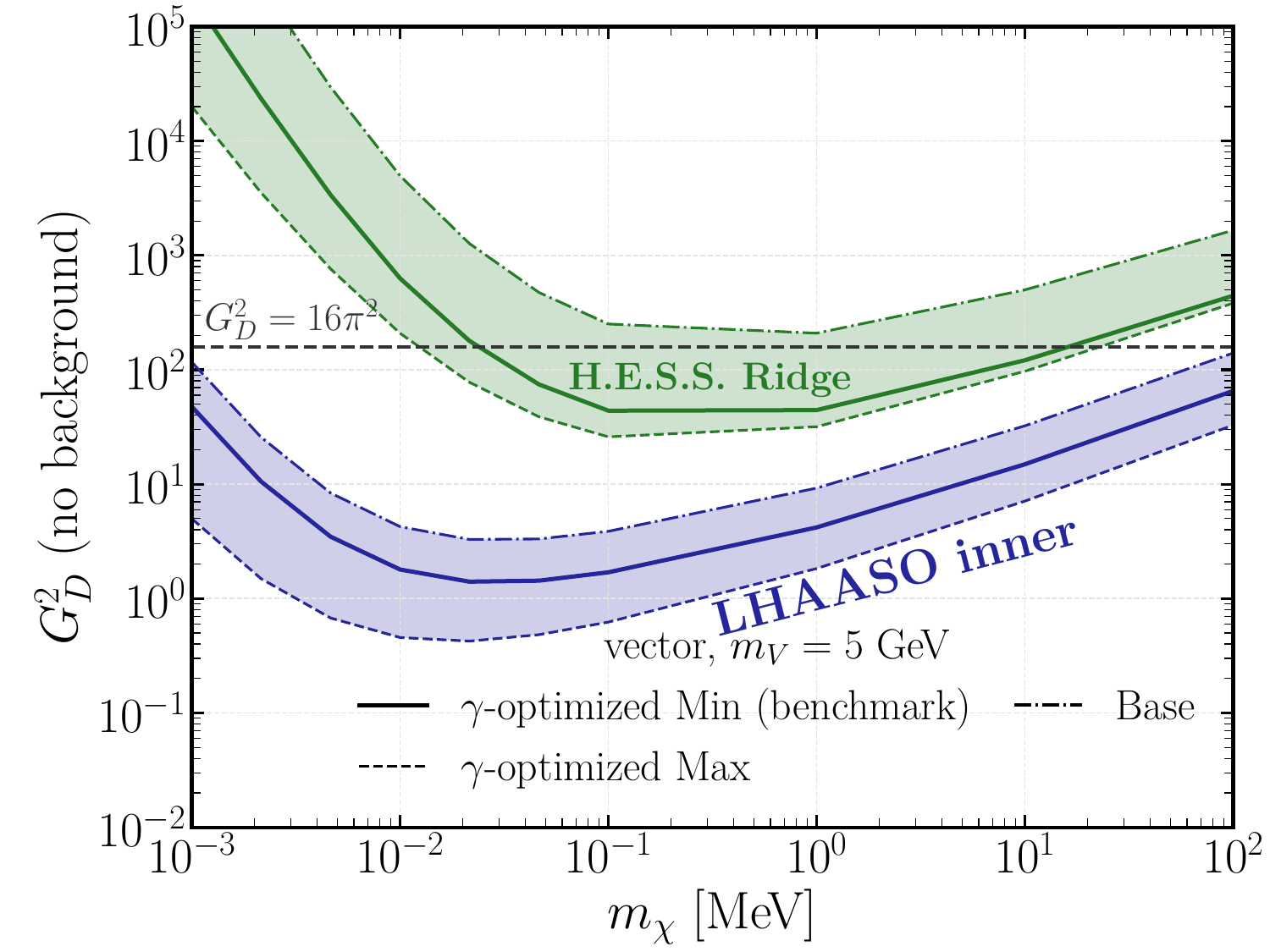}
    \includegraphics[width=0.49\textwidth]{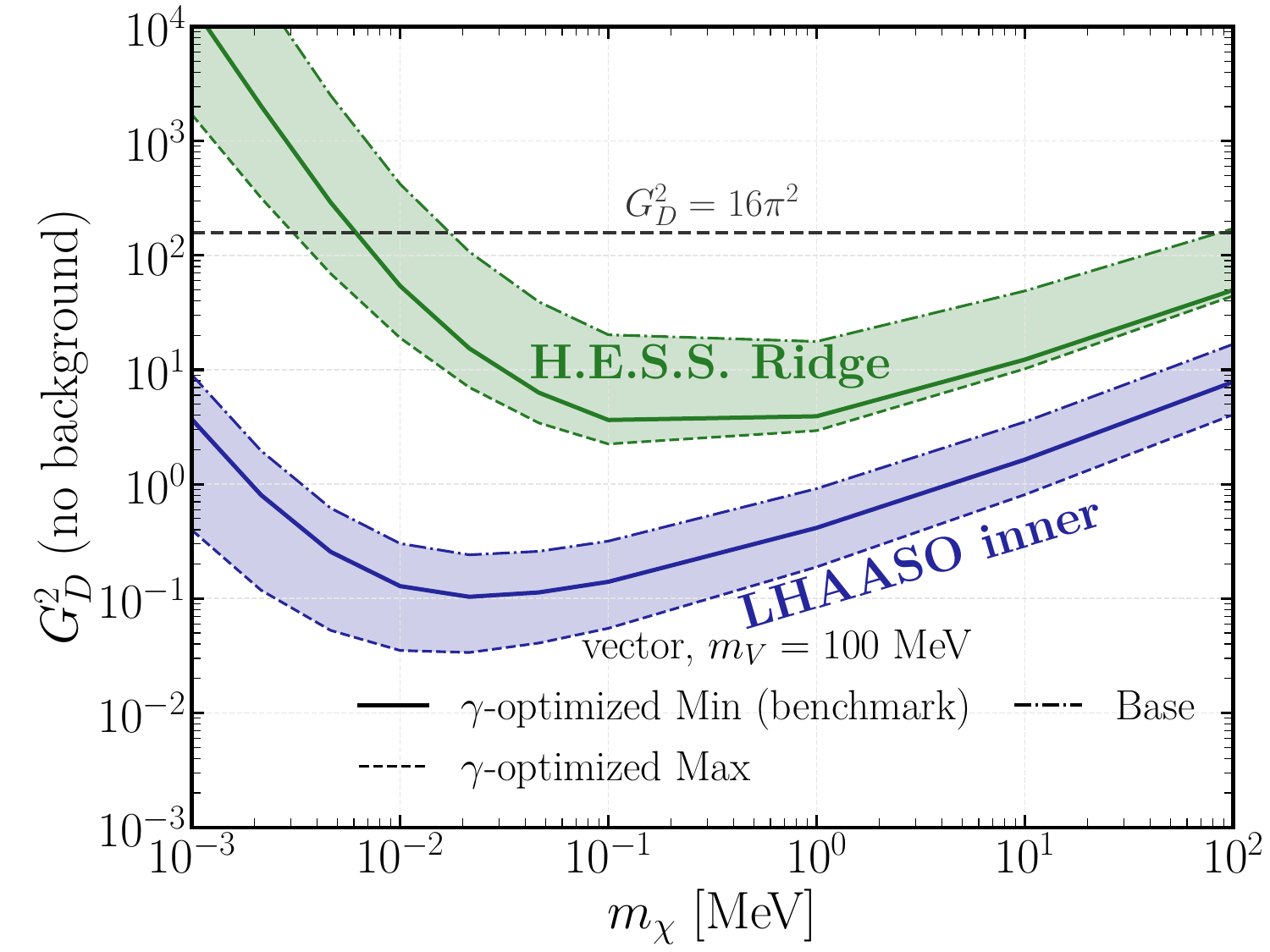}
    \caption{Same as Fig.~\ref{fig:DM_uncerts}, for different CR propagation models and the NFW profile. Thick solid lines: $\gamma$-optimized Min model (benchmark); dashed: $\gamma$-optimized Max model; dash-dotted: Base model.}
\label{fig:CR_uncerts}
\end{figure*}

Figure~\ref{fig:CR_uncerts} shows the limits obtained when the $\gamma$-optimized Min model is replaced by the $\gamma$-optimized Max and Base models, with the NFW profile fixed. Here the impact depends on the DM mass. Through the threshold of Eq.~\eqref{eq:Ethr}, light DM is only probed by the most energetic CRs, where the models differ most. The Max model, with a higher CR flux at PeV energies, strengthens the LHAASO inner-Galaxy limits by a factor of 9.5 at $m_\chi=1$~keV, decreasing to 2.0 at $100\MeV$; for the H.E.S.S. Ridge the gain is similar for the lightest DM and decreases to $\sim1.2$ at $100\MeV$. The Base model, with spatially uniform diffusion, lacks the hardening of the CR spectrum towards the inner Galaxy of the $\gamma$-optimized models, and therefore predicts fewer high-energy CRs where the DM density is largest. It weakens the LHAASO inner-Galaxy limits by a factor of 2.1--2.5 over the whole mass range, and the H.E.S.S. Ridge limits by a factor of 11 at $m_\chi=1$~keV, decreasing to 3.7 at $100\MeV$.

Overall, the LHAASO inner-Galaxy limits change by at most a factor of $\sim3$ for $m_\chi\gtrsim0.1\MeV$ under all the variations considered, and by up to a factor of $\sim10$ for the lightest DM, which is driven by the uncertain CR flux around the knee. The H.E.S.S. Ridge limits are more uncertain, by up to an order of magnitude, due to both the inner DM profile and the CR distribution near the GC. These uncertainties do not change the main conclusions of our analysis, and they could be reduced with improved measurements of the CR spectrum at PeV energies and better models of CR transport and of the DM distribution in the inner Galaxy.

\end{document}